\documentclass[a4paper,11pt]{article}
\usepackage{blindtext,titlefoot}

\usepackage{graphicx} 
\usepackage{amsthm,amsmath}
\usepackage[numbers]{natbib}
\usepackage{blindtext,titlefoot}
\usepackage{bbm}
\usepackage{amsfonts,amsthm,amssymb,amsmath}
\usepackage[title,titletoc,toc]{appendix}
\usepackage[utf8]{inputenc}
\usepackage[affil-it]{authblk}
\usepackage{xcolor}
\usepackage{tikz}
\usepackage{mathtools}
\usepackage{natbib}
\usepackage{hyperref}
\usepackage{mathrsfs}
\usepackage{enumitem}
\usepackage{comment}

\makeatletter
\newcommand*\bigcdot{\mathpalette\bigcdot@{.5}}
\newcommand*\bigcdot@[2]{\mathbin{\vcenter{\hbox{\scalebox{#2}{$\m@th#1\bullet$}}}}}

\makeatother
\newcommand{\is}{\bigcdot }

\numberwithin{equation}{section}

\DeclareMathOperator{\essinf}{ess\;inf}
\DeclareMathOperator{\esssup}{ess\;sup}

\def \Lbrack {[\![}
\def \Rbrack {]\!]}

\newtheorem{theorem}{Theorem}[section]
\newtheorem{assum}[theorem]{Assumptions}
\newtheorem{proposition}[theorem]{Proposition}
\newtheorem{remark}[theorem]{Remark}

\newtheorem{lemma}[theorem]{Lemma}

\newtheorem{corollary}[theorem]{Corollary}
\newtheorem{definition}[theorem]{Definition}

\def \Lbrack {[\![}
\def \Rbrack {]\!]}

\hypersetup{
    colorlinks=true,
    linkcolor=blue,
    filecolor=magenta,
    urlcolor=cyan,
    citecolor=red
}

\begin{document}

\title{Novel Esscher concepts for informational risks and valuation in market models under random horizon\footnote{This research is fully supported by NSERC (through grant NSERC RGPIN-2025-05129-Choulli)}}



\author{Tahir Choulli and Ella Elazkany \\ 
Mathematical and Statistical Sciences Dept.\\
University of Alberta, Edmonton, Canada}




\maketitle\unmarkedfntext{
Address correspondence to Tahir Choulli, Department of Mathematical and Statistical Sciences, University of Alberta, 632 Central Academic Building,
Edmonton, Canada; e-mail: tchoulli@ualberta.ca}

\begin{abstract} This paper considers the setting constituted by the triplet $(S,\mathbb{F},\tau)$. The pair $(S,\mathbb{F})$ is the initial model where $S$ is the discounted price process of risky assets (Stocks) and $\mathbb{F}$ is the ``public" flow of information that is available to all agents over time. $\tau$ is an arbitrary random time, which might not be seen through the flow $\mathbb{F}$, and it represents a default time in credit risk and a death time of an insured in life insurance where mortality and/or longevity risks present serious challenges.  For simplicity, we suppose $S$ follows a geometric jump-diffusion model with stochastic volatiles and $\mathbb{F}$ is generated by a Brownian motion and Poisson process. In this setting, we address how the Esscher densities and/or measures notions can evolves with the informational risk induced by $\tau$. In fact, we introduce the notion of Esscher with degrees-of-freedom associated to the informational risks generated by $\tau$. On the one hand, we give various and complete characterizations of these Esscher notions in  our informational setting. On the other hand, we fully characterize the Escher-prices bounds, for vulnerable claims (claims built on the occurrence of $\tau$), using constrained backward stochastic differential equations (CBSDEs for short hereafter) that appear to be novel as well.  
  \end{abstract}

\section{Introduction}

Throughout the paper, we suppose given the triplet $(S,\mathbb{F},\tau)$, defined on a complete probability space $(\Omega,{\mathcal{G}},P)$. Here $\mathbb{F}$ is the filtration --which is available to all agents over time-- satisfies the usual condition of completeness and right continuity, while $S$ is the discounted assets' prices. $\tau$ is an arbitrary random time, which can model the death time of an insured or the default time of a firm, and hence it might not be observable through $\mathbb{F}$. To incorporate the information in $\tau$ and $\mathbb{F}$, we consider the new flow of information, denoted throughout the paper by $\mathbb{G}$ and which results from the progressive enlargement of $\mathbb{F}$ with $\tau$. Thus, for the new informational model $(S^{\tau}, \mathbb{G},P)$, we address the evaluation problem for random-horizon-linked claims. Besides bearing financial risks, these claims depend heavily on when $\tau$ occurs compared to the finite investment horizon $T$ ($T<\infty$), and are defined via the recovery policies. As a results these vulnerable claims bear various  informational risks associated to the stochasticity of the random horizon $\tau$. For these risks, their decomposition and their classifications, we refer the reader to \cite{choulli2020martingale} for full details. The impact of these risks on deflators is addressed in \cite{choulli2022explicit}.  In this informational setting, we address the problem of evaluation of these vulnerable claims. Since there are many pricing methods, herein we focus on pricing using  the Esscher pricing concept. \\

The Esscher transform,which is also statistically known as the {\it exponential tilting}, was suggested by the Swedish actuary Esscher in \cite{Esscher}.  Then the idea of changing the probability measure to obtain a consistent positive linear pricing rule, which appeared in the actuarial literature in the context of equilibrium reinsurance markets in  \cite{{buhlmann1980economic}}, gave to the Esscher transform an important role in actuarial sciences. Afterwards, the Esscher transform was extended by Gerber and Shiu, in  \cite{gerber1994b}, to the case where the logarithm of the stock is a stochastic process with stationary and independent increments. The impact of the latter extension was huge in both actuarial sciences and finance as well,  see \cite{gerber1996}, and the references therein  to cite a few, for more details and related discussions. The third key milestone in the evolution of the Esscher concept was elaborated by B\"uhlmann-Delbaen-Embrechts-Shiryaev in   \cite{buhlmann1996no}. Here, the authors derive the {\it dynamical Esscher transform}, for discrete time models, and baptized it as {\it conditional Esscher}. This interesting formulation of the dynamical Esscher led Kallsen and Shiryaev  to extend {\it the dynamical Esscher} to the general continuous-time semimartingale setting in \cite{Kallsen}. In this latter paper, which is the first to define the general continuous-time version of the Esscher, the authors singled out two classes called exponential and linear Esscher. However, \cite{Kallsen} did not address the necessary and/or sufficient condition for the existence of those Esscher densities nor their relationship, while it focuses on giving various sufficient conditions on the cumulant process to guarantee the uniform integrability of the resulting Esscher density process. For further developments about the continuous-time Esscher  and its numerous applications, we refer the reader to \cite{benth2012risk,hubalek2006,hubalek2009,Rheinlander1} and the references therein to cite a few. Certainly the Esscher risk-neutral measures, when they exist, give a nice and clear linear arbitrage-free pricing rule in incomplete markets besides other optimal neutral-risk measures. However, as noted in \cite{bondi2020comparing} for instance, at the practical and the computational aspects the Esscher rule might not be as efficient as one can wish. In this spirit,  the second-order Esscher transform was introduced in \cite{monfort2012asset}, where they show that this two-parameters Esscher transform adds an important flexibility compared to the one parameter Esscher for pricing derivatives. For given historical dynamics which are estimated using stock return data and under no-arbitrage conditions, the two-parameters Esscher contrary to the one parameter Esscher still provides free parameters to match derivative prices. Thus, recently in \cite{choulli2025second},, the authors introduced the Esscher martingale densities/deflators of-order-two for any $d$-dimensional continuous-time semimartingale model to price adequately the risks generated by its  jumps. \\

{\it What are our achievements?}  We introduce of a new family of Esscher pricing densities/measures for the valuation of vulnerable claims taking in consideration the informational risks. Since Esscher transforms are widely used in derivative pricing because of their analytical tractability, we propose a novel class of second-order Esscher densities and measures specifically designed for the informational-risk framework. Building upon the second-order Esscher methodology introduced in \cite{choulli2025second}, we incorporate an additional degrees of freedom that enables the pricing measure to explicitly capture pure default risks. We provide a complete characterization of the resulting family of densities/measures and establish necessary and sufficient conditions for their existence as well. Rgen we focus on applying these novel Esscher measures to the pricing of vulnerable European-type claims. The presence of two free Esscher parameters, $(\psi,\varphi)$, naturally generates a family of admissible prices and therefore leads to an interval of Esscher prices. To characterize the bounds of these intervals, we employ constrained backward stochastic differential equations (CBSDEs for short hereafter). Precisely, we show that these bounds are the smallest solution to constrained linear BSDEs that appear to be new as well. Thus, westablish the well-posedness of these equations, we develop a rigorous approximation framework based on truncated value processes. For each truncation level, we construct a unique solution to an unconstrained BSDE and then prove the convergence of the associated quadruples of processes within the appropriate functional spaces. By showing that these approximating solutions converge to the solution of the constrained pricing BSDE, we obtain a rigorous existence result for the upper Esscher price process. Consequently, we provide a complete mathematical foundation for the pricing bounds in the presence of default risk, thereby supplying the existence theory that was previously unavailable in the literature.\\

This paper has four sections including this current introduction section. The second section present the mathematical model, and single out some of its preliminaries and existing results which are useful through the paper. The third section, which one the core section of this paper, introduces the novel Esscher concept for our informational setting, and elaborate the main results in this direction. The fourth section, which is the second core part of this paper, presents results on the Esscher prices bounds for vulnerable claims using the constrained BSDEs. The paper has two appendices where recall some existing useful results and we prove our technical results respectively.
\section{The mathematical model and preliminaries}\label{Section4ModelNotations}
Throughout the paper, we suppose given a filtered probability space $(\Omega, \mathcal{F},\mathbb F: (\mathcal F_t)_{t\geq 0}, \mathbb{P})$ satisfying the usual conditions, i.e. the filtration $\mathbb F$ is right-continuous and complete. On this space, we suppose defined the pair of two independent processes $(W, N)$ where $W$ is a Brownian motion and $N$ is a non-homogeneous Poisson process with intensity process $\lambda$.\\
Our financial model is modelled by one non-risk asset with price $S^{(0)}$ and one risky asset with price $S=(S_t)_{t\geq 0}$ having the following dynamics:
\begin{equation}\label{the model}
    \begin{split}
        &S^{(0)} = \exp\bigg(\int_0^\cdot r_s ds \bigg), \quad S^{(0)}_0 =1\\
&S := S_0 e^X=S_0\mathcal{E} (\widetilde{X}),\quad X:= {b}\is{t} + \sigma\is{W} + \gamma\is {N}^\mathbb F,\quad \widetilde{X}:=\widetilde{b}\is{t} + \sigma{W} + \widetilde\gamma\is{N}^{\mathbb F},\\
   &N^\mathbb F := N -\int_0^\cdot \lambda_s ds,\quad \widetilde{b} :=b+ \frac{\sigma^2}{2} +\lambda (e^{\gamma} - 1-\gamma)\quad\mbox{and}\quad  \widetilde{\gamma} := e^\gamma - 1.
   \end{split}
\end{equation}
Throughout the rest of the paper, we impose the following assumption on the model parameters
\begin{assum}\label{assum: model parameters locally bounded}
    All the parameters of the model (i.e. $\lambda$, $\sigma$, $\gamma$, $b$) are locally bounded with $\sigma$, $\lambda$ are positive processes and $\gamma>-1$.
\end{assum}
The rest of this section is divided into two subsections. In the first subsection we recall some general notations that are used throughout the paper, while the second subsection presents the model for the random time and some of its preliminaries. 
\subsection{General notation}
For any filtration  $\mathbb H\in \{\mathbb F,\mathbb G\}$ satisfying the usual condition (right continuity and completeness), we denote ${\mathcal A}(\mathbb H)$ (respectively ${\mathcal M}(\mathbb H)$) the set
of right-continuous with left limits (RCLL hereafter)and $\mathbb H$-adapted processes with $\mathbb H$-integrable variation (respectively that are $\mathbb H$-uniformly integrable martingale). The set ${\mathcal{V}}^+(\mathbb{H})$ denotes the set of all RCLL, nondecreasing, and $\mathbb{H}$-adapted processes with finite values.  
For any process $X$, we denote by $^{o,\mathbb H}X$  (respectively $^{p,\mathbb H}X$)  the
$\mathbb H$-optional (respectively $\mathbb H$-predictable) projection of $X$. For an increasing process $V$, we denote $V^{o,\mathbb H}$ (respectively $V^{p,\mathbb H}$) its dual $\mathbb H$-optional (respectively $\mathbb H$-predictable) projection. For a filtration $\mathbb H$, ${\mathcal O}(\mathbb H)$, ${\mathcal P}(\mathbb H)$ and  $\mbox{Prog}(\mathbb H)$ represent the $\mathbb H$-optional, the $\mathbb H$-predictable and the $\mathbb H$-progressive $\sigma$-fields  respectively on $\Omega\times[0,+\infty[$. For an $\mathbb H$-semimartingale $X$, we denote by $L(X,\mathbb H)$ the set of all $X$-integrable processes in the Ito's sense, and for $H\in L(X,\mathbb H)$, the resulting integral is one dimensional $\mathbb H$-semimartingale denoted by $H\is X:=\int_0^{\cdot} H_udX_u$. If ${\mathcal C}(\mathbb H)$ 
is a set of $\mathbb H$-adapted processes,
then ${\mathcal C}_{loc}(\mathbb H)$ --except when it is stated otherwise-- is the set of processes, $X$,
for which there exists a sequence of $\mathbb H$-stopping times,
$(T_n)_{n\geq 1}$, that increases to infinity and $X^{T_n}$ belongs to ${\mathcal C}(\mathbb H)$, for each $n\geq 1$.\\

We end this subsection by defining some sets of local martingale densities/deflators, which appear {\it naturally} in our analysis. 
\begin{definition}\label{DefatorIntergrable} Let $Y$ a $d$-dimensional $\mathbb{H}$-semimartingale, $Z$ be a process, and $Q$ be a probability. \\
{\rm{(a)}} $Z$ is a local martingale density/deflator for $(Y,Q,\mathbb{H})$ if $Z>0$ and both $ZY$ and $Z$ are $Q$-local martingales. The set of local martingale densities/deflators for $(Y,Q)$ will be denoted by ${\cal{Z}}_{loc}(Y,Q,\mathbb{H})$, and ${\cal{Z}}^{L\log{L}}_{loc}(Y,Q,\mathbb{H})$ denotes the set of $Z\in {\cal{Z}}_{loc}(Y,Q,\mathbb{H})$ satisfying $Z\ln(Z)$ is a $Q$-local submartingale. When  $Q=P$, we simply omit the probability and write $ {\cal{Z}}_{loc}(Y,\mathbb{H})$ and ${\cal{Z}}^{L\log{L}}_{loc}(Y,\mathbb{H})$ respectively.\\
{\rm{(b)}}  Let $L$ be a positive $\mathbb{H}$-local martingale, and let $(T_n)_{n\geq 1}$ be the sequence of $\mathbb{H}$-stopping times that increases to infinity such that $L^{T_n}\in{\cal{M}}(\mathbb{H})$ and $Q_n:=L_{T\wedge{T}_n}\cdot P$. Then  we define
\begin{equation}
\begin{split}
{\cal{Z}}_{loc}(Y,L,\mathbb{H})&:=\left\{Z:\ Z^{T_n}\in {\cal{Z}}_{loc}(Y^{T_n},Q_n,\mathbb{H}),\quad n\geq 1\right\},\\
 {\cal{Z}}^{L\log{L}}_{loc}(Y,L,\mathbb{H})&:=\left\{Z:\ Z^{T_n}\in {\cal{Z}}^{L\log{L}}_{loc}(Y^{T_n},Q_n,\mathbb{H}),\quad n\geq 1\right\}.
\end{split}\end{equation}
\end{definition}
\subsection{The random horizon model  and its parametrisation}
Thus, throughout the paper, we call the initial model (or the default/mortality free model) is the quadruplet $(\Omega,S, \mathbb{F},P)$. In addition to this initial model, we consider a random time $\tau$ which might not be  an $\mathbb{F}$-stopping time. This random  time represent the default time in credit risk theory and the death time of an insured for the setting of mortality and/or longevity risk setting. This random time is parametrized though $\mathbb F$ by the pair $(G, \widetilde{G})$, called survival probabilities or Az\'ema supermartingales, and are given by
\begin{eqnarray}\label{GGtilde}
G_t :=\ ^{o,\mathbb F}(I_{\Lbrack0,\tau\Lbrack})_t= P(\tau > t | {\mathcal F}_t) \ \mbox{ and } \ \widetilde{G}_t :=\ ^{o,\mathbb F}(I_{\Lbrack0,\tau\Rbrack})_t= P(\tau \ge t | {\mathcal F}_t).\end{eqnarray}
Furthermore, the following process $m$ given by 
\begin{equation} \label{processm}
m := G + D^{o,\mathbb F},\quad \mbox{where}\quad{D}:=I_{\Lbrack\tau,+\infty\Lbrack},
\end{equation}
is a BMO $\mathbb F$-martingale and plays important role in our analysis. As $\tau$ can not be seen before it occurrence, then the filtration that will incorporate both $\tau$ and $\mathbb{F}$ is the progressive enlargement of $\mathbb{F}$ with $\tau$, which is denoted by $\mathbb G$ and is defined as follows. 
\begin{equation}\label{processD}
\mathbb G:=({\mathcal G}_t)_{t\geq 0},\ {\mathcal G}_t:={\mathcal G}^0_{t+}\ \mbox{where} \ {\mathcal G}_t^0:={\mathcal F}_t\vee\sigma\left(D_s,\ s\leq t\right)\ .
\end{equation}
 Throughout the paper, on $\Omega\times [0,+\infty)$, we consider the $\mathbb F$-optional  $\sigma$-field  denoted by ${\mathcal O}(\mathbb F)$ and  the $\mathbb F$-progressive  $\sigma$-field denoted by $\mbox{Prog}(\mathbb F)$. Thanks to \cite[Theorem 3]{aksamit2015optional} and \cite[Theorem 2.3 and Theorem 2.11]{choulli2020martingale}, we recall 
\begin{theorem}\label{Toperator} The following assertions hold.\\
{\rm{(a)}} For any $M\in{\mathcal M}_{loc}(\mathbb F)$,  the process
\begin{equation} \label{processMhat}
{\mathcal T}(M) := M^\tau \footnote{Throughout the paper, for any random time $\sigma$ and any process $X$, we denote by $X^{\sigma}$ the stopped process given by $X^{\sigma}_t:=X_{\sigma\wedge t},\ t\geq 0$.} -{\widetilde{G}}^{-1} I_{\Rbrack 0,\tau\Rbrack} \is [M,m] +  I_{\Rbrack 0,\tau\Rbrack} \is\Big(\sum \Delta M I_{\{\widetilde G=0<G_{-}\}}\Big)^{p,\mathbb F}\end{equation}
 is a $\mathbb G$-local martingale.\\
 {\rm{(b)}}  The process 
\begin{equation} \label{processNG}
N^{\mathbb G}:=D - \widetilde{G}^{-1} I_{\Rbrack 0,\tau\Rbrack} \is D^{o,\mathbb  F}
\end{equation}
is a $\mathbb G$-martingale with integrable variation. Moreover, $H\is N^{\mathbb G}$ is a $\mathbb G$-local martingale with locally integrable variation for any $H$ belonging to
\begin{equation} \label{SpaceLNG}
{\mathcal{I}}^o_{loc}(N^{\mathbb G},\mathbb G) := \Big\{K\in \mathcal{O}(\mathbb F):\quad \vert{K}\vert G{\widetilde G}^{-1} I_{\{\widetilde{G}>0\}}\is D\in{\mathcal A}_{loc}^+(\mathbb G)\Big\}.
\end{equation}
\end{theorem}

\begin{lemma}\label{Parametrization4tau} Suppose $G>0$. Then the following assertions hold.\\
{\rm{(a)}} There exists a unique pair $(\alpha,\beta)\in{L}^1_{loc}(W,\mathbb{F})\times{L}^1_{loc}(N^{\mathbb{F}},\mathbb{F})$ such that 
\begin{equation}\label{Paramertization4m}
{m} =m_0+G_- \alpha\is{ W} + G_-\beta\is{N}^\mathbb{F},\quad \mbox{and}\quad-1<\beta\leq{G}_{-}^{-1}\ P\otimes dt\mbox{-a.e}..
\end{equation}
{\rm{(a)}} There exists a unique $\eta\in{L}^1_{loc}(N^{\mathbb{F}},\mathbb{F})$ satisfying
\begin{equation}\label{tau params}
D^{o,\mathbb{F}} - D^{p,\mathbb{F}}=  G_-\eta \bigcdot N^\mathbb{F},\quad0\leq \eta\leq 1+\beta\quad P\otimes dt\mbox{-a.e.}, 
\end{equation}
and 
\begin{equation}\label{DoF decomp}
		\widetilde{G}^{-1}{\Delta D^{o,\mathbb{F}}}= G_-^{-1}{\Delta D^{p,\mathbb{F}}} +  (1 + \beta)^{-1}\eta\Delta{N}.
		\end{equation}
\end{lemma}
The proof of the lemma is relegated to Appendix \ref{Proof4Propositions}. This lemma conveys that  $\tau$ is completely parametrized in $\mathbb{F}$ via the quadruple $(\alpha, \beta, \eta, D^{p,\mathbb{F}})$. Throughout the rest of the paper, we consider the following processes
		\begin{equation}\label{ProcessV(F)}
		V^{\mathbb{F}}:=1- \mathcal{E} \big(-\widetilde{G}^{-1} \bigcdot D^{o,\mathbb{F}} \big),\quad {\widetilde{\mathcal{E}}}:= \mathcal{E} \big(-\widetilde{G}^{-1} \bigcdot D^{o,\mathbb{F}} \big) .
		\end{equation}
\begin{lemma}\label{keylemma}	The following assertions hold.\\
{\rm{(a)}} Let $X$ be any $\mathcal{F}$-measurable integrable random variable. Then for any $t\geq 0$,
	\begin{equation}
		\mathbb{E}[X \mathbbm{1}_{\{t<\tau\}} |\mathcal{G}_t] = G^{-1}_t \mathbb{E} [X \mathbbm{1}_{t<\tau} |\mathcal{F}_t] \mathbbm{1}_{\{t<\tau\}}.
	\end{equation}
{\rm{(b)}} If $G>0$, then the following hold
		 \begin{equation}\label{G2m2D(o,F)}
			1/\mathcal{E} (G_-^{-1} \bigcdot m)= (G_0{\widetilde{\mathcal{E}}})/G = (G_0{\widetilde{\mathcal{E}}}_{-})/\widetilde{G} .
		\end{equation}
\end{lemma}
\section{Novel Esscher densities/measures for informational risks}
This section introduces the notion of Esscher with ``{\it{degree-of-freedom}}" for models stopped at the random time $\tau$, where one is facing the valuation problem of vulnerable claims. This novel notion of Esscher extends the second-order Esscher notion introduced recently in \cite{choulli2025second}, and it allows us to price adequately additional uncertainties and risks borne by the random time $\tau$. To this end, for any $d$-dimensional $\mathbb{F}$-semimartingale $Y$, we define
\begin{equation}\label{Theta(Y,F)}
\begin{split}
\Theta(Y,\mathbb{F}):=\left\{(\theta,\psi,\varphi):\ \begin{array}{lll}\ (\theta,\psi)\ \mbox{is an}\ \mathbb{R}^d\times\mathbb{R}^{d\times d}\mbox{-valued $\mathbb{F}$-predictable process}\\
\theta \in L(Y),\  \psi\ \mbox{is locally bounded},\\
\varphi\ \mbox{is a real-valued, locally bounded, and $\mathbb{F}$-optional}\end{array}\right\}.
\end{split}
\end{equation}
\begin{definition}\label{Esscher process} Let $Z$ be a RCLL and positive process $\mathbb{G}$-adapted.\\
 1) $Z$ is called a linear Esscher martingale density of order-two with degree-of-freedom for the model $(S^{\tau},\mathbb{G})$ if there exists a quadruplet $(\theta,\psi,\varphi,\Gamma^{\mathbb{G}})$ such that  \begin{equation}\label{EsscherDegree}
 \begin{split}
 &(\theta,\psi,\varphi)\in\Theta(S,\mathbb{F}),\ \Gamma^{\mathbb{G}}\ \mbox{is RCLL and $\mathbb{G}$-predictable with finite variation}, \\
&Z\in{\cal{Z}}_{loc}(S^{\tau},\mathbb{G}),\quad\mbox{and}\quad Z=\exp\left(\theta\is{S}^{\tau}+\sum(\Delta{S}^{\tau})^{tr}\psi\Delta{S}^{\tau}+\varphi\is N^{\mathbb G}-\Gamma^{\mathbb{G}}\right).
\end{split}
\end{equation}
If furthermore $Z$ is a uniformly integrable martingale, then $Q:=Z_{\infty}\cdot{P}$ is called a linear Esscher martingale measure of-order-two with degree-of-freedom $\varphi$. The set of all linear Esscher martingale densities (respectively measures) will be denoted by ${\cal{Z}}^{LE}(S^{\tau},\mathbb{G})$ (respectively  ${\cal{Q}}^{LE}(S^{\tau},\mathbb{G})$). \\
2) $Z$ is called an exponential Esscher martingale density of order-two with degree-of-freedom for the model $(S^{\tau},\mathbb{G})$ if there exists a quadruplet $(\theta,\psi,\varphi,\Gamma^{\mathbb{G}})$ such that  \begin{equation}\label{EsscherDegree}
 \begin{split}
 &(\theta,\psi,\varphi)\in\Theta(X,\mathbb{F}),\ \Gamma^{\mathbb{G}}\ \mbox{is RCLL and $\mathbb{G}$-predictable with finite variation}, \\
&Z\in{\cal{Z}}_{loc}(S^{\tau},\mathbb{G}),\quad\mbox{and}\quad Z=\exp\left(\theta\is{X}^{\tau}+\sum(\Delta{X}^{\tau})^{tr}\psi\Delta{X}^{\tau}+\varphi\is N^{\mathbb G}-\Gamma^{\mathbb{G}}\right).
\end{split}
\end{equation}
If furthermore $Z$ is a uniformly integrable martingale, then $Q:=Z_{\infty}\cdot{P}$ is called a exponential Esscher martingale measure of order-two with-degree-of-freedom. The set of all exponential Esscher martingale densities (respectively measures) will be denoted by ${\cal{Z}}^{EE}(S^{\tau},\mathbb{G})$ (respectively  ${\cal{Q}}^{EE}(S^{\tau},\mathbb{G})$). \\
3) $Z$ is called a linear Esscher (respectively, exponential Esscher) martingale density of order two for the model $(S^\tau, \mathbb G)$ if $Z$ is with zero degree of freedom.
\end{definition}
One can think of considering the process $Z':=\exp\left(\theta\is{X}^{\tau}+\sum\psi(\Delta{X}^{\tau})^2+\varphi\is D-\Gamma^{\tau}\right)$ instead. This choice sounds simple and more {\it natural}, while it has an important drawback. When $\tau$ is a stopping time, then $\mathbb{F}=\mathbb{G}$, and hence normally Esscher (respectively Esscher of-order-two) with degree-of-freedom should become Esscher (respectively Esscher -of-order-two). This is not the case unfortunately for the choice of $Z'$, as $\varphi\is{D}$ is not null, in contrast to the choice of $Z$. In fact, thanks to \cite{choulli2020martingale}, when $\tau$ is an $\mathbb{F}$-stopping time, we get $\varphi\is{N}^{\mathbb{G}}\equiv 0$,  and Esscher with any degree-of-freedom $\varphi$ becomes an Esscher. 
\begin{remark}
For the class of models considered in this paper, the Esscher density processes and the associated martingale conditions take analogous forms under both the linear and exponential Esscher transforms. To streamline the presentation, we introduce the notation $\zeta = \{\gamma, \widetilde \gamma \}$ and formulate all subsequent results in terms of $\zeta$, thereby unifying the treatment of the linear and exponential Esscher frameworks.
\end{remark}
\subsection{Main results on Esscher densities/measures}
Herein, we elaborate our main results on Esscher pricing densities and Esscher pricing measures. In fact the following theorem characterizes Esscher densities for the stopped model $(S^{\tau},\mathbb{G})$  in various manners. Among these, we give a precise relationship between Esscher densities for $(S^{\tau},\mathbb{G})$ and those of $(S,\mathbb{F})$. Then  we illustrate these results on particular cases for $(S,\mathbb{F})$ and/or the model of $\tau$. 
Below we define the set of integrands with respect to $N^{\mathbb{G}}$ such that the resulting integrals are {\it exponentially} special $\mathbb{G}$-semimartingales.
\begin{equation}\label{I(exp, N(G))}
{\mathcal{I}}^{exp}_{loc}(N^{\mathbb{G}},\mathbb{G}):=\left\{\varphi\in{\mathcal{I}}^{o}_{loc}(N^{\mathbb{G}},\mathbb{G}):\quad {{\exp(\varphi^+)}\over{\widetilde{G}}}I_{\Rbrack0,\tau\Rbrack}\is{D}^{o,\mathbb{F}}\in{\mathcal{V}}^+(\mathbb{G})\right\}.
\end{equation}
For any $\varphi\in{\mathcal{I}}^{exp}_{loc}(N^{\mathbb{G}},\mathbb{G})$, we associate to it three processes $H(\varphi)$, $J (\varphi)$ and ${\widetilde{\mathcal{E}}}^{(\varphi)}$ given by 
\begin{equation}\label{H(phi)Phi(phi)}
\begin{split}
&H(\varphi):=\bigg({\eta}(e^\varphi - 1) + 1+\beta\bigg) \exp\left(-\frac{\varphi \eta}{1+\beta}\right) ,\\ 
&J (\varphi):=\frac{{\widetilde{G}}(e^{\varphi}-1)}{G + e^{\varphi}\Delta{D}^{o,\mathbb{F}}},\quad\quad {\widetilde{\mathcal{E}}}^{(\varphi)}:={\mathcal{E}}\left({{-e^{\varphi}}\over{G+e^{\varphi}\Delta{D}^{o,\mathbb{F}}}}\is{D}^{o,\mathbb{F}}\right).\end{split}
\end{equation}
It is clear that the process $ {\widetilde{\mathcal{E}}}^{(\varphi)}$ is positive, non-increasing and bounded by one. It naturally extends $ {\widetilde{\mathcal{E}}}:={\widetilde{\mathcal{E}}}^{(0)}={\mathcal{E}}\left(-{\widetilde{G}}^{-1}\is{D}^{o,\mathbb{F}}\right)$ defined in (\ref{ProcessV(F)}).

\begin{theorem}\label{MainTheorem4Esscher}
Consider a positive and $\mathbb{G}$-adapted process, $Z^{\mathbb{G}}$, such that $Z^{\mathbb{G}}=(Z^{\mathbb{G}})^{\tau}$. Let $(H(\varphi), J (\varphi), {\widetilde{\mathcal{E}}}^{(\varphi)})$ be the functionals given by (\ref{H(phi)Phi(phi)}). Then the following assertions are equivalent.\\
{\rm{(a)}} There exists $\bar\varphi\in{\mathcal{I}}^{exp}_{loc}(N^{\mathbb{G}},\mathbb{G})$ and $Z^{\mathbb{G}}$ is Esscher density of-order-two with degree-of-freedom $\bar\varphi$.\\
{\rm{(b)}}  There exists a triplet $(\theta,\psi,\varphi)$, where $\varphi\in{\mathcal{I}}^{exp}_{loc}(N^{\mathbb{G}},\mathbb{G})$ and $(\theta, \psi)\in\Theta(S,\mathbb{F})$ such that  
\begin{equation}\label{IntegrabilityCondition4Esscher}
\int_0^{\cdot}H_s(\varphi)\exp(\zeta_s\theta_s+\zeta_s^2\psi_s)ds\in{\mathcal{A}}^+_{loc}(\mathbb{F}),\end{equation}
\begin{equation}\label{MainEquation4Esscher}
 0=\widetilde b + \alpha\sigma + \theta \sigma^2 + \lambda \widetilde \gamma\Bigl(H(\varphi)e^{\theta \zeta + \psi \zeta^2} - 1\Bigr), \quad P\otimes dt\mbox{-a.e}.,\end{equation}
and 
\begin{equation}\label{Representation4ZG}
Z^{\mathbb{G}}= {\mathcal{E}}\Biggl(\theta \sigma\is\mathcal{T}(W) + \big(H(\varphi)e^{\theta \zeta + \psi \zeta^2} - 1-\beta\big)\is\mathcal{T}(N^\mathbb{F})\Biggr)\mathcal{E} \bigg(J (\varphi) \is N^\mathbb{G} \bigg).\end{equation}
{\rm{(c)}}  There exists a unique pair $(Z^{\mathbb{F}},\Phi^{\mathbb{F}})$, where  $Z^{\mathbb{F}}$ is an Esscher density of-order-two for $(S,\mathbb{F})$ with parameters $(\theta^{\mathbb{F}},\psi^{\mathbb{F}})\in\Theta(S,\mathbb{F})$ and $\Phi^{\mathbb{F}}\in{\mathcal{I}}^{o}_{loc}(N^{\mathbb{G}},\mathbb{G})$, such that 
\begin{equation}
\begin{split}
&-{{{\widetilde{G}}}\over{G}}<\Phi^{\mathbb{F}},\quad \Phi^{\mathbb{F}}\Delta{D}^{o,\mathbb{F}}<\widetilde{G},\quad \ln\left(({\widetilde{G}}+G\Phi^{\mathbb{F}})({\widetilde{G}}-\Phi^{\mathbb{F}}\Delta{D}^{o,\mathbb{F}})^{-1}\right)\in{\mathcal{I}}^{o}_{loc}(N^{\mathbb{G}},\mathbb{G}),\\
&{{G}\over{\widetilde{G}}}\vert\Phi^{\mathbb{F}}\vert{Z}^{\mathbb{F}}\is V^{\mathbb{F}}\in{\mathcal{A}}^+_{loc}(\mathbb{F}),\quad\mbox{and}\quad Z^{\mathbb{G}}={{(Z^{\mathbb{F}})^{\tau}}\over{{\mathcal{E}}(G^{-1}_{-}\is{m})^{\tau}}}{\mathcal{E}}(\Phi^{\mathbb{F}}\is{N}^{\mathbb{G}}).\end{split}\label{EssherF2G}
\end{equation}
{\rm{(d)}}  There exists a unique triplet $({\widehat{Z}}^{\mathbb{F}},\widehat{V}, \widehat\varphi)$, where  $\widehat\varphi\in{\mathcal{I}}^{exp}_{loc}(N^{\mathbb{G}},\mathbb{G})$, ${\widehat{Z}}^{\mathbb{F}}$ is an Esscher density of-order-two for $(S,\mathbb F)$ with parameters $(\widehat\theta,\widehat\psi)\in\Theta(S,\mathbb{F})$, and $\widehat{V}\in{\mathcal{V}}^+(\mathbb{F})$, such that 
 \begin{equation}\label{EsscherG2F(bis)}
{{{\widehat{Z}}^{\mathbb{F}}}\over{G}}\exp(\varphi-\widehat{V})\is D^{o,\mathbb{F}}\in{\mathcal{A}}^+_{loc}(\mathbb{F}),\quad\mbox{and}\quad   Z^{\mathbb{G}}={{G_0({\widehat{Z}}^{\mathbb{F}})^{\tau}}\over{G^{\tau}}}\exp(\widehat\varphi\is D-\widehat{V}^{\tau}).
  \end{equation}
Furthermore, the three sets of parameters  $(\theta,\psi, \varphi)$, $(\theta^{\mathbb{F}},\psi^{\mathbb{F}},\Phi^{\mathbb{F}},Z^{\mathbb{F}})$ and $(\widehat\theta,\widehat\psi,\widehat\varphi,\widehat{V},\widehat{Z}^{\mathbb{F}})$ are related via
\begin{equation}\label{Relationship4Parameters}
\begin{split}
\bar\varphi=\widehat\varphi=\varphi,\quad \widehat\theta&=\theta^{\mathbb{F}}=\theta + \frac{\alpha}{\sigma}, \quad {\gamma^2}\widehat\psi={\gamma^2}\psi^{\mathbb{F}}={\gamma^2}\psi-\gamma\frac{\alpha}{\sigma} + \ln\bigg(H(\varphi) \bigg),\quad  e^{-\widehat{V}}= {\widetilde{\mathcal{E}}}^{(\varphi)},\\
Z^{\mathbb{F}}&=\widehat{Z}^{\mathbb{F}}={\mathcal{E}}\Biggl(\sigma\theta^{\mathbb{F}}\is{W}+\left(e^{\gamma\theta^{\mathbb{F}}+\gamma^2\psi^{\mathbb{F}}}-1\right)\is{N}^{\mathbb{F}}\Biggr),\quad \Phi^{\mathbb{F}}=J (\varphi).
 \end{split}
\end{equation}
\end{theorem}
The proof of the theorem is relegated to Subsection \ref{Proof4mainTheorem} for the sake of easy exposition. The theorem conveys three equivalent, complete and explicit characterization of Esscher density of-order-two with degree-of-freedom $\varphi$.  The first characterization, which is given by assertion (b), states the fundamental point-wise equation associated to the new Esscher notion. Furthermore, this characterization describes $Z^{\mathbb{G}}$ in a multiplicative manner by singling out the part that corresponds to the {\it pure mortality density/martingale} ${\mathcal{E}}(J (\varphi)\is{N}^{\mathbb{G}})$. For more details about the classification of $\mathbb{G}$-martingales and their properties, we refer the reader to \cite{choulli2020martingale}.   The second characterization (see  assertion (c)) links $\mathbb{G}$-Esscher to $\mathbb{F}$-Esscher in a unique manner. This gives the characterization for Esscher {\it version} in the spirit of \cite{choulli2022explicit}, where $\mathbb{G}$-deflators are disintegrated multiplicatively according to the types of risks they interfere with. The last characterization (see assertion (d)) conveys that one can use $\exp(\theta\is X^{\tau}+\sum\psi(\Delta{X}^{\tau})^2+\varphi\is{D}+\Gamma^{\tau})$ as an Esscher density with degree-of-freedom at the expenses of dropping the predictably of $\Gamma$ and allow it to be $\mathbb{F}$-optional instead. Below, we discuss how Theorem \ref{MainTheorem4Esscher} becomes when the degree-of-freedom $\varphi$ and/or when the second Esscher parameter $\psi$  are null.
\begin{corollary}\label{EsscherDegreeofFreedom} Suppose that $G>0$. Then the following assertions hold.\\
{\rm{(a)}} $Z^{\mathbb{G}}$ is an Esscher density of-order-two if and only if there exists $(\theta,\psi)\in\Theta(S,\mathbb{F})$ satisfying (\ref{IntegrabilityCondition4Esscher}), 		\begin{equation}
			0=\widetilde b + \alpha\sigma + \widetilde\gamma \beta \lambda + \theta \sigma^2 + \lambda \widetilde\gamma(e^{\theta \zeta + \psi \zeta^2} - 1)(1+\beta),
		\end{equation}
		and 
		\begin{equation}
			Z^{\mathbb{G}}= {\mathcal{E}}\Biggl(\theta \sigma \bigcdot \mathcal{T}(W) + \big(e^{\theta \zeta + \psi \zeta^2} - 1\big)(1+\beta)\is\mathcal{T}(N^\mathbb{F})\Biggr).
		\end{equation}
{\rm{(b)}} $Z^{\mathbb{G}}$ is an Esscher density if and only if there exists $\theta\in{L}(X,\mathbb{F})$ such that 
	\begin{equation}
		0=\widetilde b + \alpha \sigma + \widetilde\gamma \beta \lambda + \theta \sigma^2 + \lambda \widetilde\gamma(e^{\theta \zeta} - 1)(1+\beta),\quad P\otimes dt\mbox{-a.e.}, 
	\end{equation}
	and 
	\begin{equation}
		Z^{\mathbb{G}}= {\mathcal{E}}\Biggl(\theta \sigma \bigcdot \mathcal{T}(W) + \big(e^{\theta \zeta } - 1\big)(1+\beta)\is\mathcal{T}(N^\mathbb{F})\Biggr).
	\end{equation}
	{\rm{(c)}} $Z^{\mathbb{G}}$ is an Esscher density with degree-of-freedom $\varphi$ if and only if there exists $\theta\in{L}(X,\mathbb{F})$ such that 
		\begin{equation}\label{MartingaleEquation}
			0=\widetilde b + \alpha\sigma + \widetilde\gamma \beta\lambda + \theta \sigma^2 + \lambda \widetilde\gamma(e^{\theta \zeta} - 1)H(\varphi),\quad P\otimes dt\mbox{-a.e.}, 
		\end{equation}
		and 
		\begin{equation}\label{RepresentationEquation}
			Z^{\mathbb{G}}= {\mathcal{E}}\Biggl(\theta \sigma \bigcdot \mathcal{T}(W) + \big(e^{\theta \zeta} - 1\big)H(\varphi)\is\mathcal{T}(N^\mathbb{F})\Biggr)\mathcal{E} \bigg(J (\varphi) \is N^\mathbb{G} \bigg).\end{equation}
		where $H(\varphi)$ is given by (\ref{H(phi)Phi(phi)}). 
\end{corollary}

Below, we discuss several particular cases for the pair $(S^{\tau}, \mathbb{G})$, which are frequently studied in the literature of credit risk and/or mortality and longevity risks (Life insurance in general). 

\begin{corollary} Suppose that $G>0$, and let $\varphi\in{\mathcal{I}}^{exp}_{loc}(N^{\mathbb{G}},\mathbb{G})$ Then the following assertions hold.\\
{\rm{(a)}} Suppose $\tau$ avoids all $\mathbb{F}$-stopping times. Then $Z^{\mathbb{G}}$ is an Esscher density of-order-two with degree-of-freedom $\varphi$ if and only if there exists a pair $(\theta,\psi)$ of $\mathbb{F}$-predictable processes satisfying (\ref{IntegrabilityCondition4Esscher}),
		\begin{equation}
			0=\widetilde b + \alpha\sigma + \theta \sigma^2 + \lambda \widetilde\gamma\left(e^{\theta \zeta + \psi \zeta^2} (1 +\beta) - 1\right),\quad P\otimes dt\mbox{-a.e.}, 
		\end{equation}
		and 
		\begin{equation}
			Z^{\mathbb{G}}= {\mathcal{E}}\Biggl(\theta \sigma \bigcdot \mathcal{T}(W) + \big(e^{\theta \zeta + \psi \zeta^2} - 1\big)(1+\beta)\is\mathcal{T}(N^\mathbb{F})\Biggr)\mathcal{E} \bigg((e^\varphi - 1) \is N^\mathbb{G} \bigg).
		\end{equation}
{\rm{(b)}} Suppose that any $\mathbb{F}$-martingale is a $\mathbb{G}$-martingale (The $\mathcal{H}$-hypothesis or immersion property). Then $Z^{\mathbb{G}}$ is an Esscher density of-order-two with degree-of-freedom $\varphi$ if and only if there exists a pair $(\theta,\psi)\in\Theta(S,\mathbb{F})$ satisfying (\ref{IntegrabilityCondition4Esscher}), 
		\begin{equation}
			0=\widetilde b + \theta \sigma^2 + \lambda \widetilde\gamma(e^{\theta \zeta + \psi \zeta^2} - 1)e^{-\varphi \eta}\bigg({\eta}(e^\varphi - 1) + 1\bigg),\quad P\otimes dt\mbox{-a.e.}, 
		\end{equation}
		and 
		\begin{equation}
			Z^{\mathbb{G}}= {\mathcal{E}}\Biggl(\theta \sigma \bigcdot \mathcal{T}(W) + \big(e^{\theta \zeta + \psi \zeta^2} - 1\big)e^{-\varphi \eta}({\eta}e^\varphi - \eta + 1)\is\mathcal{T}(N^\mathbb{F})\Biggr)\mathcal{E} \bigg(\frac{G_- (e^{\varphi}-1)}{G + e^{\varphi}\Delta{D}^{o,\mathbb{F}}} \is N^\mathbb{G} \bigg).
		\end{equation}
{\rm{(c)}} Suppose that $\tau$ is independent of $\mathbb{F}$. Then $Z^{\mathbb{G}}$ is an Esscher density of-order-two with degree-of-freedom $\varphi$ if and only if there exists a pair $(\theta,\psi)\in\Theta(S,\mathbb{F})$ fulfilling (\ref{IntegrabilityCondition4Esscher}), 
		\begin{equation}
			0=\widetilde b + \theta \sigma^2 + \lambda \widetilde \gamma(e^{\theta \zeta + \psi \zeta^2} - 1),\quad P\otimes dt\mbox{-a.e.}, 
		\end{equation}
		and 
		\begin{equation}
			Z^{\mathbb{G}}= {\mathcal{E}}\Biggl(\theta \sigma \bigcdot \mathcal{T}(W) + \big(e^{\theta \zeta + \psi \zeta^2} - 1\big)\is\mathcal{T}(N^\mathbb{F})\Biggr)\mathcal{E} \bigg(\frac{G_- (e^{\varphi}-1)}{G + e^{\varphi}\Delta{D}^{p,\mathbb{F}}} \is{N}^\mathbb{G} \bigg).
		\end{equation}
{\rm{(d)}}  Suppose that all $\mathbb{F}$-martingales are continuous. Then $Z^{\mathbb{G}}$ is an Esscher density of-order-two with degree-of-freedom $\varphi$ if and only if 		
\begin{equation}
			Z^{\mathbb{G}}= {\mathcal{E}}\Biggl(-{{\widetilde b+\alpha\sigma}\over{\sigma}}\bigcdot \mathcal{T}(W) \Biggr)\mathcal{E} \bigg(\frac{G_- (e^{\varphi}-1)}{G + e^{\varphi}\Delta{D}^{p,\mathbb{F}}} \is{N}^\mathbb{G} \bigg).
		\end{equation} 
\end{corollary}
\begin{proof} 1) Remark that $\tau$ avoids $\mathbb{F}$-stopping times if and only if $\Delta{D}^{o,\mathbb{F}}\equiv0$, or equivalently  $\Delta{D}^{p,\mathbb{F}}\equiv0$ and  $\eta \equiv 0$. Thus, under this condition, we have 
		$$H(\varphi)=1+\beta, \quad\mbox{and}\quad J (\varphi)=e^{\varphi}-1.$$
Thus, by inserting these in (\ref{MainEquation4Esscher}) and (\ref{Representation4ZG}) assertions (a) follows immediately.\\
2) The immersion assumption is equivalent to $m=m_0$, or equivalently $\beta\equiv 0$, and in this case we get
		$$H(\varphi)=e^{-\eta\varphi}\left(1+\eta(e^{\varphi}-1)\right), \quad\mbox{and}\quad J (\varphi)={{G_{-}(e^{\varphi}-1)}\over{G+e^{\varphi}\Delta{D}^{o,\mathbb{F}}}}.$$ 
Hence, by inserting these in (\ref{MainEquation4Esscher}) and (\ref{Representation4ZG}) assertions (b) follows immediately.\\
3)  Under teh assumption hat $\tau$ is independent to $\mathbb{F}$, we get $G={}^{p,\mathbb{F}}(G)$ and $\widetilde{G}=G_{-}=1-D^{p,\mathbb{F}}=1-D^{p,\mathbb{F}}$ and $m=m_0$. Thus, in this case we get 
$$
\eta\equiv 0,\quad \alpha=\beta=0,\quad{H}(\varphi)=1,\quad J (\varphi)=\frac{G_- (e^{\varphi}-1)}{G + e^{\varphi}\Delta{D}^{p,\mathbb{F}}}.
$$
Thus by inserting these in  (\ref{MainEquation4Esscher}) and (\ref{Representation4ZG}), assertion (c) follows immediately. \\
4) When all $\mathbb{F}$-martingale are continuous, then this implies that $\mathbb{F}$ is generated by the Brownian motion only. In this case, one can put $\gamma=0$ in (\ref{MainEquation4Esscher})  and notice that 
$$\theta=-(\mu+\alpha\sigma)/\sigma^2,\quad \widetilde{G}=G_{-},\quad D^{o,\mathbb{F}}=D^{p,\mathbb{F}}.$$
Thus, by combining these and inserting them in  (\ref{MainEquation4Esscher}) and (\ref{Representation4ZG}), we get assertion (d). This ends the proof of the corollary.
\end{proof}

We end this subsection by addressing the existence of Esscher pricing measures of-order-two with degree-of-freedom under $\mathbb{G}$ versus those of-order-two under $\mathbb{F}$ . 
\begin{theorem}\label{EsscherMeasures} Suppose that $G>0$. Then the following assertions are equivalent.\\
{\rm{(a)}} $(S,\mathbb{F})$ admits an Esscher pricing measure of-order-two.\\
 {\rm{(b)}} $(S^{\tau},\mathbb{G})$ admits an Esscher pricing measure of-order-two with degree-of-freedom $\varphi$, for any  $\varphi$ belonging to ${\mathcal{I}}^{exp}_{loc}(N^{\mathbb{G}},\mathbb{G})$.\\
{\rm{(c)}} There exists $\varphi\in{\mathcal{I}}^{exp}_{loc}(N^{\mathbb{G}},\mathbb{G})$ such that $1/{\widetilde{\mathcal{E}}}^{(\varphi)}$ is bounded, and $(S^{\tau},\mathbb{G})$ admits an Esscher pricing measure of-order-two with degree-of-freedom $\varphi$. 
\end{theorem}
 The proof of this theorem is relegated to the next subsection (Subsection \ref{Proof4mainTheorem}). On the one hand, Theorem \ref{EsscherMeasures} conveys that the degree of freedom $\varphi$ does not affect the existence of Esscher measure, somehow, as long as $\varphi$ belongs to the set ${\mathcal{I}}^{exp}_{loc}(N^{\mathbb{G}},\mathbb{G})$. On the other hand, when we let $\varphi$ spanning this latter set, we can guarantee the existence of Esscher for the initial model $(S,\mathbb{F})$. In order to see the importance of this latter fact, we consider the case of zero degree of freedom which boils down to having densities of the form of $Z^{\mathbb{G}}:=Z^{\tau}/{\mathcal{E}}(G^{-1}_{-}\is{m})^{\tau}$, where $Z$ is the Esscher for $(S,\mathbb{F})$. In this case, we recall from \cite{choulli2022explicit}, that we could not obtain the uniform integrability of $Z$ from the uniform integrability of $Z^{\mathbb{G}}$.
\subsection{Proof of Theorems \ref{MainTheorem4Esscher} and \ref{EsscherMeasures} }\label{Proof4mainTheorem}
The proof of this theorem is based on some technical results summarized in a lemma and a proposition.
\begin{lemma}\label{ExponentialIntegrability4phi} Let $\varphi$ be an arbitrary $\mathbb{F}$-optional process, and $J (\varphi)$ be given by (\ref{H(phi)Phi(phi)}). Then the following assertions hold.\\
{\rm{(a)}} $\varphi\in{\mathcal{I}}^{exp}_{loc}(N^{\mathbb{G}},\mathbb{G})$ iff $J (\varphi)$ is $N^{\mathbb{G}}$-integrable and $\ln\left(({\widetilde{G}}+GJ (\varphi))({\widetilde{G}}-J (\varphi)\Delta{D}^{o,\mathbb{F}})^{-1}\right)\in{\mathcal{I}}^{o}_{loc}(N^{\mathbb{G}},\mathbb{G})$ if and only if $J (\varphi)\in{\mathcal{I}}^{o}_{loc}(N^{\mathbb{G}},\mathbb{G})$ and $\ln\left(({\widetilde{G}}+GJ (\varphi))({\widetilde{G}}-J (\varphi)\Delta{D}^{o,\mathbb{F}})^{-1}\right)\in{\mathcal{I}}^{o}_{loc}(N^{\mathbb{G}},\mathbb{G})$.\\
{\rm{(b)}} If $\varphi\in{\mathcal{I}}^{exp}_{loc}(N^{\mathbb{G}},\mathbb{G})$, then the process $ U^{(\varphi)}$ defined by,
\begin{equation}\label{U(phi)}
U^{(\varphi)}:={{\exp(-{\varphi}{\widetilde{G}}^{-1}\Delta{D}^{o,\mathbb{F}})(e^{\varphi}-1)-\varphi}\over{\widetilde{G}}}\is {D}^{o,\mathbb{F}}+\sum\left(\exp(-{\varphi}{\widetilde{G}}^{-1}\Delta{D}^{o,\mathbb{F}}) -1+{{\varphi}\over{\widetilde{G}}}\Delta{D}^{o,\mathbb{F}}\right),
\end{equation}
belongs to $ {\mathcal{V}}^+(\mathbb{F})$ and satisfies
\begin{equation}\label{Jumps4U(phi)}
\begin{split}U^{(\varphi)} =&\underbrace{\left(H(\varphi)(1+\beta)^{-1} - 1\right)\is N}_{=:U_1}+ \underbrace{ (e^\varphi - 1 -\varphi){\widetilde{G}}^{-1} I_{\{\widetilde{G} = G\}} \is D^{o,\mathbb{F}}}_{=:U_2}\\
	&+\underbrace{ \sum \bigg(\exp\left(-G_{-}^{-1}\varphi\Delta D^{p,\mathbb{F}}\right)  \bigg(G_{-}^{-1} (e^\varphi - 1)\Delta{D}^{p,\mathbb{F}}  + 1 \bigg) - 1 \bigg)}_{=:U_3},\end{split}
\end{equation}
where $U_2$ is continuous nondecreasing and $U_3$ is $\mathbb{F}$-predictable nondecreasing. \\
{\rm{(c)}} If $\varphi\in{\mathcal{I}}^{exp}_{loc}(N^{\mathbb{G}},\mathbb{G})$, then there exists an $\mathbb{F}$-predictable $\widetilde{U}^{(\varphi)}\in {\mathcal{V}}^+(\mathbb{F})$ such that 
\begin{equation}\label{DEcompo4U(phi)}
{1\over{1+\Delta{U}^{(\varphi)}}}\is{U}^{(\varphi)}=\widetilde{U}^{(\varphi)}+\left(1-{{1+\beta}\over{H(\varphi)}}\right)\is{N}^{\mathbb{F}}.
\end{equation}
 \end{lemma}
The proof of the lemma is relegated to Appendix \ref{Proof4Propositions}, while herein we state some intermediate results which are useful for the proof of Theorem  \ref{MainTheorem4Esscher}.

\begin{proposition}\label{Phi-N(G)} Let $(\theta,\ \psi)\in\Theta(S,\mathbb{F})$, $\varphi$ be an $\mathbb{F}$-optional process, and $(H(\varphi), J (\varphi), {\widetilde{\mathcal{E}}}^{(\varphi)})$ be given by (\ref{H(phi)Phi(phi)}). Then the following assertions hold.\\
{\rm{(a)}}  $\varphi\in{\mathcal{I}}^{exp}_{loc}(N^{\mathbb{G}},\mathbb{G})$ if and only if there exists a unique $\Phi\in{\mathcal{I}}^{o}_{loc}(N^{\mathbb{G}},\mathbb{G})$ such that 
\begin{equation}\label{Exponential2StochExponential}
\begin{split}
&-\widetilde{G}G^{-1}<\Phi,\quad \Phi\Delta{D}^{o,\mathbb{F}}<\widetilde{G},\quad   \varphi=\ln\left({{{\widetilde{G}}+G\Phi}\over{{\widetilde{G}}-\Phi\Delta{D}^{o,\mathbb{F}}}}\right)\in{\mathcal{I}}^{o}_{loc}(N^{\mathbb{G}},\mathbb{G}),\\
&\mbox{and}\quad \exp(\varphi\is{N}^{\mathbb{G}})={\mathcal{E}}(\Phi\is{N}^{\mathbb{G}}){\mathcal{E}}\left(U^{(\varphi)}\right)^{\tau},
\end{split}
\end{equation}
where $U^{(\varphi)}$ is given by (\ref{U(phi)}).  \\
{\rm{(b)}} If $\varphi\in{\mathcal{I}}^{exp}_{loc}(N^{\mathbb{G}},\mathbb{G})$, then we have 
\begin{equation}\label{Equality4(c)and(d)}
{\widetilde{\mathcal{E}}}^{\tau}{\mathcal{E}}(J (\varphi)\is{N}^{\mathbb{G}})=({\widetilde{\mathcal{E}}}^{(0)})^{\tau}{\mathcal{E}}(J (\varphi)\is{N}^{\mathbb{G}})=({\widetilde{\mathcal{E}}}^{(\varphi)})^{\tau}\exp(\varphi\is{D}),
\end{equation}
{\rm{(c)}} Suppose that $(\theta,\ \psi)\in\Theta(S,\mathbb{F})$ such that $\int_0^{\cdot}\exp(\theta_s\zeta_s+\zeta^2_s\psi_s)ds\in{\mathcal{A}}_{loc}^+(\mathbb{F})$, and consider 
\begin{equation}\label{ThetaTildePsiTilde}
\begin{split}
M^\mathbb G&:= \theta \sigma \bigcdot \mathcal{T}(W) + \big(e^{\theta \zeta + \psi \zeta^2} - 1\big)(1+\beta) \bigcdot \mathcal{T}(N^\mathbb{F})\in{\mathcal{M}}_{loc}(\mathbb{G}),\\
\kappa(\theta,\psi)&:= \theta\mathfrak{b} - \theta\zeta\lambda + \theta \sigma\alpha + \frac{\theta^2 \sigma^2}{2} + \big(e^{\theta\zeta + \psi\zeta^2} - 1\big) (1+\beta) \lambda, \\
\theta^\mathbb F&:=\theta+{{\alpha}\over{\sigma}},\quad \zeta^2\psi^\mathbb F:={\zeta^2}\psi-\zeta\frac{\alpha}{\sigma} + \ln\big(1+\beta \big),\\
M^\mathbb F&:= \sigma\theta^\mathbb F\is{W} + \big(e^{\theta^\mathbb F\zeta + \psi^\mathbb F \zeta^2} - 1\big)\is{N}^\mathbb{F}\in{\mathcal{M}}_{loc}(\mathbb{F}).
\end{split}
\end{equation}
Then the following equalities hold.
\begin{equation}\label{theta-X-Exponenetial}
\exp\Bigl( \theta \is Y^\tau +\sum \psi(\Delta{Y}^\tau)^2\Bigr)= \mathcal{E}\left( M^{\mathbb G}+ \int_0^{\tau\wedge\cdot}\kappa_s(\theta,\psi) ds\right)= {{\mathcal{E}\left(M^\mathbb F\right)^{\tau}}\over{{\mathcal{E}}(G_{-}^{-1}\is{m})^{\tau}}} \exp\left( \int_0^{\tau\wedge\cdot}\kappa_s(\theta,\psi) ds\right).
\end{equation}
where $Y$ denotes the pair of processes $X$ or $\widetilde X$ and $\mathfrak{b}= \{b,\widetilde b\}$ are their corresponding drifts.\\
{\rm{(d)}} Let  $(\theta,\ \psi,\varphi)\in\Theta(S,\mathbb{F})\times {\mathcal{I}}^{exp}_{loc}(N^{\mathbb{G}},\mathbb{G})$ such that $\int_0^{\cdot}\exp(\theta_s\gamma_s+\gamma^2_s\psi_s)ds\in{\mathcal{A}}_{loc}^+(\mathbb{F})$. Then (\ref{IntegrabilityCondition4Esscher}) holds if and only if $(1+\Delta{M}^{(\theta,\psi)})\is(U^{(\varphi)})^{\tau}\in{\mathcal{A}}^+_{loc}(\mathbb{G})$. Furthermore, under (\ref{IntegrabilityCondition4Esscher}), we always have 
\begin{equation}\label{Integrability4BigPhi}
J (\varphi){{H(\varphi)}\over{1+\beta}}\Delta{N}\exp(\gamma\theta+\gamma^2\psi)\in {\mathcal{I}}^o_{loc}(N^{\mathbb{G}},\mathbb{G}).\end{equation}
\end{proposition}
For the sake of easy exposition, we relegate the proof of this proposition to Appendix \ref{Proof4Propositions}, and below we characterize {\it exponentially special} $\mathbb{G}$-semimartingale that take the form in (\ref{EsscherDegree}). Besides the important role of this characterization in the proof of Theorem \ref{MainTheorem4Esscher}, it is interesting in itself.
\begin{proposition}\label{ExponentialSpecial4G} Let $(\theta,\psi,\varphi)\in\Theta(S,\mathbb{F})\times{\mathcal{I}}^{exp}_{loc}(N^{\mathbb{G}},\mathbb{G})$. Then $\exp(\theta\is{X}^{\tau}+\sum\psi(\Delta{X}^{\tau})^2+\varphi\is{N}^{\mathbb{G}})$ is a $\mathbb{G}$-special semimartingale if and only if the triplet $(\theta,\psi,\varphi)$ satisfies (\ref{IntegrabilityCondition4Esscher}).
\end{proposition}

The proof of this proposition is relegated to Appendix \ref{Proof4Propositions}, while herein we prove Theorem  \ref{MainTheorem4Esscher}. 
\begin{proof}[Proof of Theorem \ref{MainTheorem4Esscher}]  
Remark that the equivalence  (c) $\Longleftrightarrow$ (d) follows immediately from Proposition \ref{Phi-N(G)}-(b).  Hence, the remaining part of this poof is divided into three parts. The first part proves a crucial fact, while parts 2 and 3 prove the equivalences (a) $\Longleftrightarrow$ (b)  and (c) $\Longleftrightarrow$ (b)  respectively.\\
{\bf Part 1.} Herein, we consider $(\theta,\psi,\varphi,K)\in\Theta(S,\mathbb{F})\times {\mathcal{I}}^{exp}_{loc}(N^{\mathbb{G}},\mathbb{G})\times{\mathcal{A}}_{loc}(\mathbb{G})$ such that  $K$ is $\mathbb{G}$-predictable and (\ref{IntegrabilityCondition4Esscher}) holds,  and prove that the process 
$$Z:= \exp\left(\theta \is X^\tau + \sum\psi(\Delta{X}^\tau)^2+\varphi \is N^\mathbb{G} -K\right)=:\exp\left(Y+\varphi \is N^\mathbb{G} -K\right)$$ is a $\mathbb{G}$-local martingale if and only if $Z$ takes the form given in the right hand side of (\ref{Representation4ZG}). To this end, we combine assertions (a) and (c) of Proposition \ref{Phi-N(G)} and get 
\begin{equation}\label{MultiplicativeDEcopmpo}
\begin{split}
Z&= \mathcal{E}(M^{(\theta, \psi)}) \mathcal{E}(U^{(\varphi)})^{\tau}\exp\left( \int_0^{\tau\wedge\cdot}\kappa_s ds - K\right)\mathcal{E} \big(J (\varphi) \is{N}^\mathbb{G} \big)\\
=&\underbrace{ \mathcal{E}\left(M^{(\theta, \psi)}+(1+\Delta{M}^{(\theta, \psi)})\is(U^{(\varphi)})^{\tau} \right)\exp\left( \int_0^{\tau\wedge\cdot}\kappa_s ds - K\right)}_{=:Z'}\mathcal{E} \big(J (\varphi) \is{N}^\mathbb{G} \big),\end{split}\end{equation}
where the last equality is a direct consequence of Yor's equality and the fact that $U^{(\varphi)}$ has a finite variation. Therefore, as $\exp\left( \int_0^{\tau\wedge\cdot}\kappa_s ds - K\right)$ is RCLL predictable with finite variation and hence locally bounded, and $Z'J (\varphi)\in {\mathcal{I}}^{o}_{loc}(N^{\mathbb{G}},\mathbb{G})$ is equivalent to  (\ref{Integrability4BigPhi}), which is induced by (\ref{IntegrabilityCondition4Esscher}) (see Proposition \ref{Phi-N(G)}-(d)), we obtain $Z'\is\mathcal{E} \big(J (\varphi) \is{N}^\mathbb{G} \big)\in{\mathcal{M}}_{loc}(\mathbb{G})$.  Then direct It\^o calculation combined with (\ref{MultiplicativeDEcopmpo}) shows that $Z$ is a $\mathbb{G}$-local martingale if and only if 
\begin{equation}\label{MartingaleCondition1}
	\mathcal{E}\bigg(M^{(\theta, \psi)} + (1 +\Delta M^{(\theta, \psi)}) \is(U^{(\varphi)})^{\tau} \bigg) \exp\left( \int_0^{\tau\wedge\cdot}\kappa_s ds - K\right) \in \mathcal{M}_{loc}(\mathbb{G}).
\end{equation}
Thus, due to Proposition  \ref{Phi-N(G)}-(d), we get $(1 +\Delta M^{(\theta, \psi)}) \is(U^{(\varphi)})^{\tau} \in{\mathcal{A}}^+_{loc}(\mathbb{G})$, and we put 
\begin{equation}\label{V(phi)}
V^{(\varphi)}:= (1 +\Delta M^{(\theta, \psi)}) \is(U^{(\varphi)})^{\tau}
\quad\mbox{and}\quad M^{\mathbb{G}}:=M^{(\theta, \psi)} +V^{(\varphi)}-(V^{(\varphi)})^{p,\mathbb{G}} .
\end{equation}
On the one hand, we easily derive $1+\Delta{M}^{(\theta,\psi)}=1+\big(e^{\theta \gamma + \psi \gamma^2} - 1\big)\Delta{N}^{\tau}$, and in virtue of Lemma \ref{ExponentialIntegrability4phi}-(b) see (\ref{Jumps4U(phi)}), we obtain
\begin{equation}\label{V(phi)Calculations}
V^{(\varphi)}= U_2^{\tau} +  U_3^{\tau} + e^{\theta \gamma + \psi \gamma^2}\left(H(\varphi)(1+\beta)^{-1} - 1\right)\is{N}^{\tau}.\end{equation}
On the other hand, by combining this with Lemma \ref{Lemma4PoissonIntegrands}-(b) and $N^{\tau}-(N^{\tau})^{p,\mathbb{G}}=(1+\beta)\is{\mathcal{T}}(N^{\mathbb{F}})$, we deduce that \begin{equation}\label{CompensatedV(phi)}
V^{(\varphi)}-(V^{(\varphi)})^{p,\mathbb{G}} =\left(H(\varphi)-1-\beta\right) e^{\theta \gamma + \psi \gamma^2} \is{\mathcal{T}}(N^{\mathbb{F}})\in{\mathcal{M}}_{loc}(\mathbb{G}).
\end{equation}
This proves that $M^{\mathbb{G}}$ takes the form of 
\begin{equation}\label{form4MG}
 M^{\mathbb{G}}=\theta \sigma \bigcdot \mathcal{T}(W) + \big(H(\varphi)e^{\theta \gamma + \psi \gamma^2} - 1-\beta\big)\is \mathcal{T}(N^\mathbb{F}).
 \end{equation}
 Therefore, due to the quasi-left-continuity of $M^{(\theta, \psi)}$ and $V^{(\varphi)}-(V^{(\varphi)})^{p,\mathbb{G}}$  --see the above equality--and Yor's formula, we derive 
\begin{equation}\label{MG}
\mathcal{E}\bigg(M^{(\theta, \psi)} + (1 +\Delta M^{(\theta, \psi)}) \is(U^{(\varphi)})^{\tau} \bigg)=\mathcal{E}\bigg(M^{\mathbb{G}}\bigg){\mathcal{E}}((V^{(\varphi)})^{p,\mathbb{G}}).\end{equation}
Thus, as $M^{\mathbb{G}}$ is a $\mathbb{G}$-local martingale and  both processes $\kappa^{\tau}$ and $K$ are $\mathbb{G}$-predictable with finite variation, we deduce from (\ref{MG}) that $Z$ is a $\mathbb{G}$-local martingale if and only if 
\begin{align*}
K= \int_0^{\tau\wedge\cdot}\kappa_s ds -\ln\mathcal{E}\left((V_2^{\tau})^{p,\mathbb{G}}\right).
\end{align*}
Thus, by combining this with (\ref{MG}), (\ref{form4MG}) and (\ref{MultiplicativeDEcopmpo}), we deduce under the condition (\ref{MartingaleCondition1}) $Z$ is a $\mathbb{G}$-local martingale if and only if it has the form in (\ref{Representation4ZG}). This ends the first part. \\
\textbf{Part 2.} In this part we prove (a) $\Longleftrightarrow$ (b). To this end, we assume that assertion (a) holds. This yields the existence of a quadruplet $(\overline{\theta},\overline{\psi},\overline{\varphi},K)$ which belongs to $\Theta(S,\mathbb{F})\times{\mathcal{I}}^{exp}_{loc}(N^{\mathbb{G}},\mathbb{G})\times{\mathcal{A}}_{loc}(\mathbb{G})$ such that $K$ is $\mathbb{G}$-predictable, $K=K^{\tau}$,
\begin{equation*}
Z^{\mathbb{G}}=\exp\left(\overline\theta\is{X}^{\tau}+\sum\overline\psi(\Delta{X}^{\tau})^2+\overline\varphi\is{N}^{\mathbb{G}}+K\right),\quad\mbox{and}\quad (Z^{\mathbb{G}},Z^{\mathbb{G}}S^{\tau})\in{\mathcal{M}}_{loc}(\mathbb{G})\times{\mathcal{M}}_{loc}(\mathbb{G}).
\end{equation*}
 On the one hand, thanks to Proposition \ref{ExponentialSpecial4G}, we deduce that the triplet  $(\overline{\theta},\overline{\psi},\overline{\varphi})$ satisfies  (\ref{IntegrabilityCondition4Esscher}). On the other hand, by combining this latter fact with Proposition \ref{Phi-N(G)}-(c) and Part 1, we conclude that $Z^{\mathbb{G}}$ fulfills (\ref{Representation4ZG}). Furthermore, it is clear that in this case, $Z^{\mathbb{G}}S^{\tau}\in{\mathcal{M}}_{loc}(\mathbb{G})$ if and only if  (\ref{MainEquation4Esscher}) holds. This proves that  (a) $\Longrightarrow$ (b). To prove the converse, we assume that assertion (b) holds and combine (\ref{theta-X-Exponenetial}), Proposition \ref{Phi-N(G)}-(a), and Lemma  \ref{ExponentialIntegrability4phi}-(b) to derive
\begin{align}
&Z^{\mathbb{G}}\\
&= {\mathcal{E}}\Biggl(\theta \sigma\is\mathcal{T}(W) + \big(H(\varphi)e^{\theta \gamma + \psi \gamma^2} - 1-\beta\big)\is\mathcal{T}(N^\mathbb{F})\Biggr)\mathcal{E} \bigg(J (\varphi) \is N^\mathbb{G} \bigg)\nonumber\\
&= {\mathcal{E}}\Biggl(\theta \sigma\is\mathcal{T}(W) + \big(H(\varphi)e^{\theta \gamma + \psi \gamma^2} - 1-\beta\big)\is\mathcal{T}(N^\mathbb{F})\Biggr){\mathcal{E}}\Biggl({{-1}\over{1+\Delta{U}^{(\varphi)}}}\is{U}^{(\varphi)}\Biggr)^{\tau}\exp(\varphi\is{N}^{\mathbb{G}}) .\nonumber\\
&={\mathcal{E}}\Biggl(\theta \sigma\is\mathcal{T}(W) + \big(H(\varphi)e^{\theta \gamma + \psi \gamma^2} - 1-\beta\big)\is\mathcal{T}(N^\mathbb{F})\Biggr){\mathcal{E}}\Biggl({{-H(\varphi)+1+\beta}\over{H(\varphi)}}\is{N}^{\mathbb{F}}\Biggr)^{\tau}{\mathcal{E}}(-\widetilde{U}^{(\varphi)})^{\tau}\exp(\varphi\is{N}^{\mathbb{G}})\nonumber\\
&={\mathcal{E}}\Biggl(\theta \sigma\is\mathcal{T}(W) + (1+\beta)\big(e^{\theta \gamma + \psi \gamma^2} -1\big)\is\mathcal{T}(N^\mathbb{F})+K^{\mathbb{G},1}\Biggr){\mathcal{E}}(-\widetilde{U}^{(\varphi)})^{\tau}\exp(\varphi\is{N}^{\mathbb{G}})\nonumber\\
&=\exp\Bigl(\theta\is{X}^{\tau}+\sum\psi(\Delta{X}^{\tau})^2+\varphi\is{N}^{\mathbb{G}}+K^{\mathbb{G},2}\Bigr),\label{FinalEquality}
\end{align}
where $K^{\mathbb{G},i}$ $i=1,2$ are $\mathbb{G}$-predictable with finite variation processes.  The second equality is due to a combination of Proposition \ref{Phi-N(G)}-(a) and ${\mathcal{E}}(U^{(\varphi)})^{-1}={\mathcal{E}}(-(1+\Delta{U}^{(\varphi)})^{-1}\is{U}^{(\varphi)} )$, while the third equality is direct consequence of Lemma \ref{ExponentialIntegrability4phi}-(b).  The fourth equality follows from Yor's formula and the fact that $dN^{\tau}=(1+\beta)d{\mathcal{T}}(N^{\mathbb{F}})+(1+\beta)d(t\wedge\tau)$, while the last equality is due to (\ref{theta-X-Exponenetial}) (see Proposition  \ref{Phi-N(G)}-(d)). Hence, by combining (\ref{FinalEquality}) with the fact that $Z^{\mathbb{G}}$ is a $\mathbb{G}$-local martingale, and $Z^{\mathbb{G}}S^{\tau}$ is also a $\mathbb{G}$-local martingale due to (\ref{MainEquation4Esscher}), assertion (a) follows, and this ends part 2. \\
\textbf{Part 3.} In this part we prove  (b) $\Longleftrightarrow$ (c). Thus, suppose assertion (b) hols, and consider
$$\theta^{\mathbb{F}}:=\theta + \frac{\alpha}{\sigma}, \quad {\gamma^2}\psi^{\mathbb{F}}:={\gamma^2}\psi-\gamma\frac{\alpha}{\sigma} + \ln\bigg(H(\varphi) \bigg),\quad \widehat{Z}^{\mathbb{F}}:={\mathcal{E}}\Biggl(\sigma\theta^{\mathbb{F}}\is{W}+\left(e^{\gamma\theta^{\mathbb{F}}+\gamma^2\psi^{\mathbb{F}}}-1\right)\is{N}^{\mathbb{F}}\Biggr).$$
Therefore, on the one hand, as $\int_0^{\cdot}(\sigma_s\theta^{\mathbb{F}}_s)^2 ds\leq 2\int_0^{\cdot}(\sigma_s\theta^{\mathbb{F}}_s)^2 ds+\int_0^{\cdot}(\sigma_s\theta^{\mathbb{F}}_s)^2 ds\in{\mathcal{A}}^+_{loc}(\mathbb{F})$ and $$\sum\vert\psi^{\mathbb{F}}\vert(\Delta{X})^2\leq \gamma^2\vert\psi\vert\is{N}+\vert\alpha\gamma\vert\sigma^{-1}\is{N}+\vert\ln(H(\varphi)\vert\is{N}\in{\mathcal{V}}^+(\mathbb{F}),$$ we deduce that   $( \theta^{\mathbb{F}},\psi^{\mathbb{F}})\in\Theta(S,\mathbb{F})$. On the other hand, due to (\ref{IntegrabilityCondition4Esscher}) and (\ref{MainEquation4Esscher}), the pair $( \theta^{\mathbb{F}},\psi^{\mathbb{F}})$ fulfills 
\begin{equation}\label{mgEquationIntergrability}
\begin{split}
&\int_0^{\cdot}\exp(\gamma_s\theta^{\mathbb{F}}_s+\gamma^2_s\psi^{\mathbb{F}}_s)ds<\infty,\quad\quad P\otimes{dt}-a.e.,\\
\mbox{and}\quad &\widetilde{b}+ \sigma^2\theta^{\mathbb{F}} + \lambda{\widetilde\gamma}\left(\exp(\gamma\theta^{\mathbb{F}} + \gamma^2\psi^{\mathbb{F}})-1\right) = 0,\quad\quad P\otimes{dt}-a.e..
\end{split}
 \end{equation}
Hence, \cite[Theorem 3.13, or Corollary 3.16]{choulli2025second} allows us to conclude that ${\widehat{Z}}^{\mathbb{F}}$ is an Esscher density of-order-two  for $(S,\mathbb{F})$. Furthermore, thanks to \cite[]{choulli2022explicit} or direct calculations show that $({\widehat{Z}}^{\mathbb{F}})^{\tau}/{\mathcal{E}}(G_{-}^{-1}\is{m})^{\tau}=Z^{\mathbb{G}}/{\mathcal{E}}(J (\varphi)\is{N}^{\mathbb{G}})$. This proves that assertion (c) holds, and the proof of (b) $\Longrightarrow$ (c) is complete. To prove the converse implication, we assume that assertion (c) holds. Thus, by applying  \cite[Theorem 3.13]{choulli2025second}, we deduce the existence of a pair $(\theta^{\mathbb{F}}, \psi^{\mathbb{F}})\in\Theta(S,\mathbb{F})$ satisfying  (\ref{mgEquationIntergrability})). Thanks to \cite[Proposition 3.3]{choulli2022explicit} and (\ref{Paramertization4m}), we deduce that 
\begin{equation}\label{Equation400}
{{({Z}^{\mathbb{F}})^{\tau}}\over{{\mathcal{E}}(G^{-1}_{-}\is{m})^{\tau}}}={\mathcal{E}}\left((\sigma\theta^{\mathbb{F}}-\alpha)\is{\mathcal{T}}({W})+(e^{\gamma\theta^{\mathbb{F}}+\gamma^2\psi^{\mathbb{F}}}-1-\beta)\is{\mathcal{T}}({N}^{\mathbb{F}})\right),
\end{equation}
 and we put
\begin{equation}\label{parameters(theta,psi,phi)}
\theta:=\theta^{\mathbb{F}}-\frac{\alpha}{\sigma},\quad  {\gamma^2}\psi:={\gamma^2}\psi^{\mathbb{F}}+\gamma\frac{\alpha}{\sigma}-\ln\bigg(H(\varphi) \bigg),\quad \varphi:=\ln\left({{{\widetilde{G}}+\Phi^{\mathbb{F}}G}\over{{\widetilde{G}}-\Phi^{\mathbb{F}}\Delta{D}^{o,\mathbb{F}}}}\right).\end{equation}
Then, here again, it is easy to check that $(\theta,\psi)\in \Theta(S,\mathbb{F})$ (see above for details). By inserting (\ref{parameters(theta,psi,phi)}) in (\ref{Equation400}),  and combining the obtained process with (\ref{Paramertization4m}) afterwards, we deduce that (\ref{Representation4ZG}) holds. Similarly, by inserting (\ref{parameters(theta,psi,phi)}) in (\ref{mgEquationIntergrability}), we obtain (\ref{MainEquation4Esscher}) and  (\ref{IntegrabilityCondition4Esscher}). Hence, assertion (b) holds, and the proof of the theorem is complete. 
\end{proof}
The remaining part of this subsection addresses the proof of Theorem \ref{EsscherMeasures}, which relies on the following easy lemma. 
 \begin{lemma}\label{lemma5.1bis} Consider $\eta$ defined in (\ref{tau params}). There exists a continuous nondecreasing and $\mathbb{F}$ adapted process $C$ such that $C_0=0$, and  
 \begin{equation}
 D^{o,\mathbb{F}}=D^{o,\mathbb{F}}_0+C+\sum\Delta{D}^{p,\mathbb{F}}+G_{-}\eta\is N.
 \end{equation}
 \end{lemma}
 The proof of this is straightforward from Lemma \ref{Parametrization4tau}, and the fact that $I_{\{\Delta D^{p,\mathbb F} =0\}}\is{D}^{p,\mathbb{F}}$  is nondecreasing and continuous $\mathbb{F}$-adapted process. 
 \begin{proof}[Proof of Theorem \ref{EsscherMeasures} ] Remark that the implication (b) $\Rightarrow$ (c) is obvious. Thus, the remaining part of this proof has  two steps, where we prove the two implication (a) $\Rightarrow$ (b) and (c) $\Rightarrow$ (a) respectively\\
 {\bf Step 1.}  Suppose that assertion (a) holds, and $(S,\mathbb F)$ admits an Esscher pricing measure $\widehat Q$ of-order-two with parameters $(\theta,\psi)$. Then denote the density process of the resulting measure by $Z$, which is a uniformly integrable martingale. For any $\varphi\in{\widetilde{\mathcal{E}}}^{(\varphi)}$, thanks to Theorem \ref{MainTheorem4Esscher}-(b), the positive $\mathbb G$-local martingale 
 $$Z^{\mathbb{G}}:={{Z^{\tau}}\over{{\mathcal{E}}(G_{-}^{-1}\is{m})^{\tau}}}{\mathcal{E}}(J (\varphi)\is{N}^{\mathbb{G}}),$$ 
 is an Escher density of-order-two with degree-of-freedom $\varphi$. Thus, assertion (b) will follow as soon as we prove that  $Z^{\mathbb{G}}$ is in fact a $\mathbb{G}$-martingale uniformly integrable. To this end, we use the equivalent representation of Theorem \ref{MainTheorem4Esscher}-(c), and we derive
  \begin{equation}
  \begin{split}
   E[Z^{\mathbb{G}}_T]&=E\left[{{{\widetilde{\mathcal{E}}}^{(\varphi)}_{\tau\wedge T}}\over{G_{\tau\wedge T}}}Z^{(\theta,\psi)}_{\tau\wedge T}\exp(\varphi_{\tau}I_{\{0<\tau\leq T\}})\right]\\
   &=1-G_0+ 
   G_0E\left[{{{\widetilde{\mathcal{E}}}^{(\varphi)}_{\tau}e^{\varphi_{\tau}}}\over{G_{\tau}}}Z^{(\theta,\psi)}_{\tau}\exp(\varphi_{\tau})I_{\{0<\tau\leq T\}}\right]+G_0E\left[{{{\widetilde{\mathcal{E}}}^{(\varphi)}_T}\over{G_T}}Z^{(\theta,\psi)}_ T I_{\{\tau>T\}}\right]\nonumber\\
   &=1-G_0+G_0E\left[{\widetilde{\mathcal{E}}}^{(\varphi)}_TZ^{(\theta,\psi)}_ T\right]-G_0E\left[\int_{0}^{T}Z^{(\theta,\psi)}_t{d}{\widetilde{\mathcal{E}}}^{(\varphi)}_t\right]=1+ G_0E\left[\int_{0}^{T}{\widetilde{\mathcal{E}}}^{(\varphi)}_{t-}dZ^{(\theta,\psi)}_t\right].\label{ImportantInequa1}
  \end{split} \end{equation}
  Remark that $0<{\widetilde{\mathcal{E}}}^{(\varphi)}\leq1$, and hence ${\widetilde{\mathcal{E}}}^{(\varphi)}_{-}\is{Z}^{(\theta,\psi)}$ is also a uniformly integrable martingale. This combined with (\ref{ImportantInequa1}) imply that  $E[Z^{\mathbb{G}}_T]=1$ and we deduce that $Z^{\mathbb{G}}$  is a uniformly integrable $\mathbb{G}$-martingale. Hence assertion (b)  holds, and this ends the first step.\\
  {\bf Step 2.}  Hereto,  we prove assertion (b) $\Rightarrow$ (c). To this end, we assume that assertion (b) holds. Hence, assertion (c) will follows as soon as we prove that there always exists $\varphi \in{\widetilde{\mathcal{E}}}^{(\varphi)}$ such that $1/{\widetilde{\mathcal{E}}}^{(\varphi)}$ is bounded. To prove this latter fact, we start by using Lemma \ref{lemma5.1bis}  and considering $\varphi$ given by 
  \begin{equation}\label{phi-Process}
  e^{\varphi}:={{G}\over{(1+C)^2}}I_{\{\widetilde{G}=G\}}+{{G}\over{(1+D^{p,\mathbb{F}})(1+D^{p,\mathbb{F}}_{-})}}I_{\{\Delta{D}^{p,\mathbb{F}}\not=0\}}+Ge^{-N}(1-e^{-1})I_{\{\eta\Delta{N}\not=0\}}.
  \end{equation} 
On the one hand, we remark that $\exp(\varphi^+)\widetilde{G}^{-1}I_{\Rbrack0,\tau\Rbrack}\is{D}^{o,\mathbb{F}}\in{\mathcal{V}}^+(\mathbb{G})$ if and only if  $\exp(\varphi)\widetilde{G}^{-1}I_{\Rbrack0,\tau\Rbrack}\is{D}^{o,\mathbb{F}}\in{\mathcal{V}}^+(\mathbb{G})$, and this latter fact follows from 
\begin{equation}\label{boundedness4phi}
{{e^{\varphi}}\over{\widetilde{G}}}I_{\Rbrack0,\tau\Rbrack}\is{D}^{o,\mathbb{F}}={{I_{\Rbrack0,\tau\Rbrack}}\over{(1+C)^2}}\is{C}+{{GI_{\{\Delta{D}^{p,\mathbb{F}}\not=0\}}I_{\Rbrack0,\tau\Rbrack}}\over{\widetilde{G}(1+D^{p,\mathbb{F}})(1+D^{p,\mathbb{F}}_{-})}}\is {D}^{p,\mathbb{F}}+\eta{{G}\over{\widetilde{G}}}e^{-N}(1-e^{-1})\is{N}^{\tau}\leq 3.
\end{equation} 
Furthermore, we have 
\begin{equation}\label{Bound4phi}
\begin{split}
{{G\vert\varphi\vert}\over{\widetilde{G}}}\is{D}^{o,\mathbb{F}}&=\vert\varphi\vert\is{C}+{{G\vert\varphi\vert}\over{\widetilde{G}}}I_{\{\Delta{D}^{p,\mathbb{F}}\not=0\}}\is{D}^{p,\mathbb{F}}+G_{-}{{G\vert\varphi\vert}\over{\widetilde{G}}}\eta{I}_{\{\eta\Delta{N}\not=0\}}\is{N}\\
&\leq \underbrace{\left(-\ln(G)+2\ln(1+C)\right)\is{C}}_{=:V_1}+  \underbrace{\left(-\ln(G)+2\ln(1+{D}^{p,\mathbb{F}})\right)\is{D}^{p,\mathbb{F}}}_{=:V_2}\\
&+{{\eta}\over{1+\beta}}(1+\beta-\eta)\left(N-\ln(1+\beta-\eta)\right)\is{N}.
\end{split}
\end{equation} 
It is clear that both $V_1$ and $V_2$ are nondecreasing RCLL and $\mathbb{F}$-predictable, and hence they are locally bounded. It is also easy to check that ${{\eta}\over{1+\beta}}(1+\beta-\eta)\left(N-\ln(1+\beta-\eta)\right)$ is locally bounded. Hence, by combining these remarks, with (\ref{Bound4phi}), we conclude that $\varphi\in{\mathcal{I}}^{o}_{loc}(N^{\mathbb{G}},\mathbb{G})$, and by suing this combined with (\ref{boundedness4phi}) we conclude that $\varphi$ belongs to ${\mathcal{I}}^{exp}_{loc}(N^{\mathbb{G}},\mathbb{G})$. On the other hand, thanks to It\^o and the fact that $\ln(1+x)\leq x$ for any $x>-1$, we derive
\begin{equation*}
\begin{split}
{\widetilde{\mathcal{E}}}^{(\varphi)}&=\exp\biggl(-{{e^{\varphi}}\over{G}}I_{\{\widetilde{G}=G\}}\is{D}^{o,\mathbb{F}}+\sum\ln\left(1-{{e^{\varphi}\Delta{D}^{o,\mathbb{F}}}\over{G+e^{\varphi}\Delta{D}^{o,\mathbb{F}}}}\right)\biggr)\\
&=\exp\biggl(-{{e^{\varphi}}\over{G}}\is{C}+\sum\ln\left(1-{{e^{\varphi}\Delta{D}^{p,\mathbb{F}}}\over{G+e^{\varphi}\Delta{D}^{p,\mathbb{F}}}}\right)+\ln\left(1- {{\eta{e}^{\varphi}}\over{1+\beta-\eta+e^{\varphi}\eta}}\right)\is{N}\biggr)\\
&\geq\exp\biggl(-{{e^{\varphi}}\over{G}}\is{C}-\sum{{e^{\varphi}\Delta{D}^{p,\mathbb{F}}}\over{G}}- {{\eta{e}^{\varphi}}\over{G}}\is{N}\biggr)\\
&=\exp\biggl(-{1\over{(1+C)^2}}\is{C}-\sum{{\Delta{D}^{p,\mathbb{F}}}\over{(1+D^{p,\mathbb{F}})(1+D^{p,\mathbb{F}}_{-})}}- e^{-N}(1-e^{-1})I_{\{\eta\not=0\}}\is{N}\biggr)\\
&\geq\exp\biggl({1\over{1+C}}-1+{1\over{1+D^{p,\mathbb{F}}}}-1+e^{-N}-1\biggr)\geq e^{-3}.
\end{split}
\end{equation*} 
This proves that assertion (c) holds, and ends the second step.\\
{\bf Step 3.}  This step proves (c) $\Rightarrow $ (a). Thus, we assume that assertion (c) holds, and consider $\varphi\in {\mathcal{I}}^{exp}_{loc}(N^{\mathbb{G}},\mathbb{G})$ such that $1/{\widetilde{\mathcal{E}}}^{(\varphi)}\leq c_1$ for some constant $c_1>1$, and $(S^{\tau},\mathbb{G})$ admits an Esscher pricing measure of-order-two with degree-of-freedom $\varphi$. Let $Z^{\mathbb{G}}$ be its density process, which is uniformly integrable $\mathbb{G}$-martingale. Thus, on the one hand, it is clear that  $Z^{\mathbb{G}}/{\widehat{\mathcal{E}}}^{\tau}$ is a uniformly integrable $\mathbb{G}$-submartingale. On the other hand, thanks to Theorem  \ref{MainTheorem4Esscher}-(c), there exists an Esscher density of-order-two $Z$, for the model $(S,\mathbb{F})$, such that 
  $$
  Z^{\mathbb{G}}=G_0{{({\widetilde{\mathcal{E}}}^{(\varphi)})^{\tau}Z^{\tau}}\over{G^{\tau}}}\exp(\varphi\is{D}).$$
 As a result, we get $(Z^{\mathbb{G}}/({\widetilde{\mathcal{E}}}^{(\varphi)})^{\tau})I_{\Lbrack0,\tau\Lbrack}=G_0(Z/G)I_{\Lbrack0,\tau\Lbrack}$, which is also a uniformly integrable process (of class under $\mathbb{G}$). Then, for any constant $c>0$ and any $\mathbb{F}$-stopping time $\sigma$, we derive
   \begin{equation}
  \begin{split}
  E\left[\vert{Z}_{\sigma}\vert{I}_{\{\vert{Z}_{\sigma}\vert>c\}}\right]&= E\left[{{\vert{Z}_{\sigma}\vert}\over{G_{\sigma}}}I_{\{\tau>\sigma\}}{I}_{\{\vert{Z}_{\sigma}\vert>c\}}\right]\\
  &\leq E\left[{{\vert{Z}_{\sigma}\vert}\over{G_{\sigma}}}I_{\{\tau>\sigma\}}{I}_{\{G_{\sigma}^{-1}\vert{Z}_{\sigma}\vert>c\}}\right]=E\left[{{\vert{Z^{\mathbb{G}}}_{\sigma}}\over{{\widetilde{\mathcal{E}}}^{(\varphi)}_{\sigma}G_0}}\vert{I}_{\{\tau>\sigma\}}{I}_{\{\vert{Z^{\mathbb{G}}}_{\sigma}({\widetilde{\mathcal{E}}}^{(\varphi)}_{\sigma}G_0)^{-1}\vert>c\}}\right].
    \end{split} \end{equation}
   This proves that $Z$ is is uniformly integrable $\mathbb{F}$-martingale, and hence $(S,\mathbb{F})$ admits an Esscher pricing measure of-order-two. This ends the third step and completes the proof of the theorem is complete.  \end{proof}
\section{Esscher bounds for vulnerable claims}
Throughout this section we consider the model defined in \eqref{the model}, and in addition to Assumption \ref{assum: model parameters locally bounded}, we assume the following
\begin{assum}\label{assum: gamma non-zero}
   There exists $C\in(0,\infty)$ such that $|\sigma|^{-1}+|\gamma|^{-1}\leq C$ $P\otimes dt$-a.e..
\end{assum}
By working with discounted price processes for the stocks,  there is no loss of generality in assuming that the interest rate is null (i.e. $r =0$). Thus, throughout this section we assume that $r\equiv0$.\\

For the rest of this section we consider the following sets
\begin{equation}\label{Psi-Psi(n)Phi-Phi(n)}
\begin{split}
\Psi&:=\left\{\psi:\ \psi\ \mbox{is predictable and bounded}\right\},\quad\Psi_n:=\left\{\psi\in\Psi:\ \vert\psi\vert\leq n\right\},\quad n\in\mathbb{N}.\\
\Phi&:=\left\{\varphi:\ \varphi \ \mbox{is $\mathbb{F}$-optional and bounded} \right\},\quad\Phi_n:=\left\{\varphi\in\Phi:\ \vert\varphi\vert\leq n\right\},\quad n\in\mathbb{N}.\end{split}
\end{equation}
For the reader's convenience, we recall the following functions defined on $\Phi$ as follows: For any $\varphi\in \Phi$,
\begin{equation}
    \begin{split}
        H(\varphi):= \bigg(\eta(e^\varphi - 1) +1 +\beta \bigg) \exp\bigg(-\frac{\eta \varphi}{1+\beta} \bigg)\quad \text{and}\quad  J(\varphi) := \frac{\widetilde G(e^\varphi - 1)}{G+e^\varphi \Delta D^{o,\mathbb{F}}}
    \end{split}
\end{equation}
and the Esscher parameters $(\theta, \psi,\varphi)$ solve the following martingale condition
\begin{equation}
    \widetilde b +\theta \sigma^2 + \alpha \sigma + \lambda \widetilde \gamma \bigg(H(\varphi)e^{\theta\zeta + \psi \zeta^2} -1 \bigg) = 0 \quad \mathbb P \otimes dt-a.e.
\end{equation}

\begin{lemma}\label{lemma: Esscher and Girsanov} Suppose Assumptions \ref{assum: model parameters locally bounded} and \ref{assum: gamma non-zero} are satisfied. For any $(\psi,\varphi )\in\Psi\times\ \Phi$ and $\zeta\in\{\gamma,\widetilde\gamma\}$, by $\theta(\psi, \varphi)$ we denote the unique root for 
\begin{equation}\label{eq: M.C.}
 \widetilde b +\theta \sigma^2 + \alpha \sigma + \lambda \widetilde \gamma \bigg(H(\varphi)e^{\theta \zeta + \psi \zeta^2} -1 \bigg) = 0 \quad \mathbb P \otimes dt-a.e.,
\end{equation}
and consider the following processes  
\begin{equation}\label{Ella1}
\begin{split}
\overline{\mathcal D}^{(\psi,\varphi)}&= \mathcal E \bigg( \theta (\psi, \varphi) \sigma \is \mathcal T (W) + \big[H(\varphi) e^{\theta (\psi, \varphi) \zeta + \psi \zeta^2} - (1+\beta)\big]\is \mathcal T (N^\mathbb F) \bigg) \mathcal E\bigg( J (\varphi) \is N^\mathbb{G} \bigg)\\
 W^{(\psi,\varphi)}&:=W^\tau- \theta(\psi, \varphi) \sigma \is (t \wedge \tau)={\cal{T}}(W)-\sigma\theta(\psi,\varphi))\is(\tau\wedge\cdot),\quad W^{(0)}:= W^{(0,0)},\\
  \widetilde N^{(\psi,\varphi)} &:=   (1+\beta)^{-1}\is(N^{\mathbb{F}})^\tau + \frac{ \widetilde{b} +\theta(\psi,\varphi)\sigma^2}{\widetilde\gamma(1+\beta)}  \is (t \wedge \tau)\\
  &={\cal{T}}(N^{\mathbb{F}})-\lambda\left(H(\varphi))(1+\beta)^{-1}\exp(\theta(\psi,\varphi)\zeta+\zeta^2\psi)-1\right)\is(\tau\wedge\cdot),\quad\quad \widetilde N^{(0)}:= \widetilde N^{(0,0)}.
\end{split}
\end{equation}
Then the following assertions hold.\\
{\rm{(a)}} For any $\psi\in\Psi$ and $\varphi\in \Phi$ we have $\overline{\mathcal{D}}^{(\psi,\varphi)}\in{\mathcal{M}}(\mathbb{G}, \mathbb P)$, and 
\begin{equation}\label{eq: Q(psi, phi)}
\mathbb Q^{(\psi, \varphi)}:=\overline{\mathcal{D}}^{(\psi,\varphi)}_T\cdot \mathbb P\quad\mbox{is a well defined probability measure}.
\end{equation}
 In particular, we have $\overline{\mathcal{D}}^{(0,0)}\in{\mathcal{M}(\mathbb{G}, \mathbb P)}$ for $(\psi, \varphi)=(0,0)\in\Psi\times \Phi$, and  
\begin{equation}\label{Qzero}
\mathbb Q_0:=\overline{\mathcal{D}}^{(0,0)}_T\cdot \mathbb P\quad\mbox{is a well defined probability measure}.
\end{equation}
Moreover, we denote the expectation under measure $\mathbb Q_0$ by $\mathbb E_0[\cdot]$.\\
{\rm{(b)}} For any $(\psi, \varphi)\in\Psi\times \Phi$, the two processes $W^{(\psi,\varphi)}$ and $\widetilde N^{(\psi,\varphi)}$ belong to ${\mathcal{M}}_{loc}(\mathbb G, \mathbb Q^{(\psi,\varphi)})$.\\
{\rm{(c)}} For $p\geq 1$ and $n\in\mathbb{N}$, there exists $C_n\in(0,\infty)$ such that  
\begin{equation}\label{Reverse_Holder_4D(psi,phi)}
\begin{split}
&\vert \theta(\psi, \varphi)\vert + e^\varphi  +\mathbb{E}_0\left[\left({{\mathcal D^{(\psi,\varphi)}_T}\over{\mathcal D^{(\psi, \varphi)}_t}}\right)^p\Big|{\mathcal{G}}_t\right]\leq C_n,\quad P\mbox{-a.s.},\quad\mbox{for any}\quad \psi\in\Psi_n,\  \varphi\in\Phi_n
\end{split}
\end{equation}
where
\begin{equation}\label{D(psi,phi)}
\begin{split}
&{\mathcal D}^{(\psi, \varphi)}:={{\overline{\mathcal D}^{(\psi, \varphi)}}\over{\overline{\mathcal D}^{(0,0)}}}={\mathcal{E}}\left(L^\theta\right) \mathcal E\bigg( J (\varphi) \is N^\mathbb{G} \bigg),\\
& L^\theta:= \Delta_{\theta}\sigma \is W^{(0)} + \bigg[\frac{H(\varphi)}{1+\beta} e^{\Delta_{\theta}\zeta + \psi \zeta^2}-1 \bigg](1+\beta )\is {\widetilde{N}}^{(0)},\ \Delta_{\theta}:=\theta(\psi,\varphi) - \theta(0,0)
\end{split}
\end{equation}

{\rm{(d)}} The $\mathbb{G}$-Esscher density has the following decomposition
\begin{equation}\label{eq: Z^G decomposition}
    \overline{\mathcal{D}}^{(\psi, \varphi)} = (\mathcal{D}^\mathbb{F})^\tau \ \widetilde {\mathcal{D}} \ \mathcal E \bigg( J(\varphi) \is N^\mathbb{G} \bigg)
\end{equation}
where
\begin{equation}
    \widetilde {\mathcal{D}} = \frac{1}{\mathcal E (G_-^{-1} \is m)^\tau}
\end{equation}
is the density of the following equivalent martingale measure under filtration $\mathbb{G}$,
\begin{equation}\label{eq: Q tilde}
    \widetilde {\mathbb Q} = \widetilde {\mathcal D}_T \is \mathbb P
\end{equation}
\end{lemma}

The proof of this lemma can be obtained by following the footsteps of the proof of \cite[lemma 4.3]{choulli2025second}, while herein we state a results that shows that the representation of \cite{choulli2020martingale} remains valid under the change of probability $\mathbb Q_0$. 

\begin{lemma}\label{lemma: Q-martingale representation} For any $M\in{\cal{M}}_{loc}(\mathbb Q_0,\mathbb{G})$, the following representation hold 
$$
M=M_0+Z\is{W}^{(0)}+U\is{\widetilde{N}}^{(0)} +V\is {N}^{\mathbb{G}}.$$
\end{lemma}
\begin{proof}
Let $M\in \mathcal M_{loc}(\mathbb Q_0, \mathbb G)$, then $L^\mathbb G:=\overline{D}^{(0,0)} M \in\mathcal M_{loc}(\mathbb P, \mathbb G)$ where we recall that
\begin{equation}
    \overline{D}^{(0,0)} = \mathcal E\bigg(\theta_0 \sigma \is \mathcal T(W) +(e^{\theta_0 \zeta} -1)(1+\beta)\is \mathcal T(N^\mathbb F) \bigg),\quad \theta_0:=\theta(0,0).
\end{equation}
Therefore, we need to calculate
\begin{equation}\label{eq: M in terms of L^G and D^(0,0)}
    M = \frac{L^\mathbb G}{\overline{D}^{(0,0)}} = \frac{L^\mathbb G_0}{\overline{D}^{(0,0)}_0} + \frac{1}{\overline{D}^{(0,0)}_-}\is L^\mathbb G +L^\mathbb G_-\is \frac{1}{\overline{D}^{(0,0)}}+\frac{1}{\overline{D}^{(0,0)}_-} \is [\overline{M}^{(0,0)}, L^\mathbb G]
\end{equation}
where $\overline{M}^{(0,0)}$, obtained using Yor's formula applied to $1/\overline{D}^{(0,0)}$, and which is equal
\begin{equation}
    \begin{split}
        \overline{M}^{(0,0)} = -\theta_0 \sigma \is \mathcal T(W) - (e^{\theta_0 \zeta} -1) (1+\beta) \is \mathcal T(N^\mathbb F) + \theta^2_0 \sigma^2 \is(\tau \wedge t) + (e^{\theta_0 \zeta}-1)^2 e^{-\theta_0 \zeta} \is N^\tau
    \end{split}
\end{equation}
By \cite{choulli2022explicit} for any $L^\mathbb G \in \mathcal M_{loc}(\mathbb P, \mathbb G)$, there exists a pair $(M^\mathbb F, \overline{V})\in \mathcal M_{lox}(\mathbb P, \mathbb F) \times \mathcal I_{loc}^o(N^\mathbb G, \mathbb G)$ such that
\begin{equation}
    (L^\mathbb G )^\tau = L_0^\mathbb G + G_-^{-2} \is \mathcal T(M^\mathbb F) + \overline{V} \is N^\mathbb G
\end{equation}
By the martingale representation theorem of $\mathbb F$-martingales, the last decomposition becomes
\begin{equation}
    (L^\mathbb G )^\tau = L_0^\mathbb G + G_-^{-2}\overline{Z} \is \mathcal T(W) + G_-^{-2}\overline{U} \is \mathcal T(N^\mathbb F)  + \overline{V} \is N^\mathbb G
\end{equation}
Now, we plug the last two processes in \eqref{eq: M in terms of L^G and D^(0,0)} and obtain
\begin{equation}
    \begin{split}
        M &= \frac{1}{\overline{D}^{(0,0)}_-} \bigg[\big(G_-^{-2} \overline{Z} - L^\mathbb G_- \theta \sigma\big) \is \mathcal T(W) +\big(G_-^{-2}U-L_-^\mathbb G (e^{\theta \zeta}-1)(1+\beta)\big) \is \mathcal T(N^\mathbb F)\\
        &\quad +\bigg(L^\mathbb G_-(e^{\theta \zeta}-1)^2e^{-\theta \zeta} - (e^{\theta \zeta}-1)G_-^{-2}\frac{U}{1+\beta} +(e^{\theta \zeta}-1) +(e^{\theta \zeta}-1)^2 e^{-\theta \zeta} G_-^{-2} \frac{U}{1+\beta}\bigg) \is N^\tau\\
        &\quad +\bigg(1 - (e^{\theta \zeta} -1)\Delta N + (e^{\theta \zeta} -1)^2 e^{-\theta \zeta} \Delta N \bigg) \overline{V}\is N^\mathbb G + (L_-^\mathbb G \theta^2 \sigma^2 -\theta \sigma G_-^{-2} \overline{Z})\is (\tau \wedge t)
        \bigg]
    \end{split}
\end{equation}
Thanks to equations \eqref{Ella1} in Lemma \ref{lemma: Esscher and Girsanov} the latter equation becomes
\begin{equation}
    \begin{split}
        M =M_0+Z\is{W}^{(0)}+U\is{\widetilde{N}}^{(0)} +V\is {N}^{\mathbb{G}},
    \end{split}
\end{equation}
where 
\begin{equation}
    \begin{split}
        M_0&:= \frac{L^\mathbb G_0}{\overline{D}^{(0,0)}_0},\quad {Z}:= \frac{1}{\overline{D}^{(0,0)}_-} \bigg(G_-^{-2} \overline{Z} - L^\mathbb G_- \theta \sigma \bigg)\\
        U&:= \frac{1}{\overline{D}^{(0,0)}_-} \bigg(G_-^{-2} \overline{U} - L_-^\mathbb G(e^{\theta \zeta} -1)e^{-\theta \zeta}(1+\beta) - (e^{\theta\zeta}-1)e^{\theta \zeta}G_-^{-2}\overline{U}\bigg)\\
        V&:= \bigg(1 - (e^{\theta \zeta} -1)\Delta N + (e^{\theta \zeta} -1)^2 e^{-\theta \zeta} \Delta N \bigg) \overline{V}
    \end{split}
\end{equation}
This ends the proof of the lemma.
\end{proof}
Herein, we define the Esscher price processes bounds for any claim. To this end, we consider ${\mathcal{Q}}(\zeta)$ and ${\mathcal{Z}}(\zeta)$, for $\zeta\in\{\gamma,\widetilde\gamma\}$, given by
\begin{equation}\label{Psi(martingales)}
 {\mathcal{Q}}(\zeta):=\left\{\mathbb Q^{(\psi,\varphi)}:=\mathcal D^{(\psi, \varphi)}_T\cdot{\mathbb Q}_0:\  \mathcal D^{(\psi, \varphi)}\in {\mathcal{Z}}(\zeta)\right\}\quad\mbox{and}\quad {\mathcal{Z}}(\zeta):=\left\{\mathcal D^{(\psi, \varphi)}:\ \psi\in\Psi, \varphi \in\Phi\right\}.
\end{equation}
\begin{definition}\label{optprice} Let $\zeta\in \{\gamma,\widetilde\gamma\}$ and $\xi \in L^0(\mathcal F_{T\wedge \tau})$ be the payoff of an arbitrary claim satisfying

\begin{equation}\label{ClaimAssumption}
\mathbb{E}^{\mathbb Q_0}\left[\vert\xi\vert\right]\leq \sup_{R^{\psi, \varphi}\in{\mathcal{Q}}(\zeta)}\mathbb{E}^{R^{\psi, \varphi}}\left[\vert\xi\vert\right]=\sup_{\mathcal D^{(\psi, \varphi)}\in{\mathcal{Z}}(\zeta)}\mathbb{E}_0\left[\mathcal D^{\psi,\varphi}_T\vert\xi\vert\right]<\infty.
\end{equation}
We denote by $\Rbrack{Y}^{\rm{inf,exp}},Y^{\rm{up,exp}} \Lbrack$ (respectively $\Rbrack{Y}^{\rm{inf,lin}},Y^{\rm{up,lin}} \Lbrack$) the exponential-Esscher  (respectively the linear-Esscher) pricing stochastic interval for the claim $\xi$, where 
\begin{equation}\label{EsscherInterval}
\begin{split}
&Y^{\rm{inf,exp}}:= \underset{\psi\in\Psi, \varphi\in\Phi:\ \mathbb Q^{(\psi,\varphi)}\in{\mathcal{Q}}(\gamma)}{\essinf}(Y^{(\psi, \varphi)}),\quad  Y^{\rm{up,exp}}:= \underset{\psi\in\Psi, \varphi\in\Phi:\ \mathbb Q^{(\psi,\varphi)}\in{\mathcal{Q}}(\gamma)}{\esssup} (Y^{(\psi, \varphi)}),\\
&Y^{\rm{inf,lin}}:= \underset{\psi\in\Psi, \varphi\in\Phi:\ \mathbb Q^{(\psi,\varphi)}\in{\mathcal{Q}}(\widetilde\gamma)}{\essinf}(Y^{(\psi, \varphi)}),\quad  Y^{\rm{up,lin}}:=  \underset{\psi\in\Psi, \varphi\in\Phi:\ \mathbb Q^{(\psi,\varphi)}\in{\mathcal{Q}}(\widetilde\gamma)}{\esssup} (Y^{(\psi, \varphi)}),\\
&Y^{(\psi, \varphi)}_t:= \mathbb{E}_0 \left[{{\mathcal D^{(\psi, \varphi)}_T}\over{\mathcal D^{(\psi, \varphi)}_t}}e^{-\int_t^Tr_sds}\xi \Big| \mathcal{G}_t \right],\quad{\mathcal D}^{(\psi, \varphi)}\in{\mathcal{Z}}(\zeta),
\end{split}
\end{equation}
\end{definition}

The rest of this section is divided into three subsections. The first subsection gives some definitions and notation regarding BSDEs with constraints that we {\it naturally} encounter in our analysis. The second subsection is devoted to our main results of this section, while the last subsection proves these main results.

\subsection{Constrained BSDEs: Definitions  and notation}
Throughout this subsection, we consider a probability measure $Q$ on $(\Omega,{\mathcal{G}})$,  and we denote by $W^Q$ and $\widetilde{N}^Q$ the $Q$-Brownian motion and the compensated Poisson process under $Q$ respectively. For an y $p\in(1,\infty)$, throughout the rest of the paper we consider spaces, $\mathbb{S}^p(Q)$, $\mathbb{L}^p(W,Q)$, $\mathbb{L}^p(N,Q)$ and ${\mathcal{A}}^2(Q)$ and their norms given below.  For any unexplained notion, we refer to Section \ref{Section4ModelNotations}.
\begin{equation}\label{Spaces4BSDEs}
\begin{split}
&\mathbb{S}^p(Q):=\Biggl\{X\ \mbox{RCLL $\&$ $\mathbb{F}$-adapted process}:\ \Vert{X}\Vert_{\mathbb{S}^p(Q)}<\infty\Biggr\},\ \Vert{X}\Vert_{\mathbb{S}^p(Q)}^p:=\mathbb{E}^{Q}\Bigl [\sup_{0\leq t\leq T\wedge \tau} \vert{X}_t\vert^p\Bigr], \\
&\mathbb{L}^p(N,Q):= \left\{\varphi\in L(\widetilde{N}^Q,Q):\ \Vert\varphi\Vert_{\mathbb{L}^p(N,Q)}<\infty \right\},\quad \Vert\varphi\Vert_{\mathbb{L}^p(N,Q)}^p:=\mathbb{E}^Q \Bigl[(|\varphi|^2\is{N})_T^{p/2}\Bigr],\\
  &\mathbb{L}^p(W,Q):=\Bigl \{\varphi\in L(W^Q,Q):\  \Vert\varphi\Vert_{\mathbb{L}^p(W,Q)}<\infty \Bigr\},\quad \Vert\varphi\Vert_{\mathbb{L}^p(W,Q)}^p:=\mathbb{E}^Q \Bigl[\left(\int_0^{T\wedge \tau} |\varphi_t|^2{dt}\right)^{p/2}\Bigr],\\
  &\mathbb{D}^p(Q):= \Bigl \{\varphi \in \mathbb F \text{-optional}: \ \Vert\varphi\Vert_{\mathbb{D}^p(Q)}<\infty \Bigr\},\quad \Vert\varphi\Vert_{\mathbb{D}^p(Q)}^p:= \mathbb{E}^Q \Bigl[\left(\int_0^{T\wedge\tau} |\varphi_t|^2 dD \right)^{p/2} \Bigr],\\
 &{\mathcal{A}}^p(Q):=\Bigl\{V\in{\mathcal{A}}_{loc}(Q):\ \Vert{V}\Vert_{{\mathcal{A}}^p(Q)}<\infty\Bigr\},\quad \Vert{V}\Vert_{{\mathcal{A}}^p(Q)}^p:=\mathbb{E}^Q\left[(\mbox{Var}_T(V))^p\right].
 \end{split}
\end{equation}
Below we introduce {\it constrained} BSDEs for models with jumps.
\begin{definition}{\rm Let  $\xi\in L^p(Q)$ and consider $f$ and $\Sigma$
	two $\mathbb{F}$-optional multi-component functionals in the variables $(t,\omega, y,z,u,v)$ such that $f$ is real-valued and Lipschitz in the variables $(y,z,u,v)\in\mathbb{R}^4$ while $\Sigma(t,\omega,y,z,u)\in \mathbb R^d$  for almost every $(\omega,t)$ and for any $(y,z,u,v)$. \\
{\rm{(a)}} We call $(Y,Z,U,V,K)$ an $L^p(Q)$-solution to the constrained BSDE$(f,\xi,\Sigma)$ \linebreak (CBSDE$(f,\xi,\Sigma)$ hereafter for short) if the following properties hold: $(Y,Z,U,V,K)$ belongs to $\mathbb{S}^p(Q)\times\mathbb{L}^p(W,Q)\times\mathbb{L}^p(N,Q)\times \mathbb{D}^p(Q)\times {\mathcal{A}}^p(Q)$, $K\in {\mathcal{V}}^+$ and is predictable,
 \begin{equation}\label{BSDEwithCobstraints}
 \begin{split}
 &Y_t=\xi+\int_t^Tf(s,Y_s,Z_s,U_s, V_s)ds-\int_t^T Z_s dW^Q_s-\int_t^TU_sd\widetilde{N}^Q_s-\int_t^TV_s dN^\mathbb G+\int_t^T{K}_sds,\quad{Q}\mbox{-a.s.,}\\
 &\mbox{and}\quad \Sigma(t,Y_t,Z_t,U_t, V_t)=0,\quad Q\otimes{dt}\mbox{-a.e.}.
 \end{split}
 \end{equation}
 {\rm{(b)}}  $(\widehat{Y},\widehat{Z},\widehat{U},\widehat{V},\widehat{K})$ is called the smallest $L^p(Q)$-solution to the CBSDE$(f,\xi,\Sigma)$, i.e.\ (\ref{BSDEwithCobstraints}), if it is a solution to this CBSDE$(f,\xi,\Sigma)$, and $\widehat{Y}\leq Y$ for any other solution $(Y,Z,U,V,K)$ to (\ref{BSDEwithCobstraints}).}
 \end{definition}
It is clear, from this definition, that a reflected BSDE is a particular case of a constrained BSDE, where the constraint factor $\Sigma$ depends on the value process $Y$ only (i.e. $\Sigma(t,y,z,u,v)=\Sigma(t,y)$). In this case, the constraint naturally generates the Skorokhod condition $\int_0^T \Sigma(s,Y_s)dK_s=0$ $Q$-a.s..

To our knowledge, CBSDEs appeared for the first time in \cite{cvitanic1998backward}, for the Brownian setting only. The obtained CBSDE, in these papers, was motivated by the problem of super-replication when there are constraints on the portfolio. Hence, the constraints in the original problem of super-replication naturally translate into constraints on the solution of the resulting BSDE.  The main challenge that was noticed by the early days of the CBSDEs lies in the existence of a solution to those CBSDEs. In fact, in \cite{cvitanic1998backward}, the authors gave a counter-example showing that the existence might fail in general. Thus, still in the Brownian setting and {\it under the assumption that the CBSDE under consideration indeed has a solution}, Peng proved in \cite{peng1999monotonic} that the smallest solution exists. Many extensions were attempted afterwards, for these we refer the reader to \cite{elie2014adding,kharroubi2010backward,peng2010reflected,Xu} and the references therein to cite a few. However, in all these extensions, the assumption that the constrained BSDE should admit a solution persists and was not overcome at all, and hence in our setting these works can not be applicable at all.

\subsection{Main results on the  bounds for Esscher prices}
This subsection states our main results on the bounds for the second-order Esscher pricing intervals. In particular, we prove that the upper and lower bounds of the Esscher pricing interval are solutions to constrained BSDEs, with different constraints, but having the same driver which is linear in the solution's variables $(y,z,u,v)$. To this end, we start with an important lemma which describes the dynamics of the option price process $Y$.
\begin{proposition}\label{prop: general BSDE}
    Suppose that \eqref{assum: model parameters locally bounded} holds. Let $p\in(1,\infty)$, $\xi\in{L}^{0}(P)$ satisfying (\ref{MainAssumption4Xi}), and consider
   \begin{equation}\label{Lprocess}\Delta_{\theta}:=\theta(\psi,\varphi)-\theta(0,0).\end{equation}
 For any $\psi\in\Psi$ and $\varphi\in\Phi$, $\theta(\psi, \varphi)$ denote the unique root to (\ref{eq: M.C.}), and  $Y^{(\psi,\varphi)}$ is defined in (\ref{EsscherInterval}). Then there exists a unique triple $( Z^{(\psi,\varphi)}, U^{(\psi,\varphi)}, V^{(\psi,\varphi)}) \in  \mathbb{L}^p(W,\mathbb Q_0) \times \mathbb{L}^p(N,\mathbb Q_0)\times \mathbb{D}^p(\mathbb Q_0)$, such that the quadruple $(Y^{\psi, \varphi}, Z^{(\psi,\varphi)}, U^{(\psi,\varphi)}, V^{(\psi,\varphi)})$ belongs to the space $ \mathbb{S}^p(\mathbb Q_0) \times \mathbb{L}^p(W,\mathbb Q_0) \times \mathbb{L}^p(N,\mathbb Q_0)\times \mathbb{D}^p(\mathbb Q_0)$  and is the unique solution to
      \begin{equation}\label{BSDE 4Y(psi,phi)}
        \begin{split}
            dY&=ZdW^{(0)}+Ud{\widetilde{N}}^{(0)}+VdN^\mathbb{G}
            -f^{(\psi,\varphi)}(Z,U,V)dA^\tau ,\\
            Y_T&=\xi.
        \end{split}
    \end{equation}
    where
     \begin{equation}\label{Process A}
    A_t := t +D^{p,\mathbb{F}}_t, \quad  a_t:=\frac{dt}{dA_t},\quad t\geq0\quad\quad\mbox{and}
\end{equation}
    \begin{equation}\label{eq: f}
        \begin{split}
            f^{(\psi,\varphi)}(z,u,v) &= -{{\sigma\Delta_{\theta}}\over{\widetilde\gamma}}\left({\widetilde\gamma}z-\sigma(1+\beta)^{-1}u \right)a  + \frac{(e^\varphi -1)(G_- - \Delta D^{p,\mathbb F})}{G_- [G_- +\Delta D^{p,\mathbb F} (e^\varphi -1)]}v(1-a)\\
            &\quad + \exp\bigg(\Delta_\theta \zeta +\psi \zeta^2-\varphi \frac{\eta}{1+\beta}\bigg)\bigg[\frac{\eta (e^\varphi - 1)(1+\beta - \eta)}{1+\beta} \bigg] \big(e^{\theta_0 \zeta} - \frac{1}{1+\beta} \big)  \lambda \ v\  a.
        \end{split}
    \end{equation}
   \end{proposition}
   
   \begin{proof} 
   Before proceeding with the proof, we denote by $\widetilde L$ the following process
\begin{equation}\label{ProcessLtilde}
\begin{split}
{\mathcal{D}}^{(\psi, \varphi)}&:={{\overline{\mathcal{D}}^{(\psi, \varphi)}}\over{ \overline{\mathcal D}^{(0,0)}}}={\cal{E}}(\widetilde L),\qquad\mbox{where}\\
\widetilde L=&\Delta_{\theta}\sigma\is{W}^{(0)}+\left(H(\varphi)\exp(\Delta_{\theta}\zeta+\zeta^2\psi)-1-\beta\right)\is {\widetilde{N}}^{(0)}\\
&+\bigg[1 + \bigg(\frac{H(\varphi)}{1+\beta}e^{\Delta_\theta \zeta + \psi \zeta^2}-1 \bigg)\Delta N \bigg]J(\varphi)\is{N}^{\mathbb{G}}.
\end{split}
\end{equation}
   Now, remark that $Y^{(\psi,\varphi)}$ is a special semimartingale under $(\mathbb Q_0,\mathbb{G})$, and hence there exists a pair $(M,B)$ such that $M\in{\cal{M}}_{loc}(\mathbb Q_0,\mathbb{G})$, $B$ is $\mathbb{G}$-predictable with finite variation, $M_0=B_0=0$, and  
   $$ Y^{(\psi,\varphi)}=Y^{(\psi,\varphi)}_0+M+B.$$
   Then, on the one hand,  by applying Lemma \ref{lemma: Q-martingale representation}  directly to $M$, we deduce the existence of a triplet $(Z,U,V)$ such that 
   \begin{equation}\label{Y(psi,phi)Decompo}
   Y^{(\psi,\varphi)}=Y^{(\psi,\varphi)}_0+Z\is{W}^{(0)}+U\is{\widetilde{N}}^{(0)} +V\is {N}^{\mathbb{G}}+B.
   \end{equation}
   On the other hand, since $Y^{(\psi,\varphi)}\in{\cal{M}}_{loc}(\mathbb Q^{(\psi,\varphi)},\mathbb{G})$, Girsanov's theorem allows us to conclude that $B=-\langle M, \widetilde L\rangle^{(\mathbb Q_0,\mathbb{G})}$. Thus the rest of this proof focuses on calculating this latter process. To this end we calculate $[M, \widetilde L]$ and get 
       \begin{equation}\label{equa1}
        \begin{split}
      [M, \widetilde L]&=\Delta_{\theta}\sigma{Z}\is(\tau\wedge\cdot)+U{{H(\varphi)e^{\Delta_{\theta}\zeta+\zeta^2\psi}-1-\beta}\over{(1+\beta)^2}}\is{N}^{\tau}+ \bigg[\frac{H(\varphi)}{1+\beta} e^{\Delta_\theta \zeta + \psi \zeta^2} -1 \bigg]\Delta N V \is N^\mathbb G\\
      &\qquad+ (1+\Delta L)J(\varphi) \bigg(\frac{U}{1+\beta}\Delta N \is N^\mathbb G + V \is [N^\mathbb G, N^\mathbb G] \bigg).
         \end{split}
    \end{equation}
   Remark that we have 
   \begin{equation*}
        \begin{split}
[N^{\mathbb{G}},N^{\mathbb{G}}]
={{G-\Delta{D}^{o,\mathbb{F}}}\over{\widetilde{G}}}\is {N}^{\mathbb{G}}+{{G}\over{\widetilde{G}^2}}I_{\Rbrack0,\tau\Rbrack}\is{D}^{o,\mathbb{F}}
  \end{split}
    \end{equation*}
By inserting this latter equality in (\ref{equa1}) and using
\begin{equation*}
\begin{split}
 \widetilde N^{(0)} ={\cal{T}}(N^{\mathbb{F}})-\lambda\left(\exp(\theta(0,0)\zeta)-1\right)\is(\tau\wedge\cdot)
\end{split}
\end{equation*}
afterwards, we obtain 
 \begin{equation}\label{equa2}
        \begin{split}
      [M, \widetilde L]&=\Delta_{\theta}\sigma{Z}\is(\tau\wedge\cdot)+\lambda{{U}\over{1+\beta}}\left(H(\varphi)e^{\Delta_{\theta}\zeta+\zeta^2\psi}-1-\beta\right)e^{\theta(0,0)\zeta}\is(\tau\wedge\cdot)\\
      &\quad + V\exp\bigg(\Delta_\theta \zeta +\psi \zeta^2-\varphi \frac{\eta}{1+\beta}\bigg)\bigg[\frac{\eta (e^\varphi - 1)(1+\beta - \eta)}{1+\beta} \bigg] \big(e^{\theta_0 \zeta} - \frac{1}{1+\beta} \big)  \lambda\is(\tau\wedge\cdot) \\
      &\quad+V\frac{(e^\varphi -1)(G_- - \Delta D^{p,\mathbb F})}{G_- [G_- +\Delta D^{p,\mathbb F} (e^\varphi -1)]}I_{\Rbrack0,\tau\Rbrack}\is{D}^{p,\mathbb{F}}\\&\quad +\mathbb Q_0\mbox{-local martingale},
         \end{split}
    \end{equation}
   
Now, thanks to the martingale equation, we have the following equality
$$
-\sigma^2{{\Delta_{\theta}}\over{\widetilde\gamma}}=\lambda{e}^{\theta(0,0)\zeta}\left(H(\varphi)e^{\Delta_{\theta}\zeta+\zeta^2\psi}-1-\beta\right).$$
Thus, a combination of the above equalities gives us the driver in \eqref{eq: f}. This ends the proof.
\end{proof}   
For the remaining of this subsection, we impose the following restriction on the default time $\tau$.
\begin{assum}\label{Assum: avoid F s.t}
    Throughout this subsection we will impose the following assumption on $\tau$,
    \begin{equation}\label{condition: avoid F s.t}
        \Delta D^{o,\mathbb F} = 0,\quad \mbox{or equivalently}\quad G=\widetilde{G}.
    \end{equation}
\end{assum}
For the remaining of this paper, we suppose Assumption \ref{Assum: avoid F s.t} holds. The next corollary derives the BSDE under this special condition on $\tau$.
\begin{corollary}
Suppose Assumption \ref{assum: model parameters locally bounded}, \ref{assum: gamma non-zero} and \ref{Assum: avoid F s.t} hold. Let $p\in(1,\infty)$, $\xi\in{L}^{0}(P)$ satisfying (\ref{MainAssumption4Xi}). For any $\psi\in\Psi$ and $\varphi\in\Phi$, $\theta(\psi, \varphi)$ denote the unique root to (\ref{eq: M.C.}), and  $Y^{(\psi,\varphi)}$ is defined in (\ref{EsscherInterval}). Then, the following assertions hold.\\
{\rm{(a)}} For any $\varphi\in \Phi$,
\begin{equation}
    \theta(\psi, \varphi)=\theta (\psi, 0) = \theta(\psi)
\end{equation}

{\rm{(b)}} There exists a unique triple $( Z^{(\psi,\varphi)}, U^{(\psi,\varphi)}, V^{(\psi,\varphi)}) \in  \mathbb{L}^p(W,\mathbb Q_0) \times \mathbb{L}^p(N,\mathbb Q_0)\times \mathbb{D}^p(\mathbb Q_0)$, such that the quadruple $(Y^{\psi, \varphi}, Z^{(\psi,\varphi)}, U^{(\psi,\varphi)}, V^{(\psi,\varphi)})$ belongs to the space $ \mathbb{S}^p(\mathbb Q_0) \times \mathbb{L}^p(W,\mathbb Q_0) \times \mathbb{L}^p(N,\mathbb Q_0)\times \mathbb{D}^p(\mathbb Q_0)$  and is the unique solution to
          \begin{equation}\label{BSDE 4Y(psi,phi) eta = 0}
        \begin{split}
            dY&=ZdW^{(0)}+Ud{\widetilde{N}}^{(0)}+VdN^\mathbb{G}\\
            &\quad-{{\sigma\Delta_{\theta}}\over{\widetilde\gamma}}\left({\widetilde\gamma}Z-\sigma(1+\beta)^{-1}U\right)d(\tau\wedge t)
            -V (e^\varphi -1)G_-^{-1}dD^{p,\mathbb F},\\
            Y_T&=\xi.
        \end{split}
    \end{equation}
\end{corollary}
\begin{proof}
The proof of assertion (a) follows by substituting $\eta = 0$ in the martingale equation.
\begin{equation}\label{M.E. for phi = 0}
    \widetilde b +\theta \sigma^2 + \alpha \sigma + \lambda \widetilde \gamma \beta + \lambda \widetilde \gamma (1+\beta)\bigg(e^{\theta \zeta + \psi \zeta^2} -1 \bigg) = 0
\end{equation}
and it is obvious that $\theta$ depends only on $\psi$.
    The proof of assertion (b) follows directly from Proposition \ref{prop: general BSDE}, and we omit it here.
\end{proof}
Below we state our main theorem on the upper bound for the Esscher price processes.
\begin{theorem}\label{BSDE4Y(up)}  Suppose Assumptions \ref{assum: model parameters locally bounded}, \ref{assum: gamma non-zero} and \ref{Assum: avoid F s.t} hold. Let $p\in(1,\infty)$ and $\xi\in{L}^{0}(P)$ such that 
\begin{equation}\label{MainAssumption4Xi}
\mathbb{E}_0\left[(\xi^-_{T\wedge\tau})^p\right]+\sup_{\sigma\in{\cal{T}}_{T\wedge\tau},\psi\in\Psi,\phi\in\Phi}\mathbb{E}_0\left[{{D^{(\phi,\psi)}_T}\over{D^{(\phi,\psi)}_{\sigma}}}(\xi^+)^p\right]<\infty,
\end{equation}
where ${\cal{T}}_{T\wedge\tau}$ is the set of all $\mathbb{G}$-stopping times that bounded from above by $T\wedge\tau$.\\
Let $(Y^{(up,exp)}, Y^{(up,lin)},Y^{\psi})$, for any $(\psi, \varphi)\in\Psi\times\Phi$, be defined in (\ref{EsscherInterval}), and $Y^{(n)}$ be given by  
\begin{equation}\label{Y(n)4up}
Y^{(n)}:=\underset{\psi\in\Psi_n, \varphi \in \Phi_n}{\esssup}(Y^{\psi, \varphi}),\quad \mbox{for any}\quad n\in\mathbb{N}.
\end{equation}
If $Y^{(up)}\in\{Y^{(up,exp)},Y^{(up,lin)}\}$, then the following assertions hold.\\
{\rm{(a)}} There exists $(Z^{(up)},U^{(up)},V^{(up)},K^{(up)}, \widetilde K^{(up)})$ such that the tuple $(Y^{(up)},Z^{(up)},U^{(up)},V^{(up)},K^{(up)}, \widetilde K^{(up)})$ belongs to $ \mathbb{S}^p(\mathbb Q_0) \times \mathbb{L}^p(W,\mathbb Q_0) \times \mathbb{L}^p(N,\mathbb Q_0)\times \mathbb{D}^p(\mathbb Q_0) \times \big({\mathcal{A}}^{+,p}(\mathbb Q_0)\big)^2$ and is the smallest solution to the constrained BSDE
\begin{equation}\label{ConstrainedBSDE1}
\begin{split}
dY_s&=\left({{\widetilde{b}_s-\widetilde\gamma_s\lambda_s }\over{ \sigma_s}}Z_s+\lambda_s{U}_s\right)ds + V_s G^{-1}_{s-}dD^{p,\mathbb{F}}_s +Z_s{d}W^\tau_s+{U}_sd({N}^{\mathbb{F}})^{\tau}+ V_s dN^\mathbb{G}_s -dK_s - d\widetilde K_s,\quad Y_T=\xi,\\
&\widetilde\gamma_s{Z}_s-\sigma (1+\beta)^{-1}_s{U}_s\geq 0,\quad P\otimes{ds}\mbox{-a.e.},\quad \int_0^{T\wedge\tau} (\widetilde\gamma_s{Z}_s-\sigma(1+\beta)^{-1}_s{U}_s)dK_s=0,\ P-a.s..\\
& V_s \leq 0, \quad P \otimes {dD^{p,\mathbb{F}}_s-a.e}, \quad \int_0^{T\wedge\tau} {V}_sd{\widetilde K}_s=0\quad P\mbox{-.a.s}.
\end{split}
\end{equation}

{\rm{(b)}} There exists a unique triple $(Z^{(n)},U^{(n)}, V^{(n)})\in \mathbb{L}^p(W,\mathbb Q_0) \times \mathbb{L}^p(N,\mathbb Q_0)\times \mathbb{D}^p(\mathbb Q_0)$ such that the triplet $(Y^{(n)}, Z^{(n)},U^{(n)}, V^{(n)})$ belongs to $ \mathbb{S}^p(\mathbb Q_0) \times \mathbb{L}^p(W,\mathbb Q_0) \times \mathbb{L}^p(N,\mathbb Q_0) \times \mathbb{D}^p(\mathbb Q_0)$ and is the unique solution to the following BSDE
\begin{equation}\label{BSDE(n)}
\begin{split}
Y_t=&\xi+\int_{t\wedge \tau}^{T\wedge \tau}{{\theta_s(-n)-\theta_s(0)}\over{\widetilde\gamma_s}}\sigma_s(\widetilde\gamma_s{Z}_s-\sigma_s(1+\beta_s)^{-1}{U}_s)^+ ds\\
&+  \int_{t\wedge \tau}^{T\wedge \tau}{{\theta_s(0)-\theta_s(n)}\over{\widetilde\gamma_s}}\sigma_s(\widetilde\gamma_s{Z}_s-\sigma_s(1+\beta_s)^{-1}{U}_s)^-ds\\
&+\int_{t\wedge \tau}^{T\wedge \tau}\bigg((e^n-1) V^+_s + (1 - e^{-n}) V^-_s \bigg) G^{-1}_{s-} dD^{p,\mathbb F}_s \\
&-\int_{t\wedge \tau}^{T\wedge \tau} Z_s{d}W^{(0)}_s-\int_{t\wedge \tau}^{T\wedge \tau}{U}_sd\widetilde N^{(0)}_s - \int_{t\wedge \tau}^{T\wedge \tau}{V}_sd N^\mathbb G_s .
\end{split}
\end{equation}

{\rm{(c)}} For any $n,m\in\mathbb{N}$, ${Y}^{(n+m)}-{Y}^{(n)}$ is a nonnegative $\mathbb Q_0$-supermartingale, 
\begin{equation}\label{Increasiness4Y(n)}
Y^{(n)}\leq{Y}^{(n+1)}\leq{Y}^{up},\quad\mbox{and}\quad Y^{(n)}\ \mbox{converges pointwise to}\ Y^{up}.
\end{equation}

{\rm{(d)}} $(Y^{(n)}, Z^{(n)},U^{(n)}, V^{(n)})$  converges to $(Y^{(up)},Z^{(up)},U^{(up)}, V^{(n)})$ in $ \mathbb{S}^p(\mathbb Q_0) \times \mathbb{L}^p(W^0,\mathbb Q_0) \times \mathbb{L}^p(\widetilde{N}^0,\mathbb Q_0)\times \mathbb D^p(\mathbb Q_0)$, and 
$K^{(n)}:=\int_0^{\cdot}(\theta_s(0)-\theta_s(n))\widetilde\gamma_s^{-1}\sigma_s(\widetilde\gamma_s{Z}_s^{(n)}-\sigma_s{U}_s^{(n)})^-a_sdA_s$ converges to $K^{(up)}$ in the space ${\mathcal{A}}^{+,p}(\mathbb Q_0)$ and $\widetilde K^{(n)}:=\int_0^{\cdot}(e^n-1)\big(V^{(n)}_s\big)^+ \widetilde G^{-1}_s (1-a_s) dA_s$ converges to $\widetilde K^{(up)}$ in the space ${\mathcal{A}}^{+,p}(\mathbb Q_0)$. As a result, there exists a nonnegative and predictable processes $k^{(up)}$ and $\widetilde k^{(up)}$ such that $K^{(up)}=\int_0^{\cdot}k^{(up)}_s dA_s$ and $\widetilde K^{(up)}=\int_0^{\cdot}\widetilde k^{(up)}_s dA_s$.
\end{theorem}
The proof of the theorem is relegated to Subsection \ref{Proofs4Section3}. Assertion (a) completely characterizes the upper bound $Y^{up}$ as the smallest solution a {\it constrained} BSDE with  Skorokhod condition.\\ 

Below, we address the lower Esscher bound process, and we prove that it is also fully characterized by a constrained BSDE. Both CBSDEs (i.e.\ CBSDEs for upper bound and lower bound) have the same driver, but different constraints and different martingale structures naturally.
\begin{theorem}\label{BSDE4Y(inf)}  Suppose Assumptions \ref{assum: model parameters locally bounded}, \ref{assum: gamma non-zero} and \ref{Assum: avoid F s.t} hold. Let $p\in(1,\infty)$ and $\xi\in{L}^{0}(P)$ such that
\begin{equation*}\label{MainAssumption4Xi(inf)}
{\mathbb{E}}_0\left[(\xi^+)^p\right]+\sup_{\sigma\in{\cal{T}}_{T\wedge\tau},\psi\in\Psi,\phi\in\Phi}{\mathbb{E}}_0\left[{{D^{(\phi,\psi)}_T}\over{D^{(\phi,\psi)}_{\sigma}}}(\xi^-)^p\right]<\infty.
\end{equation*}
For $(\psi, \varphi) \in\Psi\times \Phi$, let $(Y^{(inf,exp)}, Y^{(inf,lin)},Y^{\psi})$ be defined in (\ref{EsscherInterval}), and $\overline{Y}^{(n)}$ be given by
\begin{equation*}\label{Ybar(n)}
 \overline{Y}^{(n)}:=\underset{\psi\in\Psi_n, \varphi \in \Phi_n}{\essinf}(Y^{(\psi, \varphi)}),\quad \mbox{for any}\quad n\in\mathbb{N}.\end{equation*}
If $Y^{(inf)}\in\{Y^{(inf,lin)},Y^{(inf,exp)}\}$, then the following assertions hold.\\
{\rm{(a)}}  There exists $(Z^{(inf)},U^{(inf)},V^{(inf)},K^{(inf)})$ such that $(Y^{(inf)},Z^{(inf)}, U^{(inf)}, V^{(inf)},K^{(inf)})$ belongs to $ \mathbb{S}^p(\mathbb Q_0) \times \mathbb{L}^p(W,\mathbb Q_0) \times \mathbb{L}^p(N,\mathbb Q_0)\times \mathbb{D}^p(\mathbb Q_0)\times{\mathcal{A}}^{+,p}(\mathbb Q_0)$ and is the smallest solution to the constrained BSDE
\begin{equation*}
\begin{split}
&dY_s=\left({{\widetilde{b}_s-\widetilde\gamma_s\lambda_s }\over{ \sigma_s}}Z_s+\lambda_s{U}_s\right)ds+V_s G^{-1}_{s-}dD^{p,\mathbb F}_s+Z_s{d}W_s^\tau+{U}_sd(N_s^\mathbb F)^\tau+V_sdN^\mathbb G_s+dK_s + d\widetilde K_s,\\
&Y_T=\xi,\quad \mathbb P-a.s.,\\
&\widetilde\gamma_s{Z}_s-\sigma_s(1+\beta_s)^{-1}{U}_s\leq 0,\quad \mathbb P\otimes{ds}\mbox{-a.e.},\quad \int_0^{T\wedge \tau}(\widetilde\gamma_s{Z}_s-\sigma_s(1+\beta_s)^{-1}{U}_s)dK_s=0,\ \mathbb P-a.s.,\\
&V_s \geq 0\quad \mathbb P \otimes dD^{p,\mathbb F}\text{-a.e,} \quad \int_0^{T\wedge \tau} V_s \widetilde K_s = 0 \quad \mathbb P\text{-a.s.}
\end{split}
\end{equation*}
{\rm{(b)}} There exists a unique triple $(\overline{Z}^{(n)},\overline{U}^{(n)}, \overline{V}^{(n)})$ such that $(\overline{Y}^{(n)}, \overline{Z}^{(n)},\overline{U}^{(n)},\overline{V}^{(n)})$ belongs to $\mathbb{S}^p(\mathbb Q_0)\times \mathbb{L}^p(W,\mathbb Q_0) \times \mathbb{L}^p(N,\mathbb Q_0)\times \mathbb{D}^p(\mathbb Q_0)$,  and is the unique solution to the BSDE:
\begin{equation*}\label{BSDE(n)4infinimum}
\begin{split}
Y_t&=\xi-\int_{t\wedge \tau}^{T\wedge \tau}\bigg({{\theta_s(0)-\theta_s(n)}\over{\widetilde\gamma_s}}\sigma_s(\widetilde\gamma_s{Z}_s-\sigma_s(1+\beta_s)^{-1}{U}_s)^+ +  {{\theta_s(-n)-\theta_s(0)}\over{\widetilde\gamma_s}}\sigma_s(\widetilde\gamma_s{Z}_s-\sigma_s(1+\beta_s)^{-1}{U}_s)^-\bigg)ds\\
&\quad-\int_{t\wedge \tau}^{T\wedge \tau}\bigg((e^{n}-1)V_s^+ + (1-e^{-n}) \bigg)G^{-1}_{s-}dD^{p,\mathbb F}_s -\int_{t\wedge \tau}^{T\wedge \tau}Z_s{d}W^{(0)}
_s-\int_{t\wedge \tau}^{T\wedge \tau}{U}_sd\widetilde{N}^{(0)}_s - \int_{t\wedge \tau}^{T\wedge \tau}V_s N_s^\mathbb G.
\end{split}\end{equation*}
{\rm{(c)}}  For any $n,m\in\mathbb{N}$, $\overline{Y}^{(n)}-\overline{Y}^{(n+m)}$ is a nonnegative $\mathbb Q_0$-submartingale,
\begin{equation}\label{Decreasiness4Ybar(n)}
\overline{Y}^{(n)}\geq\overline{Y}^{(n+1)}\geq{Y}^{(inf)}\quad\mbox{and}\quad \overline{Y}^{(n)}\ \mbox{converges pointwise to}\ Y^{(inf)}.
\end{equation}
{\rm{(d)}} The tuple$(\overline{Y}^{(n)}, \overline{Z}^{(n)},\overline{U}^{(n)}, \overline{V}^{(n)})$ converges to the tuple$(Y^{(inf)},Z^{(inf)},U^{(inf)},V^{(inf})$ in the space $ \mathbb{S}^p(\mathbb Q_0) \times \mathbb{L}^p(W^0,\mathbb Q_0) \times \mathbb{L}^p(\widetilde{N}^0,\mathbb Q_0)\times \mathbb D^p(\mathbb Q_0)$. Furthermore, the nondecreasing process 
$$\overline{K}^{(n)}:=\int_0^{\cdot}\left(\theta_s(0)-\theta_s(n)\right){\widetilde\gamma_s}^{-1}\sigma_s(\widetilde\gamma_s\overline{Z}_s^{(n)}-\sigma_s\overline{U}_s^{(n)})^+ds$$ converges to $K^{(inf)}$ in ${\mathcal{A}}^{+,p}(\mathbb Q_0)$-norm. As a result, there exists a predictable process $k^{(inf)}$ such that $K^{(inf)}:=\int_0^{\cdot}k^{(inf)}_sds \in{\mathcal{A}}^{+,p}(\mathbb Q_0)$.
\end{theorem}
\begin{proof}
Let $\xi$ be a claim and put $\xi':=-\xi$. For any $(\psi,\varphi)\in\Psi\times \Phi$, and we denote by $Y^{(\psi,\varphi)}(\xi)$ and $Y^{(\psi,\varphi)}(\xi')$ the process defined in (\ref{EsscherInterval}) associated with the claims $\xi$ and $\xi'$, respectively. On the one hand,  we denote the unique solution to (\ref{eq: G BSDE}) for $\xi$ (respectively $\xi'$), see Lemma \ref{lemma: BSDE for each psi phi} below, by the tuple $( Y^{(\psi,\varphi)}(\xi),Z^{(\psi,\varphi)}(\xi),U^{(\psi,\varphi)}(\xi),V^{(\psi,\varphi)}(\xi))$  (respectively $(Y^{(\psi,\varphi)}(\xi'),Z^{(\psi,\varphi)}(\xi'),U^{(\psi,\varphi)}(\xi'), V^{(\psi,\varphi)}(\xi'))$,  and thanks to the uniqueness of the solution we deduce that
$$
(Y^{(\psi,\varphi)}(\xi'),Z^{(\psi,\varphi)}(\xi'),U^{(\psi,\varphi)}(\xi'), V^{(\psi,\varphi)}(\xi')))=-( Y^{(\psi,\varphi)}(\xi),Z^{(\psi,\varphi)}(\xi),U^{(\psi,\varphi)}(\xi),V^{(\psi,\varphi)}(\xi)).
$$
On the other hand, due to the simple fact that $\underset{i\in{I}}{\essinf}(X_i)=-\underset{i\in{I}}{\esssup}(-X_i)$ for any family of random variables $(X_i)_{i\in{I}}$, we obtain
\begin{equation*}
Y^{(inf)}(\xi)=\underset{\psi\in\Psi, \varphi \in \Phi}{\essinf}(Y^{(\psi, \varphi)}(\xi))=-\underset{\psi\in\Psi, \varphi \in \Phi}{\esssup}(-Y^{(\psi, \varphi)}(\xi))=-\underset{\psi\in\Psi, \varphi \in \Phi}{\esssup}(Y^{(\psi, \varphi)}(\xi'))=-Y^{(up)}(\xi').
\end{equation*}
Therefore, by applying Theorem \ref{BSDE4Y(up)} to the claim $\xi'=-\xi$ instead and using these remarks afterwards, all assertions of the theorem follow immediately, and the proof of the theorem is complete.
\end{proof}
\subsection{Proof of Theorem \ref{BSDE4Y(up)} 
}\label{Proofs4Section3}
This subsection proves the first  main theorem of this section. To this end, we start with some intermediate results that we summarize in four lemmas. These lemmas are interesting in themselves besides conveying the main ideas behind our main results. 
\begin{lemma}\label{EtaProcess} Suppose Assumptions \ref{assum: model parameters locally bounded}, \ref{assum: gamma non-zero} and \ref{Assum: avoid F s.t}  hold. For any $\psi\in\Psi$, $\varphi\in\Phi$ and any $\zeta\in\{\gamma,\widetilde\gamma\}$, let $\theta(\psi)$ be the unique root to (\ref{M.E. for phi = 0}), and consider the functional $F$ given by 
\begin{equation}\label{PhiFunction}
F(x):=F(t,\omega,x):=\sigma_t(\omega)^2x+\lambda_t(\omega)\widetilde\gamma_t(\omega)\zeta_t(\omega)(1+\beta_t(\omega))e^x,\quad t\in[0,T],\quad \omega\in\Omega,\quad x\in\mathbb{R}.
\end{equation}
Then the following assertions hold.\\
 {\rm{(a)}}  The functional $F$ is continuous, strictly increasing, $F(\infty)=\infty$, $F(-\infty)=-\infty$, and 
 \begin{equation}\label{Phi2eta}
 \theta(\psi)={1\over{\zeta}}F^{-1}\Bigl(\zeta(-\widetilde{b}+\lambda\widetilde\gamma-\alpha \sigma)+\sigma^2\zeta^2\psi\Bigr)-\zeta\psi.
 \end{equation}
 As a result, for any $\psi\in\Psi$, there exists $C_{\psi}\in(0,\infty)$ such that $\vert\theta(\psi)\vert\leq C_{\psi}$ $P\otimes{dt}$-a.e..\\
 {\rm{(b)}}  $\theta(\psi) \widetilde\gamma^{-1}$ is decreasing ($P\otimes{dt}$-a.e.) with respect to $\psi$.\\
  {\rm{(c)}}   For any $\zeta\in\{\gamma,\widetilde\gamma\}$, we have 
        \begin{equation}\label{etaLimitsetaStar}
        \lim_{\psi \downarrow -\infty} {{\theta(\psi)}\over{\widetilde\gamma}} = {{\theta(-\infty)}\over{\widetilde\gamma}}= {{\lambda\widetilde\gamma-\alpha\sigma-\widetilde{b}}\over{\widetilde\gamma \sigma^2}}=:{{\theta^*}\over{\widetilde\gamma}},\quad \mbox{and}\quad  \lim_{\psi \uparrow \infty} {{\theta(\psi)}\over{\widetilde\gamma}} = -\infty,\quad P\otimes{dt}\mbox{-a.e.}.\end{equation}
{\rm{(d)}} For any $\zeta\in\{\gamma,\widetilde\gamma\}$ and any $\psi\in\Psi$ such that $\psi\geq0$, we have 
\begin{equation}\label{limite4eta(n)/n}\lim_{n\longrightarrow\infty}{{-\theta(n)}\over{n\widetilde\gamma}}={{\zeta}\over{\widetilde\gamma}},\quad\mbox{and}\quad {{\theta(0)-\theta(\psi)}\over{\widetilde\gamma}}\leq{{\zeta}\over{\widetilde\gamma}}\psi,\quad P\otimes{dt}\mbox{-a.e.}.
\end{equation}
\end{lemma}
The proof of the lemma mimics exactly the proof of \cite[Lemma 4.2]{choulli2025second}, and hence it will be omitted.

\begin{lemma}\label{lemma: BSDE for each psi phi}
Suppose Assumptions \ref{assum: model parameters locally bounded}, \ref{assum: gamma non-zero} and \ref{Assum: avoid F s.t}  hold, and let $p\in(1,\infty)$ and $\xi\in{L}^{0}(P)$ satisfying (\ref{MainAssumption4Xi}). For any $(\psi,\varphi)\in\Psi\times \Phi$, $\theta(\psi)$ denote the unique root to (\ref{M.E. for phi = 0}), and  $Y^{(\psi,\varphi)}$ is defined in (\ref{EsscherInterval}). Then  the following hold.\\
{\rm{(a)}} For $(\psi, \varphi)\in\Psi\times \Phi$, there exists a unique $( Z^{(\psi,\varphi)}, U^{(\psi,\varphi)}, V^{(\psi,\varphi)}) \in  \mathbb{L}^p(W,\mathbb Q_0) \times \mathbb{L}^p(N,\mathbb Q_0)\times \mathbb{D}^p(\mathbb Q_0)$, such that the quadruple $(Y^{\psi, \varphi}, Z^{(\psi,\varphi)}, U^{(\psi,\varphi)}, V^{(\psi,\varphi)})$ belongs to the space $ \mathbb{S}^p(\mathbb Q_0) \times \mathbb{L}^p(W,\mathbb Q_0) \times \mathbb{L}^p(N,\mathbb Q_0)\times \mathbb{D}^p(\mathbb Q_0)$  and is the unique solution to
    \begin{equation}\label{eq: G BSDE}
        \begin{split}
            dY&=ZdW^{(0)} + Ud\widetilde N^{(0)} +VdN^\mathbb{G} - f^{(\psi,\varphi)}(Z,U,V)dA^\tau\\
            Y_T &= \xi
        \end{split}
    \end{equation}
    Here the pair $(a,A)$ is defined in \eqref{Process A}, and the functional $f^{(\psi,\varphi)}$ is given by 
    \begin{equation}
        \begin{split}
            f^{(\psi,\varphi)}(z,u,v) :=-{{\sigma\Delta_{\theta}}\over{\widetilde\gamma}}\left({\widetilde\gamma}z-{{\sigma}\over{1+\beta}}u \right)a - v(e^\varphi -1)(1-a),\quad  \mbox{for any}\ (z,u,v)\in \mathbb R^3.
        \end{split}
    \end{equation}
Furthermore, for any $n\geq 1$, there exists $C_n\in(0,\infty)$ such that for any $(\phi,\psi)\in\Phi_n\times\Psi_n$,
\begin{equation}\label{Control4(Y,Z,U,A)(psi)}
\begin{split}
    \Vert {Y}^{(\psi,\varphi)}\Vert_{\mathbb{S}^p(\mathbb Q_0)}&+\Vert{Z}^{(\psi,\varphi)}\Vert_{\mathbb{L}^p(W,\mathbb Q_0) }+\Vert{U}^{(\psi,\varphi)}\Vert_{\mathbb{L}^p(N,\mathbb Q_0)}+ +\Vert{V}^{(\psi,\varphi)}\Vert_{\mathbb{D}^p(\mathbb Q_0)}\\
    &+\Vert\int_0^T\vert{f}^{(\psi,\varphi)}(s,Y^{(\psi,\varphi)}_s,Z^{(\psi,\varphi)}_s,U^{(\psi,\varphi)}_s)\vert ds\Vert_{L^p(\mathbb Q_0)}\leq C_n\Vert\xi\Vert_{L^p(\mathbb Q_0)}.
\end{split}
 \end{equation}
 {\rm{(b)}} Let $\psi\in\Psi$, $\varphi\in\Phi$, and $(Y^{(\psi,\varphi)},Z^{(\psi,\varphi)},U^{(\psi,\varphi)}, V^{(\psi,\varphi)})$ be the solution to (\ref{eq: G BSDE}). If we denote
 \begin{equation}\label{g(psi)} 
{g}_{\psi, \varphi}:={{\sigma}\over{\widetilde\gamma}}\Delta_{\theta}\left(\widetilde\gamma{Z^{(\psi,\varphi)}}-\sigma{{U^{(\psi,\varphi)}}\over{1+\beta}}\right)a + (e^\varphi - 1) V^{(\psi,\varphi)}  G^{-1}_{-} (1-a),
\end{equation}
then we have 
 \begin{equation}\label{Y(psi)potentials}
\begin{split}
Y^{(\psi, \varphi)}_t=\mathbb{E}_0\left[\xi+\int_{t\wedge\tau}^{T\wedge\tau} g_{\psi, \varphi}(s) dA_s\big|{\mathcal{G}}_t\right],\quad{t}\geq 0, 
\end{split}
\end{equation}
  {\rm{(c)}} Let $\rho_i := (\psi_i, \varphi_i)\in\Psi\times\Phi$, $i=1,2$ and $\Gamma$ be a predictable set and put $\rho_3:=\rho_1{I}_{\Gamma}+\rho_2{I}_{\Gamma^c}$, then
 \begin{equation}\label{psi1psi2}
 \begin{split}
 & dY^{\rho_3}={I}_{\Gamma}d(Y^{\rho_1})+{I}_{\Gamma^c}d(Y^{\rho_2}),\\
& Z^{\rho_3}= Z^{\rho_1}{I}_{\Gamma}+Z^{\rho_2}{I}_{\Gamma^c},   \quad U^{\rho_3}=U^{\rho_1}{I}_{\Gamma}+U^{\rho_2}1_{\Gamma^C}\quad \mbox{and}\quad  V^{\rho_3}=  V^{\rho_1}{I}_{\Gamma} + V^{\rho_2}1_{\Gamma^C} .
 \end{split}
 \end{equation}
  {\rm{(d)}} For any $\rho_i\in\Psi\times \Phi$, $i=1,2$, there exists $\rho_3\in\Psi\times\Phi$ such that 
  \begin{equation}\label{Domination4psi}
  \begin{split}
  &\vert\rho_3\vert\leq \max\left(\vert\rho_1\vert,\vert\rho_2\vert\right),\quad Y^{\rho_3}\geq \max\left(Y^{\rho_1},Y^{\rho_2}\right),\quad g^{\rho_3}=\max\left(g^{\rho_1},g^{\rho_2}\right),\\
    \end{split}
  \end{equation}
  As a results $\{Y^{(\psi,\varphi)}:\ (\psi,\varphi)\in\Psi\times\Phi\}$, $\{g^{\psi,\varphi}:\ (\psi,\varphi)\in\Psi\times\Phi\}$, $\{Y^{(\psi,\varphi)}:\ (\psi,\varphi)\in\Psi_n\times\Phi_n\}$ and $\{g^{\psi,\varphi}:\ (\psi,\varphi)\in\Psi_n\times\Phi_n\}$ all upward directed.
\end{lemma}
\begin{proof} The proof of the lemma is divided into three parts. The first part proves assertion (a),  the second part addresses assertions (b) and (c), while the third part proves assertion (d).\\
{\bf Part 1.} This part proves assertion (a). To this end, we start proving the following fact:
\begin{equation}\label{Control4Y(psi)}
\begin{split}
 &\mbox{For any}\ n\geq 1\ \mbox{there exists}\ C_n\in(0,\infty)\ \mbox{such that:}\\
 & \Vert\sup_{0\leq{t}\leq{T}}\vert{Y}^{(\phi,\psi)}_t\vert\Vert_{L^p(R_0)}\leq C_n\Vert\xi\Vert_{L^p(R_0)},\quad\mbox{for any}\quad (\phi,\psi)\in\Phi_n\times\Psi_n,
 \end{split}\end{equation}
Thus, consider $n\geq 1$, and use Lemma \ref{lemma: Esscher and Girsanov}-(c) -more precisely the inequality (\ref{Reverse_Holder_4D(psi,phi)}). In other words, there exists a positive constant $C_{\psi}\in(0,\infty)$ such that 
$$
\mathbb{E}_{0}\left[\left({{D^{(\phi,\psi)}_T}\over{D^{(\phi,\psi)}_t}}\right)^q\big|{\cal{F}}_t\right]\leq C_{(\psi,\phi)},\quad{P}\mbox{-a.s.}\quad t\in[0,T].$$
 Thus, by combining this fact with \cite[Theorem 2.10]{Choulli99}, $Y^{(\phi,\psi)}_T=\xi$, and $\exp(-\int_0^{\cdot}r_s ds)Y^{(\phi,\psi)}$ is a martingale under $\mathbb{Q}^{(\phi,\psi)}$, the inequality (\ref{Control4Y(psi)}) follows immediately. Then due to the assumptions \ref{assum: model parameters locally bounded} and \ref{assum: gamma non-zero}, we deduce that $Y^{\psi}$ is a special semimartingale under $\mathbb{Q}_0$, and its canonical decomposition is denoted by 
\begin{equation}\label{Doob4Y(psi)}
Y^{(\phi,\psi)}=Y^{(\phi,\psi)}_0+M^{(\phi,\psi)}+A^{(\phi,\psi)},\quad M^{(\phi,\psi)}\in{\cal{M}}_{loc}(\mathbb{Q}_0),\ A^{(\phi,\psi)}\in{\cal{A}}_{loc}(\mathbb{Q}_0)\ \mbox{and is predictable}.
\end{equation}
Thus, again Lemma \ref{lemma: Esscher and Girsanov}-(c) guarantees that $\widetilde{L}$ (the logarithm stochastic of $D^{(\phi,\psi)}$) defined in (\ref{ProcessLtilde}) belongs in fact $\mbox{bmo}_q(\mathbb{Q}_0)$. Hence, a combination of this \cite[Theorem 4.5]{Choulli99}, we conclude the existence of $\overline{C}_n\in(0,\infty)$ such that 
\begin{equation}\label{Control4Y(psi)1}
\Vert{Y}^{(\phi,\psi)}_0\Vert_{L^p(\mathbb{Q}_0)}+\Vert[M^{(\phi,\psi)},M^{(\phi,\psi)}]^{1/2}_T\Vert_{L^p(\mathbb{Q}_0)}+ \Vert\mbox{Var}_T(A^{(\phi,\psi)})\Vert_{L^p(\mathbb{Q}_0)}\leq \overline{C}_n  \Vert\sup_{0\leq{t}\leq{T}}\vert{Y}^{\psi}_t\vert\Vert_{L^p(\mathbb{Q}_0)}.
\end{equation}
By combining (\ref{Control4Y(psi)}) and (\ref{Control4Y(psi)1}), under assumptions  \ref{assum: model parameters locally bounded} and \ref{assum: gamma non-zero}, we obtain that $M^{(\phi,\psi)}\in{\cal{M}}^p(\mathbb{Q}_0)$, and due to the martingale representation  result in Lemma \ref{lemma: Q-martingale representation}  and (\ref{Control4Y(psi)1}), we obtain a unique pair $(Z^{(\phi,\psi)},U^{(\phi,\psi)},V^{(\phi,\psi)})\in\mathbb{L}^p(W,\mathbb{Q}_0) \times \mathbb{L}^p(N,\mathbb{Q}_0)\times\mathbb{D}^p(\mathbb{Q}_0,D)$ such that 
\begin{equation}\label{Y(psi)Representation}
Y^{(\phi,\psi)}=Y^{(\phi,\psi)}_0+Z^{(\phi,\psi)}\is{W}^{(0)}+U^{(\phi,\psi)}\is\widetilde{N}^{(0)}+V^{(\phi,\psi)}\is{N}^{\mathbb{G}}+A^{(\phi,\psi)}.\end{equation}
Then, as  $\exp(-\int_0^{\cdot}r_s ds)Y^{(\phi,\psi)}\in{\cal{M}}_{loc}(\mathbb{Q}^{(\phi,\psi)})$, the above equality and  Girsanov's theorem yield
\begin{equation}\label{A(psi)}
{{dA^{(\phi,\psi)}}\over{dA}}=-\left(\Delta_{\theta}{{\sigma}\over{\widetilde\gamma}}(\widetilde\gamma{Z}^{(\phi,\psi)}-{{\sigma}\over{1+\beta}}{U}^{(\phi,\psi)})-r Y^{(\phi,\psi)}\right) a - V^{(\phi,\psi)}(e^\varphi -1)(1-a).\end{equation}
On the one hand, by combining this latter claim with (\ref{Doob4Y(psi)}), (\ref{Control4Y(psi)1}) and  (\ref{Control4Y(psi)}), the inequality (\ref{Control4(Y,Z,U,A)(psi)}) follows immediately. On the other hand, by combining  (\ref{A(psi)}), (\ref{Y(psi)Representation}) and $Y^{\psi}_T=\xi$, the first statement of assertion (a) (i.e. the BSDE \eqref{eq: G BSDE} has a unique solution that is $p$-integrable) follows immediately. This ends the proof of assertion (a).\\
{\bf Part 2.}  Here we prove assertion (b). Due to assertion (a), the integrals $Z^{(\psi, \varphi)}\is W^{(0)}, U^{(\psi,\varphi)}\is \widetilde N^{(0)}$ and $ V^{(\psi,\varphi)} \is N^\mathbb G$ are $p$-integrable $\mathbb G$-martingales under $\mathbb Q_0$. Hence, by taking the conditional expectation with respect to $\mathcal G_t$ and $\mathbb Q_0$ in both sides the BSDE \eqref{eq: G BSDE} with solution $(Y^{(\psi, \varphi)}, Z^{(\psi,\varphi)}, U^{(\psi,\varphi)}, V^{(\psi,\varphi)})$ equality \eqref{Y(psi)potentials} follows. This proves assertions (b) and (c), and the rest of this part focuses on proving assertion (c). To this end, let $\rho_i := (\psi_i, \varphi_i)\in\Psi\times\Phi$, $i=1,2$ and $\Gamma$ be a predictable set and put $\rho_3:=\rho_1{I}_{\Gamma}+\rho_2{I}_{\Gamma^c}$. Consider the tuple $(\overline{Y}, \overline{Z}, \overline{U}, \overline{V})$ given by
    \begin{equation}
        \begin{split}
            &\overline{Z} = 1_\Gamma Z^{\rho_1} + 1_{\Gamma^C} Z^{\rho_2},\quad \overline{U} = 1_\Gamma U^{\rho_1} + 1_{\Gamma^C} U^{\rho_2},\quad\overline{V} = 1_\Gamma V^{\rho_1} + 1_{\Gamma^C} V^{\rho_2},\\
            &d\overline{Y} = 1_\Gamma dY^{\rho_1} + 1_{\Gamma^C} dY^{\rho_2},\quad \overline{Y}_T = \xi
        \end{split}
    \end{equation}
    Then, we derive
    \begin{equation}\label{eq: Y bar}
        \begin{split}
            d\overline{Y} = \overline{Z} dW^{(0)} + \overline{U} d\widetilde N^{(0)} + \overline{V} dN^\mathbb G -  \bigg(1_\Gamma f^{\rho_1}(Z^{\rho_1}, U^{\rho_1}, V^{\rho_1}) + 1_{\Gamma^C} f^{\rho_2}(Z^{\rho_2}, U^{\rho_2}, V^{\rho_2}) \bigg)dA^\tau 
        \end{split}
    \end{equation}
    Since $\overline{X}=X^{\rho_1}$ on $\Gamma$ and $\overline{X}=X^{\rho_2}$ on $\Gamma^C$ for $X=Z,U,V$, then \eqref{eq: Y bar} becomes
    \begin{equation}
        \begin{split}
            \overline{Y} = \overline{Z} dW^{(0)} + \overline{U} d\widetilde N^{(0)} + \overline{V} dN^\mathbb G -  \bigg(1_\Gamma f^{\rho_1}(\overline{Z}, \overline{U}, \overline{V}) + 1_{\Gamma^C} f^{\rho_2}(\overline{Z}, \overline{U}, \overline{V}) \bigg)dA^\tau , \quad \overline{Y}_T =\xi
        \end{split}
    \end{equation}
    Notice that
    \begin{equation}
        \begin{split}
            &1_\Gamma f^{\rho_1}(z,u,v) +1_{\Gamma^C} f^{\rho_2}(z,u,v)\\
            &=-\frac{\sigma}{\widetilde \gamma}\big(1_\Gamma\theta(\psi_1) + 1_{\Gamma^C}  \theta(\psi_2) \big)\left({\widetilde\gamma}z-\sigma(1+\beta)^{-1}u \right)a_t\\
            &\quad+{{\sigma\theta(0)}\over{\widetilde\gamma}}\left({\widetilde\gamma}z-\sigma(1+\beta)^{-1}u \right)a_t- (1_\Gamma e^{\varphi_1}+1_{\Gamma^C} e^{\varphi_2})v(1-a_t) + v(1-a_t)\\
            &=-\frac{\sigma}{\widetilde \gamma} \theta(\psi_3) \left({\widetilde\gamma}z-\sigma(1+\beta)^{-1}u \right)a_t+{{\sigma\theta(0)}\over{\widetilde\gamma}}\left({\widetilde\gamma}z-\sigma(1+\beta)^{-1}u \right)a_t-  e^{\varphi_3}v(1-a_t) + v(1-a_t)\\
            &= f^{\rho_3}(z,u,v).
        \end{split}
    \end{equation}    
Therefore, the BSDE becomes
\begin{equation}
        \begin{split}
            \overline{Y} = \overline{Z} dW^{(0)} + \overline{U} d\widetilde N^{(0)} + \overline{V} dN^\mathbb G - f^{(\psi_3,\varphi_3)}(\overline{Z}, \overline{U}, \overline{V})dA^\tau, \quad \overline{Y}_T =\xi
        \end{split}
    \end{equation}
    Then, by uniqueness of the solution, we obtain $(\overline{Y}, \overline{Z}, \overline{U}, \overline{V}) = (Y^{(\psi_3, \varphi_3)}, Z^{(\psi_3, \varphi_3)}, U^{(\psi_3, \varphi_3)}, V^{(\psi_3, \varphi_3)})$.
{\bf Part 3.} Herein, we prove assertion (d). For any $\big((\psi_1,\varphi_1), (\psi_2,\varphi_2)\big) \in \Psi^2\times\Phi^2$, we put  
$$\Gamma := \{g^{(\psi_1,\varphi_1)} > g^{(\psi_2,\varphi_2)}  \}.$$
 Then, in virtue of assertion (c), we get
    \begin{equation}
        \begin{split}
            1_\Gamma g^{\rho_1} +1_{\Gamma^C} g^{\rho_2}=g^{\rho_3},\quad \text{for }\rho_i\in\Psi\times \Phi,i=1,2.
        \end{split} 
    \end{equation}
  Thus, by definition of the set $\Gamma$, we obtain $g^{\rho_1} = \max(g^{\rho_1},g^{\rho_2})$ on $\Gamma$ and $g^{\rho_2} = \max(g^{\rho_1},g^{\rho_2})$ on $\Gamma^C$, or equivalently 
    \begin{equation}
        g^{\rho_3} = 1_\Gamma \max(g^{\rho_1},g^{\rho_2}) + 1_{\Gamma^C} \max(g^{\rho_1},g^{\rho_2})= \max(g^{\rho_1},g^{\rho_2})
    \end{equation}
    Now, consider the difference \(Y^{\rho_3}_t-Y^{\rho_1}_t\) and we use assertion (b) to derive
$$
Y^{\rho_3}_t-Y^{\rho_1}_t
= \mathbb{E}_0\!\Big[\int_{t\wedge \tau}^{T\wedge \tau}\big(g^{\rho_3}_s-g^{\rho_1}_s\big) dA_s\Bigm|\mathcal{G}_t\Big]\quad t\in[0,T],
$$
Note that,
$$
g^{\rho_3}_s-g^{\rho_1}_s
= \big(g^{\rho_1}_s\,1_{\Gamma}+g^{\rho_2}_s\,1_{\Gamma^C}\big)-g^{\rho_1}_s
= (g^{\rho_2}_s-g^{\rho_1}_s)\,1_{\Gamma^C}.
$$

By the definition of \(\Gamma\) we have \(g^{\rho_2}_s-g^{\rho_1}_s\geq 0\) on \(\Gamma^C\),  and hence $g^{\rho_3}_s-g^{\rho_1}_s \ge 0$ and 
$$
Y^{\rho_3}_t-Y^{\rho_1}_t
= \mathbb{E}\!\Big[\int_t^T (g^{\rho_3}_s-g^{\rho_1}_s)\,\mathrm{d}s\Bigm|\mathcal{F}_t\Big] \geq 0,
$$
This proves \(Y^{\psi_3}_t \ge Y^{\psi_1}_t\) a.s. In a similar manner, we get \(Y^{\rho_3}_t \geq Y^{\rho_2}_t\) a.s. and combining the two inequalities yields the asserted result:
$$
Y^{\rho_3}_t \geq \max(Y^{\rho_1}_t,Y^{\rho_2}_t)\quad\text{a.s.}
$$
Now, let $|\cdot|$ be any point-wise norm on $\mathbb{R}^2$. Then
$$
|\rho_3|=|\rho_1|\,1_{\Gamma}+|\rho_2|\,1_{\Gamma^C}
\le \max\{|\rho_1|,|\rho_2|\}\,1_{\Gamma}+\max\{|\rho_1|,|\rho_2|\}\,1_{\Gamma^C}
=\max\{|\rho_1|,|\rho_2|\},
$$
so $ |\rho_3|\le\max\{|\rho_1|,|\rho_2|\} $ a.s.

\end{proof}
Below we address the driver $g^{\psi, \varphi}$ of the BSDE (\ref{eq: G BSDE}) and show it can be controlled uniformly when $(\psi, \varphi)$ spans $\Psi_n\times \Phi_n$.  
\begin{lemma}\label{lemma4Gtilde} Suppose Assumptions \ref{assum: model parameters locally bounded}, \ref{assum: gamma non-zero} and \ref{Assum: avoid F s.t} hold, and for any $(n,\psi, \varphi)\in\mathbb{N}\times\Psi\times \Phi$ we consider ${g}^{\psi, \varphi}$ defined in (\ref{Domination4psi}) and $(\widetilde{g}_n,g_n,\widetilde{g})$ given by 
\begin{equation}\label{Processes(g(n)gTilde(n)gTilde}
\begin{split}
&\widetilde{g}_n(t):=\underset{\psi\in\Psi_n, \varphi \in \Phi_n}{\esssup}({g}^{\psi, \varphi}(t))\geq 0,\quad\widetilde{g}(t):=\sup_{n\geq 0}\widetilde{g}_n(t)=\underset{\psi\in\Psi, \varphi\in\Phi}{\esssup}({g}^{\psi, \varphi}(t)),\quad t\geq0,\\
&g_n(t) ={{\theta_t(-n)-\theta_t(0)}\over{\widetilde\gamma_t}}\sigma_t\left(\widetilde\gamma_t{Z}_t^{(n)}-{{\sigma_t}\over{1+\beta_t}}{U}_t^{(n)}\right)^+a_t +  {{\theta_t(0)-\theta_t(n)}\over{\widetilde\gamma_t}}\sigma_t\left(\widetilde\gamma_t{Z}_t^{(n)}-{{\sigma_t}\over{1+\beta_t}}{U}_t^{(n)}\right)^-{a}_t\\
&\hskip 1cm + \bigg((e^n-1) \big(V^{(n)}_t\big)^+ + (1 - e^{-n}) \big(V^{(n)}_t\big)^- \bigg) G^{-1}_{t-} (1-a_t)
\end{split}
\end{equation}
where $(Z^{(n)},U^{(n)}, V^{(n)})$ with $Y^{(n)}$ constitute the solution to (\ref{BSDE(n)}). Then the following assertions hold.\\
{\rm{(a)}} Let $n\in\mathbb{N}$, and $(\psi_n,\varphi_n)$  be given by 
\begin{equation}\label{eq: psi_n & phi_n}
    \psi_n := - n 1\!\!1_{\Gamma^+_n} + n 1\!\!1_{\Gamma^-_n}\in\Psi_n \quad \mbox{and} \quad \varphi_n:= n 1\!\!1_{\Lambda^+_n} - n 1\!\!1_{\Lambda^-_n}\in\Phi_n
\end{equation}
where
\begin{equation}
    \Gamma^+_n:=\{\widetilde \gamma Z^{(n)} - \sigma (1+\beta)^{-1}U^{(n)} \geq 0 \}, \quad \Gamma^-_n:= \widetilde\Omega \setminus\Gamma^+_n,\quad \Lambda^-_n:=\{V^{(n)} \leq 0 \}, \quad \Lambda^+_n:= \widetilde\Omega \setminus\Lambda^-_n
\end{equation}
Then we have 
\begin{equation}\label{Y(psi)Y(n)potentials}
\begin{split}
{Y}^{(n)}_t=\mathbb{E}_0\left[\xi+\int_{t\wedge\tau}^{T\wedge\tau} g_n(s)dA_s\big|{\mathcal{G}}_t\right]=Y^{(\psi_n,\varphi_n)}_t.
\end{split}
\end{equation}
{\rm{(b)}} For any $n\in\mathbb{N}$, we have 
\begin{equation}\label{Assumption}
  \begin{split}
\mathbb{E}_0\left[\int_0^{T\wedge\tau}\widetilde{g}_n(t)dA_t\right]&\leq\mathbb{E}_0\left[\int_0^{T\wedge\tau}\vert\underset{\psi\in\Psi_n, \varphi\in\Phi_n}{\esssup}{g}_{\psi, \varphi}(t)\vert{d}A_t\right]\\
&=\sup_{\psi\in\Psi_n, \varphi\in\Phi_n}\mathbb{E}_0\left[\int_0^{T\wedge\tau}\vert{g}_{\psi, \varphi}(t)\vert{d}A_t\right]=:C_n<\infty.
  \end{split}
\end{equation}
{\rm{(c)}} For $n\in\mathbb{N}$, $\widetilde{g}_n={g}_n,$, $P\otimes{dt}$-a.e., and hence $(g_n)_n$ increases to $\widetilde{g}$, $P\otimes{dt}$-a.e.. \end{lemma}
\begin{proof}
    {\rm{(a)}} The first equality follows by Lemma \ref{lemma: BSDE for each psi phi} assertion (b). To prove the second equality, consider the BSDE \eqref{eq: G BSDE} with solution $(Y^{(n)},Z^{(n)},U^{(n)}, V^{(n)})$,
    \begin{equation}
        \begin{split}
            dY^{(n)} = Z^{(n)} dW^{(0)} + U^{(n)} d\widetilde N^{(0)} + V^{(n)} dN^\mathbb G - g_n(Z^{(n)},U^{(n)}, V^{(n)})dA^\tau
        \end{split}
    \end{equation}
    where $g_n$ is given in \eqref{Processes(g(n)gTilde(n)gTilde}. Then, the fact that $x1\!\!1_{\{x\geq 0\}} = x^+$ and $x1\!\!1_{\{x\leq 0\}} = -x^-$ we get 
  \begin{equation}
    \begin{split}
        g_n(Z^{(n)},U^{(n)}, V^{(n)}) &= -\frac{\theta(0)}{\widetilde \gamma} \sigma \big(\widetilde \gamma Z^{(n)} - \sigma (1+\beta)^{-1} U^{(n)}\big) a \\
        &+ \frac{\sigma}{\widetilde \gamma} \big(\widetilde \gamma Z^{(n)} - \sigma (1+\beta)^{-1} U^{(n)} \big) a[\theta(-n) 1_{\Gamma^+_n} + \theta (n) 1_{\Gamma^-_n}]\\
        &\quad - V^{(n)} G^{-1}_- (1-a) +V^{(n)} G^{-1}_- (1-a) [e^n 1_{\Lambda^+_n} + e^{-n} 1_{\Lambda^-_n}]
    \end{split}
\end{equation}  
using \eqref{eq: psi_n & phi_n} we get
$$
g_n(Z^{(n)},U^{(n)}, V^{(n)}) = g_{(\psi_n, \varphi_n)}(Z^{(n)},U^{(n)}, V^{(n)})
$$
Thus, by uniqueness of the solution to the BSDE, $Y^{(n)} = Y^{(\psi_n, \varphi_n)}$.

    {\rm{(b)}} Now we prove assertion (b). We have
    \begin{equation}
        \begin{split}
            \mathbb E_0 \bigg[\int_0^{T\wedge \tau} \widetilde g_n(t)dA_t \bigg] & = \mathbb E_0 \bigg[\int_0^{T\wedge\tau}
 \underset{\psi \in \Psi_n, \varphi \in \Phi_n}{\esssup} \bigg(g_{\psi, \varphi} (t) \bigg) dA_t \bigg] \quad \mbox{(by definition of $\widetilde g$)}\\
 &\leq \mathbb E_0 \bigg[\int_0^{T\wedge\tau}
\bigg| \underset{\psi \in \Psi_n, \varphi \in \Phi_n}{\esssup} g_{\psi, \varphi} (t) \bigg| dA_t \bigg] \quad \mbox{(by $x\leq|x|$)}
        \end{split}
    \end{equation}
    By Lemma \ref{lemma: BSDE for each psi phi}-part (d), $g_{\psi, \varphi}$ is upward directed. Hence there exists a sequence $\big((\psi_k, \varphi_k\big)_k \in \Psi_k\times \Phi_k$ such that $\vert g_{\psi_k, \varphi_k}(t)\vert$ converges $\mathbb P \otimes dt$-a.e. to $\underset{\psi\in\Psi_n,\varphi\in\Phi_n}{\esssup} \vert g_{\psi, \varphi} (t) \vert$. Hence,
    \begin{equation}
        \begin{split}
            \mathbb E_0 \bigg[\int_0^{T\wedge\tau}
 \bigg| \underset{\psi \in \Psi_n, \varphi \in \Phi_n}{\esssup} g_{\psi, \varphi} (t) \bigg| dA_t \bigg] = \mathbb E_0 \bigg[\int_0^{T\wedge\tau} \underset{k\longrightarrow \infty}{\lim} \vert g_{\psi_k, \varphi_k} (t) \vert dA_t \bigg]
        \end{split}
    \end{equation}
Since the subsequence $(\psi_k, \varphi_k)$ is bounded, then there exists a random variable $\chi$ with $\mathbb E_0 [|\chi|] < \infty$ such that $\vert g_{\psi_k, \varphi_k}(t)\vert \leq \chi$ for every $k\in \mathbb N$. Together, we can apply Fatou's Lemma,
     \begin{equation}
        \begin{split}
             \mathbb E_0 \bigg[\int_0^{T\wedge\tau} \underset{k\longrightarrow \infty}{\lim} \vert g_{\psi_k, \varphi_k} (t) \vert dA_t \bigg] \leq \underset{k\longrightarrow \infty}{\lim} \mathbb E_0 \bigg[\int_0^{T\wedge\tau}  \vert g_{\psi_k, \varphi_k} (t) \vert dA_t \bigg] 
        \end{split}
    \end{equation}
    Since  $\big((\psi_k, \varphi_k\big)_k$ is a subsequence and $\Psi_{n_k}\times \Phi_{n_k} \subset \Psi_n\times \Phi_n$, then
    \begin{equation}
        \begin{split}
            \underset{k\longrightarrow \infty}{\lim} \mathbb E_0 \bigg[\int_0^{T\wedge\tau}  \vert g_{\psi_k, \varphi_k} (t) \vert dA_t \bigg] \leq \underset{\psi\in\Psi_n,\varphi\in\Phi_n}{\sup} \mathbb E_0 \bigg[\int_0^{T\wedge\tau}  \vert g_{\psi, \varphi} (t) \vert dA_t \bigg] =: C_n
        \end{split}
    \end{equation}
    By combining this inequality with Lemma \ref{lemma: BSDE for each psi phi} assertion (a), we complete the proof of assertion (b).

    {\rm{(c)}} 
To prove assertion (c) we notice that due to Lemma \ref{lemma: BSDE for each psi phi}-(c)-(d), we always have ${g}_n={g}_{\psi_n}\leq \widetilde{g}_n$, and the rest of the proof focuses on the converse inequality. Let  $(\psi_k,\varphi_k)_k\subset\Psi_n \times \Phi_n$ such that $g_{\psi_k,\varphi_k}$ converges to $\widetilde{g}_n$, $P\otimes{dt}$-a.e.. Hence, by virtue of assertion (b) (which implies that $\underset{\psi\in\Psi_n, \varphi\in\Phi_n}{\esssup}\vert{g}_{\psi,\varphi}\vert\in L^1(\mathbb Q_0\otimes dt)$) and the dominated convergence theorem, this convergence holds in $L^1(P\otimes dt)$. Thus, we derive
\begin{equation*}
\begin{split}
&\mathbb E\left[\xi+\int_{t\wedge\tau}^{T\wedge \tau} \widetilde{g}_n(s)dA_s\big|{\mathcal{G}}_t\right]\\&=\lim_{k\longrightarrow\infty} \mathbb E\left[\xi+\int_{t\wedge\tau}^{T\wedge \tau}  g_{\psi_k,\varphi_k}(s)dA_s\big|{\mathcal{G}}_t\right]\\
&=\lim_{k\longrightarrow\infty}Y^{(\psi_k,\varphi_k)}_t\leq Y^{(n)}_t=\mathbb E\left[\xi+\int_{t\wedge\tau}^{T\wedge \tau} g_{n}(u)dA_s\big|{\mathcal{G}}_t\right].
\end{split}
\end{equation*}
Then, by setting $t=0$ and taking expectation under $\mathbb Q_0$ afterwards, we get
$$\mathbb E\left[\int_0^{T\wedge\tau}  \widetilde{g}_n(s)dA_s\right]\leq \mathbb{E}\left[\int_0^{T\wedge\tau} g_{n}(s)dA_s\right].$$
Thus, by combining this with  ${g}_n\leq\widetilde{g}_n$ $P\otimes{dt}$-a.e., we obtain ${g}_n= \widetilde{g}_n$, $P\otimes dt$-a.e.. This ends the proof of the lemma.
\end{proof}
The following lemma constitutes our last main technical step for the proof of Theorem \ref{BSDE4Y(up)}, and it elaborates some inequalities for the process $\widetilde{g}$. 
\begin{lemma}\label{LemmaforgTildeControl}
Consider the process $\widetilde{g}$ defined in (\ref{Processes(g(n)gTilde(n)gTilde}). Then the following assertions hold.\\
{\rm{(a)}} For any $t\in[0,T]$, we have 
\begin{equation}\label{Domination4Gtilde}
\begin{split}
&\mathbb{E}_0\left[\int_{t\wedge\tau}^{T\wedge\tau} \widetilde g(s)dA_s\big|{\mathcal{G}}_t\right]\leq \mathbb{E}_0\left[\Delta+\xi^-\big|{\mathcal{G}}_t\right],\quad\mbox{where}\\
 \Delta&:=\underset{\psi\in\Psi, \varphi\in\Phi}{\esssup}\sup_{0\leq{t}\leq{T\wedge\tau}}\mathbb{E}_0[D^{(\psi, \varphi)}_T(D^{(\psi,\varphi)}_t)^{-1}\xi^+\big|{\mathcal{G}}_t].
\end{split}
\end{equation}

{\rm{(b)}} For $p\in(1,\infty)$, there exists a universal constant $C_p\in(0,\infty)$ such that 
\begin{equation}\label{Domination4Gtilde1}
\begin{split}
\mathbb{E}_0\left[\left(\int_0^{T\wedge\tau}\widetilde{g}(u)dA_u\right)^p\right]\leq C_p \mathbb{E}_0\left[(\xi^-)^p\right]+ C_p \sup_{(\psi,\varphi)\in\Psi\times\Phi,\ \sigma\in{\cal{T}}_{T\wedge\tau}}\mathbb{E}_0[D^{(\psi, \varphi)}_T(D^{(\psi,\varphi)}_{\sigma})^{-1}(\xi^+)^p].
\end{split}
\end{equation}
\end{lemma}

\begin{proof}[Proof of Lemma \ref{LemmaforgTildeControl}] Thanks to Lemma \ref{lemma4Gtilde}-(a), for any $n\geq 1$, we derive 
\begin{equation}\label{eq: g_n controled}
\begin{split}
\mathbb{E}_0\left[\int_{t\wedge\tau}^{T\wedge\tau} g_n(s)dA_s\big|{\mathcal{G}}_t\right]&=\mathbb{E}_0\left[-\xi+{Y}^{(n)}_t\big|{\mathcal{G}}_t\right]=\mathbb{E}_0\left[-\xi+Y^{(\psi_n, \varphi_n)}_t\big|{\mathcal{G}}_t\right]\\
&\leq \mathbb{E}_0\left[\xi^-+\underset{\psi\in\Psi,\varphi\in\Phi}{\esssup} \ \underset{0\leq t\leq T\wedge\tau}{\sup}(Y^{(\psi,\varphi)}_t)^+\big|{\mathcal{G}}_t\right],
\end{split}
\end{equation}
due to $-\xi \leq \xi^-$ and $Y^{(\psi_n, \varphi_n)}_t \leq \underset{\psi\in\Psi,\varphi\in\Phi}{\esssup} \ \underset{0\leq t\leq T\wedge\tau}{\sup}(Y^{(\psi,\varphi)}_t)^+$. Since $g_n$ is increasing, then $\lim_{n\longrightarrow \infty} g_n = \lim \sup_{n\longrightarrow \infty} g_n$. Hence by taking the limit in the above and applying Fatou we get
\begin{equation*}
    \begin{split}
        \underset{n\longrightarrow \infty}{\lim}\mathbb{E}_0\left[\int_{t\wedge\tau}^{T\wedge\tau} g_n(s)dA_s\big|{\mathcal{G}}_t\right]&= \underset{n\longrightarrow \infty}{\lim\sup}\ \mathbb{E}_0\left[\int_{t\wedge\tau}^{T\wedge\tau} g_n(s)dA_s\big|{\mathcal{G}}_t\right] \\
        &\leq \mathbb{E}_0\left[\int_{t\wedge\tau}^{T\wedge\tau} \underset{n\longrightarrow \infty}{\lim\sup}\ g_n(s)dA_s\big|{\mathcal{G}}_t\right] \\
        &= \mathbb{E}_0\left[\int_{t\wedge\tau}^{T\wedge\tau} \widetilde g(s)dA_s\big|{\mathcal{G}}_t\right]
    \end{split}
\end{equation*}
Therefore, by combining this equality with (\ref{eq: g_n controled}), the inequality (\ref{Domination4Gtilde}) follows immediately, and the proof of assertion (a) is complete. Thus, the rest of the proof focuses on proving assertion (b). 
To this end, we apply Garsia's lemma to the process $\int_0^{\cdot}\widetilde{g}(u)dA_u$, see \cite[Theorem 99]{dellacheriemeyer80}, and derive 
\begin{equation}\label{Inequa997}
\begin{split}
\mathbb{E}_0\left[\left(\int_0^{T\wedge \tau}\widetilde{g}(u)dA_u\right)^p\right]& \leq \widetilde C_p \mathbb{E}_0\left[\bigg(\xi^-+ \underset{\psi\in\Psi,\varphi\in\Phi}{\esssup} \ \underset{0\leq t\leq T\wedge\tau}{\sup} (Y^{(\psi,\varphi)}_t)^+\bigg)^p \right]\\  
&\leq 2^{p-1}\widetilde C_p \mathbb{E}_0\left[(\xi^-)^p\right]+ 2^{p-1} \widetilde C_p \mathbb{E}_0\left[\left(\underset{\psi\in\Psi,\varphi\in\Phi}{\esssup} \ \underset{0\leq t\leq T\wedge\tau}{\sup}(Y^{(\psi, \varphi)}_t)^+\right)^p\right]\\
&= C_p \mathbb{E}_0\left[(\xi^-)^p\right]+ C_p \mathbb{E}_0\left[\left(\underset{k\longrightarrow \infty}{\lim} \ \underset{0\leq t\leq T\wedge\tau}{\sup}(Y^{(\psi_k, \varphi_k)}_t)^+\right)^p\right] \\
&= C_p \mathbb{E}_0\left[(\xi^-)^p\right]+ C_p \underset{k\longrightarrow \infty}{\lim} \ \mathbb{E}_0\left[\left(\underset{0\leq t\leq T\wedge\tau}{\sup}(Y^{(\psi_k, \varphi_k)}_t)^+\right)^p\right] \\
&\leq C_p\mathbb{E}_0\left[(\xi^-)^p\right]+ C_p \underset{\psi\in\Psi,\varphi\in\Phi}{\esssup} \mathbb{E}_0\left[\left(\underset{0\leq t\leq T\wedge\tau}{\sup}(Y^{(\psi, \varphi)}_t)^+\right)^p\right] 
\end{split}
\end{equation} 
The latter inequality follows from combining  the convergence monotone theorem and Lemma \ref{lemma: BSDE for each psi phi}-(d), which states that the family $\{Y^{\psi, \varphi}:\ (\psi, \varphi) \in\Psi\times\Phi\}$ is upward direct.\\ Then the rest of the proof mimics exactly the proof of corresponding part in \cite[Lemma 4.2]{choulli2025second}. This ends the proof of the lemma.
\end{proof}
\begin{lemma}\label{lemma: N^G martingale under Q}
    Suppose Assumptions \ref{assum: model parameters locally bounded}, \ref{assum: gamma non-zero} and \ref{Assum: avoid F s.t} hold, then the process
    \begin{equation}
         N^\mathbb{G} -(e^{\varphi} - 1)G^{-1}_{-} \is D^{p,\mathbb{F}}
    \end{equation}
    is a martingale under $\mathbb Q^{(\psi,\varphi)}$.
\end{lemma}
\begin{proof}
    By Lemma \ref{lemma: Esscher and Girsanov}, we have
    \begin{equation}
        \mathbb Q^{(\psi,\varphi)} = \overline{D}^{(\psi,\varphi)} \is \mathbb P, \quad\mbox{with}\quad  \overline{D}^{(\psi,\varphi)} = \mathcal E(\overline{M}^{(\psi,\varphi)}) 
    \end{equation}
    where 
    \begin{equation}
        \overline{M}^{(\psi,\varphi)}= \theta \sigma \is \mathcal T(W) + (e^{\theta \zeta + \psi \zeta^2} -1)(1+\beta) \is \mathcal T(N^\mathbb F) + \frac{\widetilde G}{G} (e^\varphi -1)[1+(e^{\theta \zeta + \psi \zeta^2} -1)\Delta N]\is N^\mathbb G
    \end{equation}
 Let $V \in \mathcal V$ such that $N^\mathbb G - V \in \mathcal M_{loc}(\mathbb Q^{(\psi,\varphi)}, \mathbb G)$. This is equivalent to $\overline{D}^{(\psi,\varphi)} (N^\mathbb G - V)\in \mathcal M_{loc}(\mathbb P, \mathbb G)$. By integration by parts and ignoring the $(\mathbb P, \mathbb G)$-martingale terms, this fact is equivalent to 
    \begin{equation}
        - V + [\overline{M}^{(\psi,\varphi)}, N^\mathbb G] \in \mathcal M_{loc}(\mathbb Q^{(\psi,\varphi)}, \mathbb G)\quad\mbox{or equivalently} \quad V = \bigg([\overline{M}^{(\psi,\varphi)}, N^\mathbb G]  \bigg)^{p, (\mathbb P, \mathbb G)}.
    \end{equation} 
    By calculating $[\overline{M}^{(\psi,\varphi)}, N^\mathbb G] $  and inserting it in the above equality, we obtain
    \begin{equation}
        V = (e^{\varphi} - 1)G^{-1}_{-} \is D^{p,\mathbb{F}}.
    \end{equation}
    This ends the proof of the lemma.
\end{proof}
Now, we are at the stage of proving Theorem  \ref{BSDE4Y(up)}.\\
\begin{proof}[Proof of Theorem \ref{BSDE4Y(up)}]
Remark that there is no loss of generality in assuming that $r\equiv0$, which will be enforced throughout this proof. The proof of the theorem is divided into five parts. The first part proves assertion (b) and (c), while the second part proves both assertions (a) and (d).\\
{\bf Part 1.} Hereto we prove assertion (b). To this end, for $s\in [0,T\wedge\tau]$, $(z,u, v)\in\mathbb{R}^3$ and $n\in\mathbb{N}$, we put 
\begin{equation}\label{DRiver4Y(n)BSDE}
\begin{split}
     \kappa_n(s,z,u, v)&:={{\theta_s(-n)-\theta_s(0)}\over{\widetilde\gamma_s}}\sigma_s(\widetilde\gamma_s{z}-\sigma_s(1+\beta_s)^{-1}{u})^+ a_s +  {{\theta_s(0)-\theta_s(n)}\over{\widetilde\gamma_s}}\sigma_s(\widetilde\gamma_s{z}-\sigma_s(1+\beta_s)^{-1}{u})^- a_s\\
     &+ (e^n - 1) v^+ G^{-1}_{s-} (1-a_s) + (1- e^{-n}) v^- G^{-1}_{s-}  (1-a_s),
\end{split}
\end{equation}
It is clear that $ \kappa_n$ is the driver of the BSDE (\ref{BSDE(n)}). In virtue of Assumptions \ref{assum: model parameters locally bounded} and \ref{assum: gamma non-zero}, and Lemma \ref{lemma: Esscher and Girsanov}, there exists a constant $C_n\in(0,\infty)$ such that 
\begin{equation*}
\begin{split}
&\vert\kappa_n(t,z_1,u_1, v_1) - \kappa_n(t,z_2,u_2, v_2)\vert\\
&\leq{C_n}\vert (\widetilde\gamma_s{z_1}-\sigma_s(1+\beta_s)^{-1}{u_1})^+-(\widetilde\gamma_s{z_2}-\sigma_s(1+\beta_s)^{-1}{u_2})^+\vert\\
&\quad+C_n\vert (\widetilde\gamma_s{z_1}-\sigma_s(1+\beta_s)^{-1}{u_1})^--(\widetilde\gamma_s{z_2}-\sigma_s(1+\beta_s)^{-1}{u_2})^-\vert\\
&\quad + {C_n}\vert v_1^+ - v_2^+\vert+C_n\vert v_1^- - v_2^- \vert\\
&\leq{4C_n}\left(\vert{z_1}-{z_2}\vert+\vert{u_1}-{u_2}\vert+\vert{v_1}-{v_2}\vert\right).
\end{split}
\end{equation*}
This proves that the driver $\kappa_n$ is Lipschitz. Hence, we deduce that there exists a unique solution to the BSDE (\ref{BSDE(n)}), denoted by $(\widetilde{Y}^{(n)}, \widetilde{Z}^{(n)},\widetilde{U}^{(n)}, \widetilde V^{(n)})$ and belongs to the space  $ \mathbb{S}^p(\mathbb Q_0) \times \mathbb{L}^p(W,\mathbb Q_0) \times \mathbb{L}^p(N,\mathbb Q_0)\times \mathbb D^p(\mathbb Q_0)$.
Thus, assertion (b) will follow immediately as soon as we prove that
\begin{equation}\label{ReamingProof}
Y^{(n)}=\widetilde{Y}^{(n)},\ \mbox{and there exists $(\psi_n,\varphi_n)\in\Psi_n\times \Phi_n$ such that }\ \widetilde{Y}^{(n)}={Y}^{(\psi_n,\varphi_n)}.
\end{equation}\\
To this end, in virtue of Lemma \ref{EtaProcess}, we remark that  
\begin{equation}\label{h(psi)-kappa(n)}
g_{\psi, \varphi}(t,z,u, v)\leq \kappa_n(t,z,u,v),\quad\mbox{ for any}\ t\in[0,T],\ (z,u,v)\in\mathbb{R}^3,\ (\psi, \varphi) \in\Psi_n \times \Phi_n.
\end{equation}
 On the one hand, due to Lemma \ref{lemma: Esscher and Girsanov}, direct It\^o calculations allows us to deduce that for any $(\psi, \varphi)\in\Psi_n\times\Phi_n$, 
 \begin{equation}\label{Dpsi}
{\mathcal{E}}\Bigl(\sigma(\theta(\psi)-\theta(0))\is{W}^{(0)}-{{\sigma^2}\over{\widetilde\gamma\lambda}}(\theta(\psi)-\theta(0))e^{-\theta(0)\zeta}(1+\beta)\is{\widetilde N}^{(0)} + (e^\varphi -1)\is N^\mathbb{G}\Bigr)={D}^{(\psi,\varphi)}\in{\mathcal{M}}(\mathbb{G},\mathbb Q_0),
 \end{equation}
 and hence 
  \begin{equation}\label{Wpsi2}
  W^{(\psi, \varphi)}=W^{(0)}-\sigma[\theta(\psi)-\theta(0)]\is(\cdot \wedge\tau) \quad \mbox{and}\quad (1+\beta)\is{\widetilde N}^{(\psi, \varphi)}=(1+\beta)\is{\widetilde N}^{(0)}+{{\sigma^2}\over{\widetilde\gamma}}[\theta(\psi)-\theta(0)] \is(\cdot \wedge \tau).
   \end{equation}
Thanks to these two remarks, we derive 
 \begin{equation*}\label{differnce4Y(n)Y(psi)}
 \begin{split}
 &\widetilde{Y}^{(n)}_t-Y^{(\psi, \varphi)}_t\\
 &=\int_{t\wedge\tau}^{T\wedge \tau}\left(\kappa_n(s, \widetilde{Z}^{(n)},\widetilde{U}^{(n)}, \widetilde{V}^{(n)})-g_{\psi, \varphi}(s,Z^{(\psi,\varphi)}_s,U^{(\psi,\varphi)}_s, V^{(\psi,\varphi)})\right) dA_s-\int_{t\wedge\tau}^{T\wedge\tau}(\widetilde{Z}^{(n)}_s-Z^{(\psi,\varphi)}_s)dW^{(0)}_s\\
 &\quad-\int_{t\wedge\tau}^{T\wedge\tau}(\widetilde{U}^{(n)}_s-U^{(\psi,\varphi)}_s)d\widetilde{ N}^{0,0}_s - \int_{t\wedge\tau}^{T\wedge\tau}(\widetilde{V}^{(n)} - V^{(\psi,\varphi)}) dN^\mathbb{G}\\
 &\geq\int_{t\wedge\tau}^{T\wedge \tau}\left(g_{\psi,\varphi}(s, \widetilde{Z}^{(n)},\widetilde{U}^{(n)}, \widetilde{V}^{(n)})-g_{\psi, \varphi}(s,Z^{(\psi,\varphi)}_s,U^{(\psi,\varphi)}_s, V^{(\psi,\varphi)})\right) dA_s-\int_{t\wedge\tau}^{T\wedge\tau}(\widetilde{Z}^{(n)}_s-Z^{(\psi,\varphi)}_s)dW^{(0)}_s\\
 &\quad-\int_{t\wedge\tau}^{T\wedge\tau}(\widetilde{U}^{(n)}_s-U^{(\psi,\varphi)}_s)d\widetilde{N}^{0,0}_s - \int_{t\wedge\tau}^{T\wedge\tau}(\widetilde{V}^{(n)} - V^{(\psi,\varphi)}) dN^\mathbb{G}\\
 &=-\int_{t\wedge\tau}^{T\wedge\tau}(\widetilde{Z}^{(n)}_s-Z^{(\psi,\varphi)}_s)\left(dW^{(0)}_s-\sigma_s(\theta_s(\psi)-\theta_s(0))ds\right)\\
 &\quad -\int_{t\wedge\tau}^{T\wedge\tau}(\widetilde{U}^{(n)}_s-U^{(\psi,\varphi)}_s)\left(d\widetilde{N}^{0,0}_s+{{\sigma_s^2}\over{\widetilde\gamma_s}}(\theta_s(\psi)-\theta_s(0))ds\right)\\
 &\quad - \int_{t\wedge\tau}^{T\wedge\tau} (\widetilde V^{(n)} - V^{(\psi,\varphi)}) \left(dN^\mathbb{G} -(e^{\varphi_s} - 1)G^{-1}_{s-} dD^{p,\mathbb{F}}_s\right)\\
 &=-\int_{t\wedge\tau}^{T\wedge\tau}(\widetilde{Z}^{(n)}_s-Z^{(\psi,\varphi)}_s)d W^{\psi, \varphi}_s-\int_{t\wedge\tau}^{T\wedge\tau}(\widetilde{U}^{(n)}_s-U^{(\psi,\varphi)}_s)d\widetilde{ N}^{\psi, \varphi}_s \\
 &\quad- \int_{t\wedge\tau}^{T\wedge\tau} (\widetilde V^{(n)} - V^{(\psi,\varphi)}) \left(dN^\mathbb{G} -(e^{\varphi_s} - 1)G^{-1}_{s-} dD^{p,\mathbb{F}}_s\right)
 \end{split}
 \end{equation*}
 Thus, by taking conditional expectation under $\mathbb Q^{(\psi, \varphi)}$ in the above inequality and using Lemma \ref{lemma: N^G martingale under Q}, we deduce that 
 \begin{equation}
     \widetilde{Y}^{(n)}_t-Y^{(\psi, \varphi)}_t\geq 0
 \end{equation}
Hence, $\widetilde{Y}^{(n)}\geq Y^{(\psi,\varphi)}$ for any $\psi\in\Psi_n$, and hence this yields $Y^{(n)}\leq\widetilde{Y}^{(n)}$.
 
 To prove the converse, we consider the quadruple $(\widetilde{Y}^{(n)}, \widetilde{Z}^{(n)},\widetilde{U}^{(n)}, \widetilde{V}^{(n)})$ solution to (\ref{BSDE(n)}), then by Lemma \ref{lemma4Gtilde}-(a) we obtain $\kappa_n(s, \widetilde{Z}^{(n)}_s,\widetilde{U}^{(n)}_s,\widetilde{V}^{(n)}_s)=g_{\psi_n,\varphi_n}(s, \widetilde{Z}^{(n)}_s,\widetilde{U}^{(n)}_s,\widetilde{V}^{(n)}_s)$. This implies that $(\widetilde{Y}^{(n)},\widetilde{Z}^{(n)},\widetilde{U}^{(n)}, \widetilde{V}^{(n)})$ is a solution to the BSDE (\ref{eq: G BSDE}) when $(\psi, \varphi)=(\psi_n, \varphi_n)$. Therefore, the uniqueness of the solution of this latter BSDE is guaranteed by Lemma \ref{lemma: BSDE for each psi phi}, and this yields  $\widetilde{Y}^{(n)}={Y}^{(\psi_n, \varphi_n)}\leq Y^{(n)}$. This proves $\widetilde{Y}^{(n)}=Y^{(n)}$, and the proof of assertion (b) is complete. \\
{\bf Part 2.} This part proves assertion (c). Due to the definition of the essential supremum, it is clear that $(Y^{(n)})_n$ is a nondecreasing sequence and 
\begin{equation}\label{ComparisonY(n)}
Y^{(n)}\leq Y^{(n+1)}\leq Y^{up},\quad\mbox{and}\quad \sup_n Y^{(n)}\leq Y^{up}.
\end{equation}
Furthermore, thanks to Lemma \ref{lemma: BSDE for each psi phi}-(d), there exists a sequence $(\psi_k,\varphi_k)_k\subset\Psi\times\Phi$ such that 
$$Y^{up}= \underset{\psi\in\Psi, \varphi\in\Phi}{\esssup} Y^{(\psi, \varphi)}=\displaystyle\lim_{k\longrightarrow\infty}Y^{(\psi_k, \varphi_k)}_t.$$
Thus, as $(\psi_k, \varphi_k)$ is bounded, there exists $n_k\in\mathbb{N}$ such that $(\psi_k, \varphi_k)\in\Psi_{n_k}\times\Phi_{n_k}$, and we deduce 
$$Y^{up}=\lim_{k\longrightarrow\infty}Y^{(\psi_k,\varphi_k)}\leq \sup_k Y^{(n_k)}\leq \sup_n Y^{(n)}.$$
A combination of this with (\ref{ComparisonY(n)}) yields the convergence almost surely of $Y^{(n)}$ to $Y^{up}$. This proves (\ref{Increasiness4Y(n)}).  To prove the remaining claim of assertion (c), for any $n\geq 0$, we consider $(Y^{(n)}, Z^{(n)},U^{(n)}, V^{(n)})$ the unique solution to (\ref{BSDE(n)}), and  $(Y^{\psi, \varphi}, Z^{(\psi,\varphi)},U^{(\psi,\varphi)}, V^{(\psi,\varphi)})$ is the unique solution to (\ref{eq: G BSDE}). Then, thanks to assertion (a) of Lemma \ref{lemma4Gtilde},  for any $n,m\in\mathbb{N}$, we derive
\begin{equation}\label{GtildeVersusg(n)}
\begin{split}
 Y^{(n+m)}_t-Y^{(n)}_t&=E_0\left[\int_{t\wedge\tau}^{T\wedge\tau}\left(g_{n+m}(u)-g_n(u)\right)dA_u \big|{\mathcal{G}}_t\right]\\
&=E_0\left[\int_0^{T\wedge\tau}\left(g_{n+m}(u)-g_n(u)\right)dA_u \big|{\mathcal{G}}_t\right]-\int_0^{t\wedge\tau}\left(g_{n+m}(u)-g_n(u)\right)dA_u.
\end{split}
\end{equation}
Thus, in virtue of Lemma \ref{lemma4Gtilde}-(c), we conclude that is in fact a nonnegative supermartingale under $(\mathbb{G},\mathbb Q_0)$. This ends the proof of assertion (c). \\
{\bf Part 3.} In this part, we prove the following properties.
\begin{equation}\label{ConvergenceResults}
\begin{split}
&i)\quad (Y^{(n)}, Z^{(n)}, U^{(n)}, V^{(n)})\quad\mbox{  converges in}\quad \mathbb{S}^p(\mathbb Q_0)\times\mathbb{L}^p(W,\mathbb Q_0) \times \mathbb{L}^p(N,\mathbb Q_0) \times \mathbb D^p(\mathbb Q_0),\\
&ii)\quad \mbox{the limit of  $(Y^{(n)}$ coincides with the process $Y^{up}$},\\
&iii)\quad \mbox{there exists a subsequence of $(Y^{(n)}, Z^{(n)}, U^{(n)}, V^{(n)})$ that converges $P\otimes{dt}$-a.e.}.
\end{split}
\end{equation}
On the one hand, by combining Lemma \ref{LemmaforgTildeControl}-(b) and Lemma \ref{lemma4Gtilde}-(c) and the dominated convergence theorem, we deduce that under the assumption  (\ref{MainAssumption4Xi}), we have
\begin{equation}\label{NormConvergence4g(n)}
\sup_{m}\mathbb{E}_0\left[\left(\int_0^{T\wedge\tau}(g_{n+m}(u)-g_{n}(u))dA_u\right)^p\right]\longrightarrow 0\quad\mbox{and}\quad\mathbb{E}_0\left[\left(\int_0^{T\wedge\tau}(\widetilde{g}(u)-g_{n}(u))dA_u\right)^p\right]\longrightarrow 0.
\end{equation}
On the other hand, thanks to (\ref{GtildeVersusg(n)}), we get 
\begin{equation*}
\begin{split}
0\leq Y^{(n+m)}_t-Y^{(n)}_t=\mathbb{E}_0\left[\int_{t\wedge\tau}^{T\wedge\tau}\left(g_{n+m}(u)-g_n(u)\right)dA_u \big|{\mathcal{G}}_t\right]\leq\mathbb{E}_0\left[\int_0^{T\wedge\tau}\left(g_{n+m}(u)-g_n(u)\right)dA_u \big|{\mathcal{G}}_t\right].
\end{split}
\end{equation*}
Then by combining this latter inequality with Doob's inequality, we deduce the existence of a universal constant $C_p\in(0,\infty)$ such that 
\begin{equation*}
\begin{split}
\sup_{m}\Vert Y^{(n+m)}-Y^{(n)}\Vert_{{\mathcal{S}}^p}^p\leq C_p\sup_{m}\mathbb{E}_0\left[\left(\int_0^{T\wedge\tau}(g_{n+m}(u)-g_{n}(u))dA_u\right)^p\right]\longrightarrow 0.
\end{split}
\end{equation*}
This implies that $(Y^{(n)})_n$ is a Cauchy sequence in $ \mathbb{S}^p(\mathbb Q_0)$-norm and hence it converges in this norm to its limit $\widetilde{Y}\in \mathbb{S}^p(\mathbb Q_0)$. By combining this with assertion (c) proved in part 2--which states that $Y^{(n)}$ converges almost surely pointe-wise to $Y^{up}$-- and the uniqueness of the limit, we obtain the convergence of $Y^{(n)}$ almost surely pointe-wise and in $ \mathbb{S}^p(\mathbb Q_0)$-norm to $Y^{up}$. This proves the property  ii)  in (\ref{ConvergenceResults}). \\
Now we address the convergence of $(Z^{(n)}, U^{(n)}, V^{(n)})$ in the space $\mathbb{L}^p(W,\mathbb Q_0) \times \mathbb{L}^p(N,\mathbb Q_0)\times \mathbb D^p(\mathbb Q_0)$. Thus, by combining assertion (c) which guarantees that  $Y^{(n)}- Y^{(n+m)}$ is a $(\mathbb{G}, \mathbb Q_0)$-submartingale,  and \cite[Lemma 1.9]{Stricker} applied to $Y^{(n)}- Y^{(n+m)}$, we obtain 
\begin{equation}\label{Norm4Z(n)U(n)}
%
\Vert{Z}^{(n+m)}-{Z}^{(n)} \Vert^p_{L^p(W,\mathbb Q_{0,})}+\Vert{U}^{(n+m)}-{U}^{(n)}\Vert^p_{L^p(N,\mathbb Q_0)} +\Vert{V}^{(n+m)}-{V}^{(n)}\Vert^p_{\mathbb D^p(\mathbb Q_0)} \leq \Vert Y^{(n+m)}-Y^{(n)}\Vert_{{\mathcal{S}}^p(\mathbb Q_0)}^p.
\end{equation}
As a result, we conclude the existence of a pair $({Z}^{up},{U}^{up}, {V}^{up})\in\mathbb{L}^p(W,\mathbb Q_0) \times \mathbb{L}^p(N,\mathbb Q_0)\times \mathbb D^p(\mathbb Q_0)$ such that $(Y^{(n)}, Z^{(n)},U^{(n)},V^{(n)})$  converges to $({Y}^{up}, {Z}^{up},{U}^{up}, V^{up})$ in the space ${\mathcal{S}}^p(\mathbb Q_0)\times\mathbb{L}^p(W,\mathbb Q_0) \times \mathbb{L}^p(N,\mathbb Q_0)\times \mathbb D^p(\mathbb Q_0)$, and the property i) in (\ref{ConvergenceResults}) is proved.\\

Furthermore, if $(T_k)_k$ be a sequence of stopping times that increases to infinity and $N^{T_n}\leq k$, then for any $k\geq 1$ we have 
\begin{equation*}
\begin{split}
\mathbb Q_0\otimes{{dt}\over{T}}\left(\lambda{e}^{\eta(0)\zeta}\vert{U}^{up}-{U}^{(n)}\vert{I}_{\Lbrack0,T_k\Rbrack}>\epsilon\right)&\leq(T\epsilon)^{-1}\mathbb{E}_0\left[\int_0^{T_k}\lambda_s{e}^{\eta_s(0)\zeta_s}\vert{U}^{up}_s-{U}^{(n)}_s\vert{d}s\right]\\
&=(T\epsilon)^{-1}\mathbb{E}_0\left[\int_0^{T_k}\vert{U}^{up}_s-{U}^{(n)}_s\vert{d}N_s\right]\\
&\leq(T\epsilon)^{-1}\Vert{U}^{up}-{U}^{(n)}\Vert^p_{L^p(N,\mathbb Q_0)}\Vert\sqrt{N_{T_k}}\Vert_{L^p(\mathbb Q_0)}.
\end{split}
\end{equation*}
This proves that ${U}^{(n)}$ converges to ${U}^{up}$ in probability under $\mathbb Q_0\otimes({dt}/T)$, then there exists a subsequence $(Y^{(n)}, Z^{(n)},U^{(n)},V^{(n)})$ that converges to $({Y}^{up}, {Z}^{up},{U}^{up},{V}^{up})$ in both norm and $\mathbb Q_0\otimes{dt}$-a.e.. This proves the property iii) in (\ref{ConvergenceResults}), and part 3 is complete.\\
 {\bf Part 4.} Here we prove that $({Y}^{up}, {Z}^{up},{U}^{up}, V^{up})$ satisfies the constraints and the Skorokhod condition, i.e. the last two conditions in (\ref{ConstrainedBSDE1}), and prove assertion (d).
  To this end, consider the subsequence $({Z}^{(n)},{U}^{(n)}, V^{(n)})$ that converges to $({Z}^{up},{U}^{up}, V^{up})$ $\mathbb Q_0\otimes{dA_t}$-a.e., which is obtained in Part 3.  Thus, by combining this with the fact that ${{\theta_s(-n)-\theta_s(0)}/{\widetilde\gamma_s}}$ converges to $(\theta_s^*-\theta_s(0))/\widetilde\gamma_s$, $\mathbb Q_0\otimes{dt}$-a.e. a and $(1-e^{-n})$ converges to $1$ $\mathbb Q_0\otimes{dD^{p,\mathbb{F}}}$ due to Lemma \ref{EtaProcess} -(c), we deduce that 
  $$(\theta_s(-n)-\theta_s(0)){\widetilde\gamma_s}^{-1}\sigma_s(\widetilde\gamma_s{Z}_s^{(n)}-\sigma_s(1+\beta_s)^{-1}{U}_s^{(n)})^+a_s + (1-e^{-n})V^-_s G^{-1}_{s-} (1-a_s)$$
  converges to 
$$(\theta_s^*-\theta_s(0)){\widetilde\gamma_s}^{-1}\sigma_s(\widetilde\gamma_s\widetilde{Z}_s-\sigma_s(1+\beta_s)^{-1}\widetilde{U}_s)^+ a_s + \widetilde V^-_s G^{-1}_{s-}(1-a_s), \quad \mathbb Q_0\otimes{dA_t}-a.e$$

Hence,  as $g_n$ increases to $\widetilde{g}$ in both ${\mathcal{A}}^p(\mathbb Q_0)$-norm and $\mathbb Q_0\otimes{dA_t}$-a.e. (see Lemmas \ref{lemma4Gtilde}  and \ref{LemmaforgTildeControl}), and 
\begin{equation*}
\begin{split}
g_n(s)&=\underbrace{{{\theta_s(-n)-\theta_s(0)}\over{\widetilde\gamma_s}}\sigma_s(\widetilde\gamma_s{Z}_s^{(n)}-\sigma_s(1+\beta_s)^{-1}{U}_s^{(n)})^+a_s + (1-e^{-n})\big(V_s^{(n)}\big)^- G^{-1}_{s-} (1-a_s)}_{g_n^{(1)}(s)}\\
&+ \underbrace{ {{\theta_s(0)-\theta_s(n)}\over{\widetilde\gamma_s}}\sigma_s(\widetilde\gamma_s{Z}_s^{(n)}-\sigma_s(1+\beta_s)^{-1}{U}_s^{(n)})^- a_s + (e^n-1)\big(V_s^{(n)} \big)^+ G^{-1}_{s-} (1-a_s)}_{g_n^{(2)}(s)}\\
\end{split}
\end{equation*}
we obtain 
\begin{equation}\label{eq: g^(2)/n goes to 0}
\begin{split}
\mathbb{E}_0\left[\int_0^{T\wedge\tau}{{g_n^{(2)}(t)}\over{n}}dA_t\right]\leq {1\over{n}} \mathbb{E}_0\left[\int_0^{T\wedge\tau}{g}_n(u)dA_u\right]\leq {1\over{n}}\mathbb{E}_0\left[\int_0^{T\wedge\tau}\widetilde{g}(u)dA_u\right]\longrightarrow 0.
\end{split}
\end{equation}
By combining this with Fatou's lemma, the convergence a.e. of $ (\eta_s(0)-\eta_s(n))\sigma_s(\widetilde\gamma_s{Z}_s^{(n)}-\sigma_s(1+\beta_s)^{-1}{U}_s^{(n)})^-/n\widetilde\gamma_s$ to $\zeta\sigma_s(\widetilde\gamma_s\widetilde{Z}_s-\sigma_s(1+\beta_s)^{-1}\widetilde{U}_s)^-/\widetilde\gamma$, which is due (\ref{limite4eta(n)/n}) in Lemma \ref{EtaProcess}-(d), and $\zeta\sigma/\widetilde\gamma>0$, we deduce that 
\begin{equation*}
\begin{split}
\widetilde\gamma_s\widetilde{Z}_s-\sigma_s(1+\beta_s)^{-1}\widetilde{U}_s\geq0 \quad P\otimes {dt}-a.e..
\end{split}
\end{equation*}
For the second part of $g^{(2)}_n$ we have that the limit of $(e^n-1)n^{-1}(V^{(n)}_s)^+G^{-1}_{s-} (1-a_s)$ goes to infinity when $n\longrightarrow \infty$. This contradicts (\ref{eq: g^(2)/n goes to 0}). Hence, (\ref{eq: g^(2)/n goes to 0}) holds if and only if
\begin{equation}
    V_s \leq 0 \quad P \otimes dD^{p,\mathbb{F}}-a.e.
\end{equation}
As both processes $\int_0^{\cdot}g_n(t)dt$ and $\int_0^{\cdot}{g}_n^{(1)}(t)$ converge in $p$-variation-norm (i.e. ${\mathcal{A}}^p(\mathbb Q_0)$-norm), we deduce that $K_n(t):=\int_0^t {{\theta_s(0)-\theta_s(n)}\over{\widetilde\gamma_s}}\sigma_s(\widetilde\gamma_s{Z}_s^{(n)}-\sigma_s(1+\beta_s)^{-1}{U}_s^{(n)})^- a_sdA_s$ converges also in $p$-variation-norm and that $\widetilde K_n(t):= \int_0^t (e^n-1)\big(V_s^{(n)} \big)^+  G^{-1}_{s-} (1-a_s) dA_s$ also converges in $p$-variation-norm. Hence we deduce the existence of a nonnegative and adapted processes $k^{up}$ and $\widetilde k^{up}$ such that 
\begin{equation*}
\begin{split}
K^{(n)}:=\int_0^{\cdot} {{\theta_s(0)-\theta_s(n)}\over{\widetilde\gamma_s}}\sigma_s(\widetilde\gamma_s{Z}_s^{(n)}-\sigma_s(1+\beta_s)^{-1}{U}_s^{(n)})^- a_sdA_s\\
\mbox{converges in ${\mathcal{A}}^p(\mathbb Q_0)$-norm to } \int_0^{\cdot}{k}^{up}_sdA_s=:K^{up}.\end{split}
\end{equation*}
and
\begin{equation*}
\begin{split}
\widetilde K^{(n)}:=\int_0^{\cdot} (e^n-1)\big(V_s^{(n)} \big)^+ G^{-1}_{s-} (1-a_s) dA_s\quad\mbox{converges in ${\mathcal{A}}^p(\mathbb Q_0)$-norm to }\quad\int_0^{\cdot}{\widetilde k}^{up}_sdA_s=:\widetilde K^{up}.
\end{split}
\end{equation*}

Therefore, as $I_{\{\widetilde\gamma{Z}^{(n)}-\sigma(1+\beta)^{-1}{U}^{(n)}\geq\epsilon\}}$ converges $\mathbb Q_0\otimes{dt}$-almost everywhere to $I_{\{\widetilde\gamma\widetilde{Z}-\sigma(1+\beta)^{-1}\widetilde{U}\geq\epsilon\}}$ for any $\epsilon>0$, we deduce that for any $\epsilon>0$, we have 
\begin{equation*}
\begin{split}
0=\mathbb{E}_0\left[\left(I_{\{\widetilde\gamma{Z}^{(n)}-\sigma(1+\beta)^{-1}{U}^{(n)}\geq\epsilon\}}\is{K_n}\right)_{T\wedge\tau}\right]\longrightarrow \mathbb{E}_0\left[\left(I_{\{\widetilde\gamma\widetilde{Z}-\sigma(1+\beta)^{-1}\widetilde{U}\geq\epsilon\}}\is{K}^{up}\right)_{T\wedge\tau}\right].
\end{split}
\end{equation*}
This implies, when taking $\epsilon$ to zero, that 
\begin{equation*}
\begin{split}
\int_0^{T\wedge\tau} I_{\{\widetilde\gamma_s\widetilde{Z}_s-\sigma_s(1+\beta_s)^{-1}\widetilde{U}_s>0\}}d{K}^{up}_s=0\quad P\mbox{-.a.s}.\ \mbox{i.e.}\ \int_0^{T\wedge\tau} (\widetilde\gamma_s{Z}_s^{up}-\sigma_s(1+\beta_s)^{-1}{U}_s^{up})d{K}^{up}_s=0\quad P\mbox{-.a.s}..
\end{split}
\end{equation*}
Similarly,
\begin{equation*}
\begin{split}
\int_0^{T\wedge\tau} I_{\{\widetilde V_s<0\}}d{\widetilde K}^{up}_s=0\quad P\mbox{-.a.s}.\quad \mbox{or equivalently}\quad\int_0^{T\wedge\tau} {V}_s^{up}d{\widetilde K}^{up}_s=0\quad P\mbox{-.a.s}..
\end{split}
\end{equation*}
This proves simultaneously assertion (d) and the last two conditions in  (\ref{ConstrainedBSDE1}), and Part 4 is complete.\\
{\bf Part 5.} Thanks to Part 3, we consider a subsequence  $(Y^{(n)}, Z^{(n)},U^{(n)}, V^{(n)})$ that converges to $({Y}^{up}, {Z}^{up},{U}^{up}, V^{(n)})$ in both norm and $\mathbb Q_0\otimes{dt}$-a.e. and $(Z^{(n)}\is{W}^{(0)}_t,U^{(n)}\is\widetilde{ N}^{(0)}_t, V^{(n)} \is N^\mathbb{G}_t)$ converges $P$-almost surely to $(Z^{up}\is{W}^{(0)}_t,U^{up}\is\widetilde{N}^{(0)}_t, V^{up} \is N^\mathbb{G})$ for any $t\in[0,T\wedge\tau]$. Thus, by taking the limit  $\mathbb Q_0\otimes{dt}$-a.e.in the BSDE (\ref{BSDE(n)}), or equivalently in 
\begin{equation*}
\begin{split}
Y_t^{(n)}=&\xi+\int_{t\wedge\tau}^{T\wedge\tau} {{\theta_s(-n)-\theta_s(0)}\over{\widetilde\gamma_s}}\sigma_s(\widetilde\gamma_s{Z}_s^{(n)}-\sigma_s(1+\beta_s)^{-1}{U}_s^{(n)})^+ds + \int_{t\wedge\tau}^{T\wedge\tau} (1-e^{-n})\big(V_s^{(n)}\big)^- G^{-1}_{s-} dD^{p,\mathbb{F}} \\
&+  K^{(n)}_{T\wedge\tau}-K^{(n)}_{t\wedge\tau} +\widetilde K^{(n)}_{T\wedge\tau} - \widetilde K^{(n)}_{t\wedge\tau} -\int_{t\wedge\tau}^{T\wedge\tau} Z_s^{(n)}{d}W^{(0)}_s-\int_{t\wedge\tau}^{T\wedge\tau}{U}_s^{(n)}d\widetilde{N}^{(0)}_s - \int_{t\wedge\tau}^{T\wedge\tau}{V}_s^{(n)}dN_s^\mathbb{G},
\end{split}\end{equation*}
we deduce that $({Y}^{up}, {Z}^{up},{U}^{up},{K}^{up})\in\mathbb{S}^p(\mathbb Q_0)\times\mathbb{L}^p(W,\mathbb Q_0) \times \mathbb{L}^p(N,\mathbb Q_0)$, fulfills indeed (\ref{ConstrainedBSDE1}) fully. \\
 The proof that this solution is the smallest solution, to the constrained BSDE (\ref{ConstrainedBSDE1}), mimics the footsteps (see the last part of part 3) of the proof of \cite[Theorem 4.6]{choulli2025second}.\end{proof}

\begin{appendices}
\section{Some technical Lemmas}

\begin{lemma}\label{Lemma4PoissonIntegrands} The following assertions hold.\\
{\rm{(a)}} Let $\theta$ be an $\mathbb{F}$-predictable process. Then $\theta\in {L}^1_{loc}(N^{\mathbb{F}},\mathbb{F})$ if and only if $\vert\theta\vert\is N\in {\mathcal{A}}_{loc}^+(\mathbb{F})$.\\
{\rm{(b)}} Let $\phi$ be an $\mathbb{F}$-optional process such that $\vert\phi\vert\is{N}\in{\mathcal{A}}^+_{loc}(\mathbb{F})$, and suppose that $\mathbb{F}$ is generated by $(W,N)$. Then we have 
$$\phi\is{N}-\left(\phi\is{N}\right)^{p,\mathbb{F}}=\phi\is{N}^{\mathbb{F}}.$$
{\rm{(c)}} Let $\mathbb{H}$ be a filtration. If $V\in{\mathcal{V}}^+_{loc}(\mathbb{H})$ and $\Delta{V}<1$, then $(1-\Delta{V})^{-1}\is{V}\in{\mathcal{V}}_{loc}^+(\mathbb{H})$.\\
{\rm{(d)}} Let $\mathbb{H}$ be a filtration. If $V\in{\mathcal{V}}^+_{loc}(\mathbb{H})$, then $(1+\Delta{V})^{-1}\is{V}\in{\mathcal{A}}_{loc}^+(\mathbb{H})$.
\end{lemma}
\begin{proof}1) Here we prove assertion (a). It is clear that  $\theta\in {L}^1_{loc}(N^{\mathbb{F}},\mathbb{F})$ if and only if $\sqrt{\theta^2\is{N}}\in {\mathcal{A}}_{loc}^+(\mathbb{F})$. Due to $\theta^2I_{\{\vert\theta\vert\leq1\}}\is Nin \in {\mathcal{A}}_{loc}^+(\mathbb{F})$, then $\theta\in {L}^1_{loc}(N^{\mathbb{F}},\mathbb{F})$ if and only if $\sqrt{\theta^2I_{\{\vert\theta\vert>1\}}\is{N}}\in {\mathcal{A}}_{loc}^+(\mathbb{F})$. Thus, put $V:=\theta^2I_{\{\vert\theta\vert>1\}}\is{N}$ which is RCLL and nondecreasing process, we get $\sqrt{V_{-}}\is N\in{\mathcal{A}}_{loc}^+(\mathbb{F})$. Hence, by combing these with 
$$\vert\theta\vert{I}_{\{\vert\theta\vert>1\}}\is N\leq \sqrt{\theta^2I_{\{\vert\theta\vert>1\}}+V_{-}}\is N\leq \vert\theta\vert{I}_{\{\vert\theta\vert>1\}}\is N+\sqrt{V_{-}}\is{N},$$ the proof of assertion (a) follows immediately. \\
2) Thanks to the predicatble representation theorem, we deduce the existence of $\psi\in  {L}^1_{loc}(N^{\mathbb{F}},\mathbb{F})$ such that $\phi\is{N}-\left(\phi\is{N}\right)^{p,\mathbb{F}}=\psi\is{N}^{\mathbb{F}}$. As $\left(\phi\is{N}\right)^{p,\mathbb{F}}$ is continuous, we get $\varphi\Delta{N}=\psi\Delta{N}$, and hence $\varphi\is{N}=\psi\is{N}$. Therefore, assertion (b) follows immediately. \\
3) Let $V\in{\mathcal{V}}^+_{loc}(\mathbb{H})$ such that $\Delta{V}<1$. On the one hand,  for any $\delta\in(0,1)$, we $(1-\Delta{V})^{-1}I_{\{\Delta{V}\leq\delta\}} \is{V}\leq (1-\delta)^{-1} V\in{\mathcal{V}}^+_{loc}(\mathbb{H})$. On the other hand, $(1-\Delta{V})^{-1}I_{\{\Delta{V}\leq\delta\}}$ takes finitely many positive values only on each compact $[0,T]$. Hence, it is clear that, that $(1-\Delta{V})^{-1}I_{\{\Delta{V}>\delta\}} \is{V}\leq (1-\delta)^{-1} V\in{\mathcal{V}}^+_{loc}(\mathbb{H})$. This proves assertion (c), while assertion (d) is obvious due to the facts that the process  $(1+\Delta{V})^{-1}\is{V}\leq V$ and hence it belongs to ${\mathcal{V}}_{loc}^+(\mathbb{H})$, and has bounded jumps. 
\end{proof}
\begin{lemma}\label{FiniteVariationG2F} Suppose $G>0$ and let $V$ be a nonnegative and $\mathbb{F}$-adapted process.\\
 Then $V^{\tau}\in{\mathcal{V}}^+(\mathbb{G})$ if and only if $V\in{\mathcal{V}}^+(\mathbb{F})$.
\end{lemma}
\begin{proof} It is very clear that $V^{\tau}\in{\mathcal{V}}^+(\mathbb{G})$  as soon as $V\in{\mathcal{V}}^+(\mathbb{F})$. Hence the remaining part of this proof focus on the converse implication. To this end, we suppose that $V^{\tau}\in{\mathcal{V}}^+(\mathbb{G})$, and put
$$ X^{\mathbb{G}}:=e^{-V^{\tau}}I_{\Lbrack0,\tau\Lbrack}=e^{-V}I_{\Lbrack0,\tau\Lbrack}.$$ Then $ X^{\mathbb{G}}$ is bounded and RCLL process. Thus, in virtue of Dellacherie/Meyer(???), $^{o,\mathbb{F}}(X^{\mathbb{G}})=Ge^{-V}$ is a RCLL process, and hence $V$ is a RCLL and $\mathbb{F}$-adapted process. To prove that $V$ is nondecreasing with finite values, we consider any pair of bounded $\mathbb{F}$-stopping times $(\sigma_1,\sigma_2)$ such that $\sigma_1\leq \sigma_2$ $P$-a.s. Due to  $V^{\tau}\in{\mathcal{V}}^+(\mathbb{G})$, we get  $V_{\sigma_1\wedge\tau}\leq V_{\sigma_2\wedge\tau}<\infty$, $P$-a.s., or equivalently 
$$ I_{\{\sigma_2\leq\tau\}}\leq I_{\{V_{\sigma_1}\leq V_{\sigma_2}<\infty\}},\quad P\mbox{-a.s.}.$$
By taking the conditional expectations with respect to ${\mathcal{F}}_{\sigma_2}$ on both sides, we deduce that 
$$ {\widetilde{G}}_{\sigma_2}\leq I_{\{V_{\sigma_1}\leq V_{\sigma_2}<\infty\}},\quad P\mbox{-a.s.}.$$
Thus, in virtue of the assumption $G>0$, the above inequality implies that $V_{\sigma_1}\leq V_{\sigma_2}<\infty$, $P$-a.s.. This ends the proof of the lemma.
\end{proof}
\section{Proofs for Lemmas \ref{Parametrization4tau}-\ref{ExponentialIntegrability4phi} and Propositions \ref{Phi-N(G)}-\ref{ExponentialSpecial4G} }\label{Proof4Propositions} 
 \begin{proof}[Proof of Lemma \ref{Parametrization4tau}] As $G>0$, then both processes $G_{-}$ and $\widetilde{G}$ are also positive, and hence $G_{-}^{-1}$ is locally bounded and  $G_{-}^{-1}\is{m}$ is an $\mathbb{F}$-;local martingale. Thanks to the predictable representation, we deduce the existence of a pair $(\alpha,\beta)$ of $\mathbb{F}$-predictable processes such that 
 $$m=m_0+\alpha{G}_{-}\is W+\beta{G}_{-}\is{N}^{\mathbb{F}}.$$
 Furthermore, due to $-G_{-}<\Delta{m}=\widetilde{G}-G_{-}=G_{-}\beta\Delta{N}$, we obtain $-1<\beta$ $P\otimes{dt}$-a.e.. This proves assertion (a). To prove assertion (b), we similarly use the predictable representation result for the pure jumps $G_{-}\is(D^{o,\mathbb{F}}-D^{p,\mathbb{F}})\in {\mathcal{M}}_{0,loc}(\mathbb{F})$, and get the existence of an $\mathbb{F}$-predictable $\eta$ such that 
 $$D^{o,\mathbb{F}}=D^{p,\mathbb{F}}+G_{-}\eta\is{N}^{\mathbb{F}}.$$
 This proves (\ref{tau params}), and as a result we derive 
 $$\Delta{D}^{o,\mathbb{F}}=\Delta{D}^{p,\mathbb{F}}+G_{-}\eta\Delta{N}.$$
 Then by combining this with $\widetilde{G}=G_{-}+\Delta{m}=G_{-}(1+\beta\Delta{N})$, (\ref{DoF decomp}) follows immediately, and the proof of the lemma is complete.
\end{proof}
 \begin{proof}[Proof of Lemma \ref{ExponentialIntegrability4phi}] This proof has three parts, where assertions (a), (b) and (c) are proved.\\
 {\bf Part 1.} Here we prove assertion (a). To this end, we remark that, in virtue of 
 $$\varphi=\ln\left(({\widetilde{G}}+GJ (\varphi))({\widetilde{G}}-J (\varphi)\Delta{D}^{o,\mathbb{F}})^{-1}\right)$$ and the fact that $\varphi\in {\mathcal{I}}^{exp}_{loc}(N^{\mathbb{G}},\mathbb{G})$ is equivalent to $\varphi\in {\mathcal{I}}^{o}_{loc}(N^{\mathbb{G}},\mathbb{G})$ and $e^{\varphi^+}{\widetilde{G}}^{-1}I_{\Rbrack0,\tau\Rbrack}\is{D}^{o,\mathbb{F}}\in {\mathcal{V}}^+(\mathbb{G})$, we deduce that the proof of assertion (a) boils down to prove 
 \begin{equation}\label{Equivalnce4(a)}
 \exp(\varphi^+){\widetilde{G}}^{-1}I_{\Rbrack0,\tau\Rbrack}\is{D}^{o,\mathbb{F}}\in {\mathcal{V}}^+(\mathbb{G})\ \mbox{ iff}\ J (\varphi)\ \mbox{is}\ N^{\mathbb{G}}\mbox{-integrable iff}\ J (\varphi)\in {\mathcal{I}}^{o}_{loc}(N^{\mathbb{G}},\mathbb{G}).
 \end{equation}
 Thus, as $D$ is a single jump process, we remark that $ J (\varphi)$ is $N^{\mathbb{G}}$-integrable iff
 \begin{equation}\label{PhiIntegrability}
 {{\vert{J} (\varphi)\vert}\over{\widetilde{G}}}I_{\Rbrack0,\tau\Rbrack}\is{D}^{o,\mathbb{F}}=\frac{{e}^{\varphi^+}-1}{G + e^{\varphi}\Delta{D}^{o,\mathbb{F}}}I_{\Rbrack0,\tau\Rbrack}\is{D}^{o,\mathbb{F}}+\frac{1-{e}^{-\varphi^-}}{G + e^{\varphi}\Delta{D}^{o,\mathbb{F}}}I_{\Rbrack0,\tau\Rbrack}\is{D}^{o,\mathbb{F}}\in{\mathcal{V}}^+(\mathbb{G}).\end{equation}
 Thanks to $0\leq 1-e^{-\varphi^-}\leq 1$,  it is obvious that $V_1:={\widetilde{G}}^{-1}(1-e^{-\varphi^-})I_{\Rbrack0,\tau\Rbrack}\is{D}^{o,\mathbb{F}}  \in{\mathcal{A}}^+(\mathbb{G})$  such that $\Delta{V}_1<1$. Hence, in virtue of Lemma \ref{Lemma4PoissonIntegrands}-(c), we get $(1-e^{-\varphi^-})(G + e^{\varphi}\Delta{D}^{o,\mathbb{F}})^{-1}I_{\Rbrack0,\tau\Rbrack}\is{D}^{o,\mathbb{F}} \in{\mathcal{V}}^+(\mathbb{G}).$ Therefore, (\ref{PhiIntegrability}) is equivalent to $V_2:=({e}^{\varphi^+}-1)(G + e^{\varphi}\Delta{D}^{o,\mathbb{F}})^{-1}I_{\Rbrack0,\tau\Rbrack}\is{D}^{o,\mathbb{F}} \in{\mathcal{V}}^+(\mathbb{G})$, or equivalently $(1-\Delta{V_2})^{-1}\is{V}_2=({e}^{\varphi^+}-1){\widetilde{G}}^{-1}I_{\Rbrack0,\tau\Rbrack}\is{D}^{o,\mathbb{F}} \in{\mathcal{V}}^+(\mathbb{G})$, due to Lemma \ref{Lemma4PoissonIntegrands}-(c) again and $\Delta{V}_2<1$. This proves the first equivalence in (\ref{Equivalnce4(a)}). To prove the second equivalence, we suppose that $J (\varphi)$ is $N^{\mathbb{G}}$-integrable and use  Lemma \ref{FiniteVariationG2F} to deduce that  $G\vert{J} (\varphi)\vert{\widetilde{G}}^{-1}\is{D}^{o,\mathbb{F}}\in{\mathcal{V}}^+(\mathbb{F})$. Then  by combining this with the fact that the jumps of $ GJ (\varphi){\widetilde{G}}^{-1}\is{D}^{o,\mathbb{F}}$ are bounded, which follows from 
 $$
 J (\varphi){{G}\over{\widetilde{G}}}\Delta{D}^{o,\mathbb{F}}=\frac{({e}^{\varphi^+}-1)G}{\widetilde{G} + (e^{\varphi^+}-1)\Delta{D}^{o,\mathbb{F}}}\Delta{D}^{o,\mathbb{F}}-\frac{(1-{e}^{-\varphi^-})G}{G + e^{\varphi}\Delta{D}^{o,\mathbb{F}}}\Delta{D}^{o,\mathbb{F}},
 $$
 we deduce that in fact  $GJ (\varphi){\widetilde{G}}^{-1}\is{D}^{o,\mathbb{F}}\in{\mathcal{A}}_{loc}(\mathbb{F})$, and hence $J (\varphi)\in{\mathcal{I}}^{o}_{loc}(N^{\mathbb{G}},\mathbb{G}) $. This proves the second equivalence, and ends the proof of assertion (a).\\
{\bf Part 2.} Hereto, we consider $\varphi\in  {\mathcal{I}}^{exp}_{loc}(N^{\mathbb{G}},\mathbb{G})$, and proves assertion (b). On the one hand, in virtue of assertion (a) and Lemma \ref{FiniteVariationG2F}, we deduce that both $\vert{J} (\varphi)\vert{\widetilde{G}}^{-1}\is D^{o,\mathbb{F}}$ and $V:=\vert\varphi\vert{\widetilde{G}}^{-1}\is D^{o,\mathbb{F}}$ belong to ${\mathcal{V}}^+(\mathbb{F})$. Hence, $\exp(V)\in{\mathcal{V}}^+(\mathbb{F}) $, and as result, we deduce that the two processes 
$$
\sum \left(\exp(\vert\varphi\vert{\widetilde{G}}^{-1}\Delta{D}^{o,\mathbb{F}})-1\right)\ \mbox{and}\  U_1:=J (\varphi){\widetilde{G}}^{-1}\is D^{o,\mathbb{F}},$$
are well defined RCLL with finite variation processes. Furthermore, as both ${\widetilde{G}}(G + e^{\varphi}\Delta{D}^{o,\mathbb{F}})^{-1}I_{\{\varphi\geq0\}}$ and  ${\widetilde{G}}\exp({\varphi\Delta{D}^{o,\mathbb{F}}/{\widetilde{G}}})(G + e^{\varphi}\Delta{D}^{o,\mathbb{F}})^{-1}I_{\{\varphi<0\}}$ are bounded, we conclude that 
$$
U_2:=\sum {{{\widetilde{G}}}\over{G + e^{\varphi}\Delta{D}^{o,\mathbb{F}}}}\left(\exp({\varphi\Delta{D}^{o,\mathbb{F}}/\widetilde{G}})-1\right)\quad\mbox{is RCLL with finite variation}.$$
Then from combining this latter fact with Lemma \ref{Lemma4PoissonIntegrands}-(c), and the two facts 
$$
\Delta{U_3}<1,\quad (1-\Delta{U}_3)^{-1}\is{U}_3-{{\varphi}\over{\widetilde{G}}}I_{\{\widetilde{G}=G\}}\is{D}^{o,\mathbb{F}} =U^{(\varphi)},\quad\mbox{where}\quad U_3:=U_1-U_2,$$
we conclude that $U^{(\varphi)}$ is RCLL with finite variation. On the other hand, by using $ D^{o,\mathbb{F}}=I_{\{G=\widetilde{G}\}}\is  D^{o,\mathbb{F}}+\sum \Delta D^{o,\mathbb{F}}$, where $I_{\{G=\widetilde{G}\}}\is  D^{o,\mathbb{F}}$ is continuous, and inserting this equality in (\ref{U(phi)}), we get 
\begin{align*}
	U^{(\varphi)} &=  \frac{e^\varphi - 1 -\varphi}{\widetilde{G}} I_{\{\widetilde{G} = G\}} \is D^{o,\mathbb{F}}+ \sum_{s\leq \cdot} \bigg[\exp\left(-\varphi \frac{\Delta D^{o,\mathbb{F}}}{\widetilde{G}}\right) \bigg(\frac{\Delta D^{o,\mathbb{F}}  }{\widetilde{G}} (e^\varphi - 1) + 1 \bigg) - 1 \bigg] .
\end{align*}
Thus, by combining this with the facts that $e^x-1-x\geq 0$ and  $\widetilde{G}\exp(\varphi\widetilde{G}^{-1}\Delta{D}^{o,\mathbb{F}})\leq e^{\varphi}\Delta{D}^{o,\mathbb{F}}+G$, due to the convexity of $e^x$, we conclude that the process $U^{(\varphi)}\in{\mathcal{V}}^+(\mathbb{F})$. Furthermore, in virtue of $\{\Delta D^{o,\mathbb{F}} \neq 0 \} = \{\Delta D^{p,\mathbb{F}} \neq 0 \} \cup \{\eta\Delta N \neq 0 \}$, where $\{\Delta D^{p,\mathbb{F}} \neq 0 \}\cap\{\eta\Delta N \neq 0 \}=\emptyset$, we obtain the equality (\ref{U(phi)}), which we recall below.
\begin{equation*}\begin{split}
	U^{(\varphi)} =&\underbrace{\left(H(\varphi)(1+\beta)^{-1} - 1\right)\is N}_{=:U_1}+ \underbrace{ (e^\varphi - 1 -\varphi){\widetilde{G}}^{-1} I_{\{\widetilde{G} = G\}} \is D^{o,\mathbb{F}}}_{=:U_2}\\
	&+\underbrace{ \sum \bigg(\exp\left(-\varphi \frac{\Delta D^{p,\mathbb{F}}}{G_{-}}\right)  \bigg(\Delta{D}^{p,\mathbb{F}} G_{-}^{-1} (e^\varphi - 1) + 1 \bigg) - 1 \bigg)}_{=:U_3}.\end{split}
\end{equation*}
Thus, it is clear that  $U_2$ is continuous as $ I_{\{\widetilde{G} = G\}} \is D^{o,\mathbb{F}}$ is continuous. Due to $\{\Delta D^{p,\mathbb{F}} = 0\} \subseteq \{\Delta U_3 = 0\}$, we get $\{\Delta U_3 \neq 0\} \subseteq \{\Delta D^{p,\mathbb{F}} \neq 0\}$ and hence the pure jump nondecreasing process $U_3$ has accessible jumps. By combining this with the fact that the filtration $\mathbb{F}$ is quasi-left-continuous (see \cite[Definition 82, Chapter IV]{DM1972} for the definition), we easily conclude that due to \cite[Theorem 83-(a), Chapter IV]{DM1972} that $U_3$ is an $\mathbb{F}$-predictable and nondecreasing process. This proves assertion (b).\\
{\bf Part 3.} This part considers  $\varphi\in {\mathcal{I}}^{exp}_{loc}(N^{\mathbb{G}},\mathbb{G})$ and proves assertion (c). Thanks to assertion (b), we claim that $U^{(\varphi)} \in{\mathcal{V}}^+(\mathbb{F})$, and hence $(1+\Delta{U}^{(\varphi)})^{-1}\is{U}^{(\varphi)}\in {\mathcal{A}}^+_{loc}(\mathbb{F})$, due to Lemma \ref{Lemma4PoissonIntegrands}-(d). Furthermore, due to assertion (b) again (see precisely  (\ref{U(phi)})), we get 
 \begin{equation*}\begin{split}
 {1\over{1+\Delta{U}^{(\varphi)}}}\is{U}^{(\varphi)}=&\left(1-{{1+\beta}\over{H(\varphi)}}\right)\is N+ (e^\varphi - 1 -\varphi){\widetilde{G}}^{-1} I_{\{\widetilde{G} = G\}} \is D^{o,\mathbb{F}}\\
	&+ \sum \bigg(1-G_{-}\exp\left(\varphi \frac{\Delta D^{p,\mathbb{F}}}{G_{-}}\right)  \bigg(\Delta{D}^{p,\mathbb{F}}(e^\varphi - 1) + G_{-} \bigg)^{-1} \bigg).\end{split}
\end{equation*}
It is clear that $ (e^\varphi - 1 -\varphi){\widetilde{G}}^{-1} I_{\{\widetilde{G} = G\}} \is D^{o,\mathbb{F}}$ is continuous, and the third process on the right-had-side of the above equality is nondecreasing $\mathbb{F}$-predictable, see the proof of assertion (b) for details. Thus, the compensator of $(1+\Delta{U}^{(\varphi)})^{-1}\is{U}^{(\varphi)}$ exits and is given by
  \begin{equation*}\begin{split}
\widetilde{U}^{(\varphi)}=&\biggl(\left(1-{{1+\beta}\over{H(\varphi)}}\right)\is N\biggr)^{p,\mathbb{F}}+ (e^\varphi - 1 -\varphi){\widetilde{G}}^{-1} I_{\{\widetilde{G} = G\}} \is D^{o,\mathbb{F}}\\
	&+ \sum \bigg(1-G_{-}\exp\left(\varphi \frac{\Delta D^{p,\mathbb{F}}}{G_{-}}\right)  \bigg(\Delta{D}^{p,\mathbb{F}}(e^\varphi - 1) + G_{-} \bigg)^{-1} \bigg).\end{split}
\end{equation*}
 Therefore, assertion (c) follows immediately from this equality and Lemma \ref{Lemma4PoissonIntegrands}-(b), and the proof of the lemma is complete.
 \end{proof}
\begin{proof}[Proof of Proposition \ref{Phi-N(G)}]  The proof is achieved in five parts.\\
{\bf Part 1.} Hereto, we prove that if a pair $(\varphi,\Phi)\in{\mathcal{I}}^{exp}_{loc}(N^{\mathbb{G}},\mathbb{G})\times{\mathcal{I}}^{o}_{loc}(N^{\mathbb{G}},\mathbb{G})$ fulfilling the first three properties in (\ref{Exponential2StochExponential}), then it satisfies the fourth property in (\ref{Exponential2StochExponential}). \\
To this end, on the one hand, we apply It\^o to $\exp(\varphi \is N^\mathbb{G})$, and get  
\begin{equation}\label{Equation3.15}
\exp(\varphi \is N^\mathbb{G})= 1 + \exp(\varphi \is N^\mathbb{G}_{-}) \bigcdot \bigg[\varphi \is N^\mathbb{G} + \sum_{s\leq \cdot} \bigg(e^{\varphi \Delta N^\mathbb{G}} - 1 - \varphi \Delta N^\mathbb{G}\bigg) \bigg].
\end{equation}
Then in virtue of ${\widetilde{G}}\Delta{N}^\mathbb{G} = {G}\Delta{D} - I_{\Rbrack 0, \tau \Lbrack}{\Delta D^{o,\mathbb{F}}} $, we calculate
\begin{align}
	 &\sum_{s\leq \cdot}  \bigg[\exp(\varphi \Delta N^{\mathbb{G}}) -1 - \varphi \Delta N^\mathbb{G}\bigg]\nonumber\\
	 & =  \sum_{s\leq \cdot} \bigg(\exp\left(\varphi \frac{G}{\widetilde{G}}\right) -1 -\varphi \frac{G}{\widetilde{G}}\bigg)\Delta D +  \sum_{s\leq \cdot}\bigg( \exp\left(-\frac{\varphi }{\widetilde{G}}\Delta D^{o,\mathbb{F}}\right)-1 +\frac{\varphi }{\widetilde{G}}\Delta D^{o,\mathbb{F}} \bigg)I_{\Rbrack 0, \tau \Lbrack}\nonumber\\
&=  \sum_{s\leq \cdot} \bigg(\exp\left(- \frac{\varphi}{\widetilde{G}}\Delta D^{o,\mathbb{F}}\right)(e^\varphi - 1) -\varphi \bigg)\Delta D +  \sum_{s\leq \cdot}\bigg( \exp\left(-\frac{\varphi}{\widetilde{G}} \Delta D^{o,\mathbb{F}}\right) -1 +\varphi \frac{\Delta D^{o,\mathbb{F}}}{\widetilde{G}} \bigg)I_{\Rbrack 0, \tau \Rbrack}\nonumber\\
&=\Phi_1 \is D +  \sum_{s\leq \cdot} \bigg( \exp\left(-\frac{\varphi }{\widetilde{G}}\Delta D^{o,\mathbb{F}}\right) -1 +\varphi \frac{\Delta D^{o,\mathbb{F}}}{\widetilde{G}} \bigg)I_{\Rbrack 0, \tau \Rbrack},\label{Equation3.16}
\end{align}
where 
\begin{equation}\label{Phi1-Phi2}
	 \Phi_1:=  \exp\left(-\varphi {\widetilde{G}}^{-1}{\Delta D^{o,\mathbb{F}}}\right)(e^\varphi - 1)  - \varphi\quad \mbox{and} \quad \Phi_2:=\Phi_1+\varphi.
\end{equation}
By inserting these with (\ref{Equation3.16}) into (\ref{Equation3.15}), we get
 \begin{equation}\label{Phi2phi}
\exp(\varphi \is N^\mathbb{G})={\mathcal{E}}\left(\Phi_2\is{D}-{{\varphi}\over{\widetilde{G}}}I_{\Rbrack 0, \tau \Rbrack}\is{D}^{o,\mathbb{F}}+ \sum \bigg( \exp\left(-\frac{\varphi }{\widetilde{G}}\Delta D^{o,\mathbb{F}}\right) -1 +\varphi \frac{\Delta D^{o,\mathbb{F}}}{\widetilde{G}} \bigg)I_{\Rbrack 0, \tau \Rbrack}\right).
 \end{equation}
On the other hand, we apply Yor's formula to ${\mathcal{E}}(\Phi\is{N}^{\mathbb{G}}){\mathcal{E}}(U^{(\varphi)})^{\tau}$, and get 
\begin{equation*}
\begin{split}
{\mathcal{E}}(\Phi\is{N}^{\mathbb{G}}){\mathcal{E}}(U^{(\varphi)})^{\tau}&={\mathcal{E}}\left((1+\Delta{U}^{(\varphi)})\Phi\is{N}^{\mathbb{G}}+(U^{(\varphi)})^{\tau}\right)\\
&={\mathcal{E}}(\Phi_2\is{N}^{\mathbb{G}}+(U^{(\varphi)})^{\tau})= {\mathcal{E}}\left(\Phi_2\is{D}-{{\Phi_2}\over{\widetilde{G}}}I_{\Rbrack 0, \tau \Rbrack}\is{D}^{o,\mathbb{F}}+(U^{(\varphi)})^{\tau}\right).
\end{split}
\end{equation*}
Therefore, by comparing this equality to (\ref{Phi2phi}), we conclude that the last property in (\ref{Exponential2StochExponential}) follows immediately. This ends the first part. \\
{\bf Part 2.} Hereto, we prove assertion (a). Let $\Phi\in  {\mathcal{I}}^{o}_{loc}(N^{\mathbb{G}},\mathbb{G})$  satisfying  (\ref{Exponential2StochExponential}). Then,  on the one hand, it is clear that the associated $\varphi$ given by the third equality in  (\ref{Exponential2StochExponential}) belongs to ${\mathcal{I}}^{o}_{loc}(N^{\mathbb{G}},\mathbb{G})$ by assumption. On the other hand, the fact that $\varphi\in{\mathcal{I}}^{exp}_{loc}(N^{\mathbb{G}},\mathbb{G})$ follows from combining 
$$
{{e^{\varphi^+}}\over{\widetilde{G}}}I_{\Rbrack0,\tau\Rbrack}\is {D}^{o,\mathbb{F}}={{I_{\Rbrack0,\tau\Rbrack}}\over{\widetilde{G}}}\is {D}^{o,\mathbb{F}}+{{\Phi^+/\widetilde{G}}\over{1-\Phi^+\Delta {D}^{o,\mathbb{F}}{\widetilde{G}}^{-1} }}I_{\Rbrack0,\tau\Rbrack}\is {D}^{o,\mathbb{F}},
 $$
 and Lemma \ref{Lemma4PoissonIntegrands}-(c) applied to $V_1:=\Phi^+{\widetilde{G}}^{-1}I_{\Rbrack0,\tau\Rbrack}\is {D}^{o,\mathbb{F}}\in{\mathcal{V}}^+(\mathbb{G})$ as $\Delta{V}_1=\Phi^+\Delta {D}^{o,\mathbb{F}}/{\widetilde{G}}I_{\Rbrack0,\tau\Rbrack}<1$ due to the second condition in (\ref{Exponential2StochExponential}). By combining this with part 1, the proof of the ``only if" part of the equivalence in assertion (a) follows immediately. To prove the reverse implication, we consider $\varphi\in  {\mathcal{I}}^{exp}_{loc}(N^{\mathbb{G}},\mathbb{G})$, and remark that $\Phi:=J (\varphi)$ defined in (\ref{H(phi)Phi(phi)}) satisfies the first three properties in (\ref{Exponential2StochExponential}). Hence, in virtue of part 1, assertion (a) will follows immediately  as soon as we prove  that $\Phi\in {\mathcal{I}}^{o}_{loc}(N^{\mathbb{G}},\mathbb{G})$. To this end, we remark that 
\begin{equation*}
{{G}\over{\widetilde{G}}}\vert{J} (\varphi)\vert={{G\vert{e}^{\varphi}-1\vert}\over{G+e^{\varphi}\Delta{D}^{o,\mathbb{F}}}}={{G({e}^{\varphi^+}-1)}\over{G+e^{\varphi}\Delta{D}^{o,\mathbb{F}}}}+{{G(1-{e}^{-\varphi^-})}\over{G+e^{\varphi}\Delta{D}^{o,\mathbb{F}}}}\leq {{e^{\varphi^+}}\over{\widetilde{G}}}+1,
\end{equation*}
which yields $G{\widetilde{G}}^{-1}J (\varphi)\is{D}^{o,\mathbb{F}}\in {\mathcal{V}}^+(\mathbb{F})$. Furthermore, the jumps of this process are bounded due to 
$$
{{G}\over{\widetilde{G}}}J (\varphi)\Delta{D}^{o,\mathbb{F}}=G{{e^{\varphi}\Delta{D}^{o,\mathbb{F}}}\over{G+e^{\varphi}\Delta{D}^{o,\mathbb{F}}}}-{{G}\over{G+e^{\varphi}\Delta{D}^{o,\mathbb{F}}}}\Delta{D}^{o,\mathbb{F}}.$$
This proves that $G\widetilde{G}^{-1}J (\varphi)\is{D}^{o,\mathbb{F}}\in {\mathcal{A}}_{loc}(\mathbb{F})$, and this implies that $J (\varphi)\in {\mathcal{I}}^{o}_{loc}(N^{\mathbb{G}},\mathbb{G})$. \\
{\bf Part 3.} In this part, we consider $\varphi\in{\mathcal{I}}^{exp}_{loc}(N^{\mathbb{G}},\mathbb{G})$ and prove assertion (b). Thanks to assertion (a),  we derive 
$$
\exp(\varphi\is{D})=\exp(\varphi\is{N}^{\mathbb{G}})\exp\left({{\varphi}\over{\widetilde{G}}}I_{\Rbrack0,\tau\Rbrack}\is {D}^{o,\mathbb{F}}\right)={\mathcal{E}}(\Phi\is{N}^{\mathbb{G}}){\mathcal{E}}(U^{(\varphi)})^{\tau}\exp\left({{\varphi}\over{\widetilde{G}}}I_{\Rbrack0,\tau\Rbrack}\is {D}^{o,\mathbb{F}}\right). 
$$
Thus, in virtue of this latter equality, we remark that assertion (b) is equivalent to
\begin{equation}\label{assertionb}
{\widetilde{\mathcal{E}}}^{\tau}\exp\left(-{{\varphi}\over{\widetilde{G}}}I_{\Rbrack0,\tau\Rbrack}\is {D}^{o,\mathbb{F}}\right)={\mathcal{E}}(U^{(\varphi)})^{\tau}({\widetilde{\mathcal{E}}}^{(\varphi)})^{\tau}.
\end{equation}
Thus, on the one hand, by applying It\^o to $\exp\left(-\varphi{\widetilde{G}}^{-1}I_{\Rbrack0,\tau\Rbrack}\is {D}^{o,\mathbb{F}}\right)$ and using Yor's formula (i.e. ${\mathcal{E}}(X){\mathcal{E}}(Y)={\mathcal{E}}(X+Y+[X,Y])$ for any pair of semimartingales $X$ and $Y$) afterwards, we derive 
\begin{equation}\label{Equa4asserionb}
\begin{split}
&{\widetilde{\mathcal{E}}}^{\tau}\exp\left(-{{\varphi}\over{\widetilde{G}}}I_{\Rbrack0,\tau\Rbrack}\is {D}^{o,\mathbb{F}}\right)\\
&={\widetilde{\mathcal{E}}}^{\tau}{\mathcal{E}}\left(-{{\varphi}\over{\widetilde{G}}}I_{\Rbrack0,\tau\Rbrack}\is {D}^{o,\mathbb{F}}+\sum\left(\exp(-{{\varphi}\over{\widetilde{G}}}\Delta{D}^{o,\mathbb{F}})-1+{{\varphi}\over{\widetilde{G}}}\Delta{D}^{o,\mathbb{F}}\right)I_{\Rbrack0,\tau\Rbrack}\right)\\
&={\mathcal{E}}\left(-{{\varphi+\exp(-\varphi{\widetilde{G}}^{-1}\Delta{D}^{o,\mathbb{F}})}\over{\widetilde{G}}}I_{\Rbrack0,\tau\Rbrack}\is {D}^{o,\mathbb{F}}+\sum\left(\exp(-{{\varphi}\over{\widetilde{G}}}\Delta{D}^{o,\mathbb{F}})-1+{{\varphi}\over{\widetilde{G}}}\Delta{D}^{o,\mathbb{F}}\right)I_{\Rbrack0,\tau\Rbrack}\right).\end{split}
\end{equation}
On the other hand, we use Yor's formula again and get 
\begin{equation*}
\begin{split}
{\mathcal{E}}(U^{(\varphi)})^{\tau}({\widetilde{\mathcal{E}}}^{(\varphi)})^{\tau}={\mathcal{E}}\left(U^{(\varphi)}-{{e^{\varphi}\exp(-\varphi{\widetilde{G}}^{-1}\Delta{D}^{o,\mathbb{F}})}\over{\widetilde{G}}}\is {D}^{o,\mathbb{F}}\right)^{\tau}.
\end{split}
\end{equation*}
Then (\ref{Equality4(c)and(d)}) follows immediately from combining the above equality, (\ref{U(phi)}) and (\ref{Equa4asserionb}). \\
{\bf Part 4.} Herein, we consider a pair $(\theta,\ \psi)\in\Theta(S,\mathbb{F})$ satisfying $\int_0^{\cdot}\exp(\theta_s\gamma_s+\gamma^2_s\psi_s)ds\in{\mathcal{A}}_{loc}^+(\mathbb{F})$ and prove assertion (c). Thus, we start by applying It\^o to  $Z:=\exp\left( \theta \is X+ \sum\psi(\Delta{X})^2\right) $ and we derive
\begin{equation*}\label{Ito4Z}
\begin{split}
&Z_{-}^{-1}dZ\\
&=\theta{d}X+ \psi{d}\sum(\Delta{X})^2+{1\over{2}}\theta^2\sigma^2 dt+d\sum\left(\exp(\theta\Delta{X}+ \psi(\Delta{X})^2)-1-\theta\Delta{X}-\psi(\Delta{X})^2\right),\\
&=\theta{d}X+{1\over{2}}\theta^2\sigma^2 dt+d\sum\left(\exp(\theta\Delta{X}+ \psi(\Delta{X})^2)-1-\theta\Delta{X}\right),\\
&=\theta\sigma{d}W+\theta\gamma{d}N^{\mathbb{F}}+\theta\mu{d}t+{1\over{2}}\theta^2\sigma^2 dt+\left(e^{\theta\gamma+\gamma^2\psi}-1-\theta\gamma\right)dN\\
&=\theta\sigma{d}W+\left(\exp(\theta\gamma+\gamma^2\psi)-1\right)dN+\left(\theta\mu+{1\over{2}}\theta^2\sigma^2 -\lambda\theta\gamma\right)dt.
\end{split}
\end{equation*}
 Then by combining this latter equality with Theorem \ref{Toperator}-(a) and $N^{\tau}=(1+\beta)\is{\mathcal{T}}(N^{\mathbb{F}})+(1+\beta)\lambda\is{t}$ which is a direct consequence of Theorem \ref{Toperator}-(a), we obtain
 \begin{equation*}\label{Ito4Z}
\begin{split}
Z_{-}^{-1}dZ^{\tau}&=\theta\sigma{d}{\mathcal{T}}(W)+\left(\exp(\theta\gamma+\gamma^2\psi)-1\right)(1+\beta)d{\mathcal{T}}(N^{\mathbb{F}})\\
&+\Biggl(\theta\mu+{1\over{2}}\theta^2\sigma^2 -\lambda\theta\gamma+\alpha\sigma\theta+\lambda(1+\beta)\left(\exp(\theta\gamma+\gamma^2\psi)-1\right)\Biggr)d(t\wedge\tau).\\
&=dM^{(\theta,\psi)}+\kappa{d}(t\wedge\tau).
\end{split}
\end{equation*}
  Here $M^{(\theta,\psi)}$ and $\kappa$ are defined in (\ref{ThetaTildePsiTilde}). This proves the first equality in (\ref{theta-X-Exponenetial}), while the second equality is direct consequence of \cite[Proposition 3.3] {choulli2022explicit}. This ends the proof of  assertion (c).\\
{\bf Part 5:} Here, we consider $(\theta,\ \psi,\varphi)\in\Theta(S,\mathbb{F})\times {\mathcal{I}}^{exp}_{loc}(N^{\mathbb{G}},\mathbb{G})$ such that $\int_0^{\cdot}\exp(\theta_s\gamma_s+\gamma^2_s\psi_s)ds\in{\mathcal{A}}_{loc}^+(\mathbb{F})$, and we prove assertion (d). Then, in virtue of assertions (b) and (c) and Lemma \ref{ExponentialIntegrability4phi}-(b), we calculate 
\begin{equation*}\label{Ito4Z}
\begin{split}
(1+\Delta{M}^{(\theta,\psi)})\is({U}^{(\varphi)})^{\tau}=U_2^{\tau}+U_3^{\tau}+\exp(\theta\gamma+\gamma^2\psi)\left({{H(\varphi)}\over{1+\beta}}-1\right)\is{N}^{\tau},
\end{split}
\end{equation*}
where $U_2$ and $U_3$ are $\mathbb{F}$-predictable. Therefore, $(1+\Delta{M}^{(\theta,\psi)})\is({U}^{(\varphi)})^{\tau}\in {\mathcal{A}}^+_{loc}(\mathbb{G})$ if and only if $\exp(\theta\gamma+\gamma^2\psi)\left(H(\varphi)(1+\beta)^{-1}-1\right)\is{N}^{\tau}\in {\mathcal{A}}^+_{loc}(\mathbb{G})$, or equivalently
$$
\biggl(e^{\theta\gamma+\gamma^2\psi}\left({{H(\varphi)}\over{1+\beta}}-1\right)\is{N}^{\tau}\biggr)^{p,\mathbb{F}}=\lambda\int_0^{\cdot}G_{s-}\exp(\theta_s\gamma_s+\gamma^2_s\psi_s)\left(H_s(\varphi)-1-\beta_s\right)ds\in{\mathcal{A}}^+_{loc}(\mathbb{F}).$$
Finally this latter fact is equivalent to  (\ref{IntegrabilityCondition4Esscher}) due to $G_{-}(1+\beta)\leq 1$ and $\int_0^{\cdot}\exp(\theta_s\gamma_s+\gamma^2_s\psi_s)ds\in{\mathcal{A}}_{loc}^+(\mathbb{F})$. This proves the first claim of assertion (d). To prove the second claim, we assume that   (\ref{IntegrabilityCondition4Esscher}) holds. Then remark that 
\begin{equation*}
J (\varphi){{H(\varphi)}\over{1+\beta}}\Delta{N}\exp(\gamma\theta+\gamma^2\psi)=\exp\left(\theta\gamma+\gamma^2\psi\right){H(\varphi)}{{e^{\varphi}-1}\over{1+\beta-\eta+e^{\varphi}\eta}}\Delta{N}\in{\mathcal{I}}^{o}_{loc}(N^{\mathbb{G}},\mathbb{G})\end{equation*}
 is equivalent to 
\begin{equation*}
U_4:=e^{\theta\gamma+\gamma^2\psi}{H(\varphi)}{{e^{\varphi}-1}\over{1+\beta-\eta+e^{\varphi}\eta}}\Delta{N}{{G}\over{\widetilde{G}}}\is{D}^{o,\mathbb{F}}\in{\mathcal{A}}_{loc}(\mathbb{F}).\end{equation*}
To prove this latter fact, we decompose $\mbox{Var}(U_4)$ as follows
\begin{equation*}
\begin{split}
&\mbox{Var}(U_4)\\
&=G_{-}e^{\theta\gamma+\gamma^2\psi}{H(\varphi)}{{\vert{e}^{\varphi}-1\vert(1+\beta-\eta)}\over{1+\beta-\eta+e^{\varphi}\eta}}{{\eta}\over{1+\beta}}\is{N}\\
&=G_{-}e^{\theta\gamma+\gamma^2\psi}{H(\varphi)}\Biggl({{\eta(e^{\varphi^+}-1)}\over{1+\beta+\eta(e^{\varphi^+}-1)}}\left(1-{{\eta}\over{1+\beta}}\right)+{{1+\beta-\eta}\over{1+\beta-\eta+e^{\varphi}\eta}}{{\eta}\over{1+\beta}}(1-e^{-\varphi^-})\Biggr)\is{N}\\
&\leq e^{\theta\gamma+\gamma^2\psi}{H(\varphi)}\is{N}.
\end{split}
\end{equation*}
Thus, due to (\ref{IntegrabilityCondition4Esscher}) which is equivalent to $\exp(\theta\gamma+\gamma^2\psi){H(\varphi)}\is{N}\in {\mathcal{A}}^+_{oc}(\mathbb{F})$, we deduce that $\mbox{Var}(U_4)$ belongs to ${\mathcal{A}}^+_{oc}(\mathbb{F})$ and  the proof of the second claim follows immediately. This proves assertion (d), and ends the proof of the proposition.
\end{proof}
  \begin{proof}[Proof of Proposition \ref{ExponentialSpecial4G} ] Let  $(\theta,\ \psi,\varphi)\in\Theta(S,\mathbb{F})\times {\mathcal{I}}^{exp}_{loc}(N^{\mathbb{G}},\mathbb{G})$  and put 
  \begin{equation*}
  Y:=\theta\is{X}+\sum\psi(\Delta{X})^2,\quad\mbox{and}\quad{L}:=\exp(Y^{\tau}+\varphi\is{N}^{\mathbb{G}}).\end{equation*} 
  Then, thanks to It\^o, we derive   
  \begin{equation*}
  \begin{split}
  dL&=\exp(Y)d\exp(\varphi\is{N}^{\mathbb{G}})+\exp(\varphi\is{N}^{\mathbb{G}}_{-})d\exp(Y^{\tau}),\\
  &=\exp(Y){\mathcal{E}}(U^{(\varphi)})d{\mathcal{E}}(\Phi\is{N}^{\mathbb{G}})+\exp(Y){\mathcal{E}}_{-}(\Phi\is{N}^{\mathbb{G}}){\mathcal{E}}_{-}(U^{(\varphi)})d(U^{(\varphi)})^{\tau}\\
  &+L_{-}d\left(Y^{\tau}-\sum\Delta{Y}^{\tau}I_{\{\vert\Delta{Y}\vert>1\}}+{1\over{2}}\langle{Y}^c\rangle^{\tau}+\sum(e^{\Delta{Y}}-1-\Delta{Y}I_{\{\vert\Delta{Y}\vert\leq1\}})I_{\Rbrack0,\tau\Rbrack}\right),\\
   &=L_{-}\exp(\Delta{Y})(1+\Delta{U}^{(\varphi)})\Phi{d}{N}^{\mathbb{G}}+L_{-}\exp(\Delta{Y})d(U^{(\varphi)})^{\tau}\\
  &+L_{-}d\left(Y^{\tau}-\sum\Delta{Y}^{\tau}I_{\{\vert\Delta{Y}\vert>1\}}+{1\over{2}}\langle{Y}^c\rangle^{\tau}+\sum(e^{\Delta{Y}}-1-\Delta{Y}I_{\{\vert\Delta{Y}\vert\leq1\}})I_{\Rbrack0,\tau\Rbrack}\right).\\
  \end{split}
  \end{equation*}
  Therefore, we deduce that $L$ is a special $\mathbb{G}$-semimartingale if and only if 
 \begin{equation}\label{Special1}
  Y'':=e^{\Delta{Y}}(1+\Delta{U}^{(\varphi)})\Phi\is{N}^{\mathbb{G}}+\underbrace{e^{\Delta{Y}}\is(U^{(\varphi)})^{\tau}+\sum(e^{\Delta{Y}}-1-\Delta{Y}I_{\{\vert\Delta{Y}\vert\leq1\}})I_{\Rbrack0,\tau\Rbrack}}_{=:Y'}\in{\mathcal{A}}_{loc}(\mathbb{G}).
  \end{equation}
  Now, we will prove that the above fact is equivalent to both $e^{\Delta{Y}}(1+\Delta{U}^{(\varphi)})\Phi\is{N}^{\mathbb{G}}$ and $Y'$ being special $\mathbb{G}$-semimartingale, or equivalently they both belong to ${\mathcal{A}}_{loc}(\mathbb{G})$. Indeed, we denote by $\Gamma_n:=\{ e^{\Delta{Y}}(1+\Delta{U}^{(\varphi)})\leq n\}$, and we consider a sequence of $\mathbb{F}$-stopping times $(T_k)_k$ which increases to infinity  and  $(Y'')^{T_k}\in{\mathcal{A}}(\mathbb{G})$and $(\Phi\is{N}^{\mathbb{G}})^{T_k}\in{\mathcal{M}}(\mathbb{G})$. Then  $(I_{\Gamma_n}\is{Y''})^{T_k}\in{\mathcal{A}}(\mathbb{G})$, and hence
   \begin{equation*}
  \begin{split}
   E\left[ (I_{\Gamma_n}\is{Y'})_{T_k}\right]=E\left[ (I_{\Gamma_n}\is{Y''})_{T_k}\right]\leq  E\left[ (I_{\Gamma_n}\is\mbox{Var}(Y''))_{T_k}\right]\leq  E\left[\mbox{Var}(Y'')_{T_k}\right]<\infty.
   \end{split}  \end{equation*}
   This is equivalent to 
   $$ E\left[ (I_{\Gamma_n}e^{\Delta{Y}}\is(U^{(\varphi)})^{\tau})_{T_k}+ \sum(e^{\Delta{Y}}-1)I_{\{\Delta{Y}>1\}}I_{\Gamma_n}I_{\Rbrack0,\tau\Rbrack}\right]\leq C_k.$$
   Then by Fatou, we deduce immediately that $Y'\in{\mathcal{A}}_{loc}(\mathbb{G})$ and hence also $Y''-Y'\in{\mathcal{A}}_{loc}(\mathbb{G})$. This proves that $L$ is a special  $\mathbb{G}$-semimartingale if and only if 
   \begin{equation}\label{special2}
   \begin{split}
   e^{\Delta{Y}}(1+\Delta{U}^{(\varphi)})\Phi\in{\mathcal{I}}^{o}_{loc}(N^{\mathbb{G}},\mathbb{G}),\ \sum(e^{\Delta{Y}^{\tau}}-1)I_{\{\Delta{Y}>1\}}\in {\mathcal{A}}^+_{loc}(\mathbb{G}),\ e^{\Delta{Y}}\is(U^{(\varphi)})^{\tau}\in  {\mathcal{A}}^+_{loc}(\mathbb{G}).
   \end{split}
   \end{equation}
   On the one hand, it is easy to check that
      \begin{equation}\label{special3}
   \sum(e^{\Delta{Y}^{\tau}}-1)I_{\{\Delta{Y}>1\}}\in {\mathcal{A}}^+_{loc}(\mathbb{G}) \quad\mbox{if and only if} \int_0^{\cdot}e^{\theta_s\gamma_s+\gamma_s^2\psi_s}(1+\beta_s)ds\in{\mathcal{A}}^+_{loc}(\mathbb{F}).
      \end{equation}
On the other hand, we use Lemma \ref{ExponentialIntegrability4phi}-(b)  (see (\ref{Jumps4U(phi)})) and derive 
$$e^{\Delta{Y}}\is(U^{(\varphi)})^{\tau}=\exp(\theta\gamma+\gamma^2\psi)\left({{H(\varphi)}\over{1+\beta}}-1\right)
\is{N}^{\tau}+U_2^{\tau}+U_3^{\tau}.$$
This proves that 
    \begin{equation*}\label{special4}
   e^{\Delta{Y}}\is(U^{(\varphi)})^{\tau}\in {\mathcal{A}}^+_{loc}(\mathbb{G}) \quad\mbox{if and only if} \int_0^{\cdot}\exp(\theta\gamma+\gamma^2\psi)(H(\varphi)-1-\beta)ds\in{\mathcal{A}}^+_{loc}(\mathbb{F}).
      \end{equation*}
 Therefore, by combining this with (\ref{special3}) and the fact that $H(\varphi)\geq 1+\beta$, we conclude that 
       \begin{equation}\label{special4}
     \mbox{$e^{\Delta{Y}}\is(U^{(\varphi)})^{\tau}$ and  $\sum{e}^{\Delta{Y}^{\tau}}I_{\{\Delta{Y}>1\}}$ belong to}\ {\mathcal{A}}^+_{loc}(\mathbb{G})  \quad\mbox{iff} \int_0^{\cdot}e^{\theta_s\gamma_s+\gamma^2_s\psi_s}H_s(\varphi)ds\in{\mathcal{A}}^+_{loc}(\mathbb{F}).
          \end{equation}   
Again, thanks to Lemma \ref{ExponentialIntegrability4phi}-(b)  (see (\ref{Jumps4U(phi)})), we have 
 \begin{equation*}
  \begin{split}
 e^{\Delta{Y}}(1+\Delta{U}^{(\varphi)})\Phi&=e^{\Delta{Y}}(1+\Delta{U}_3)\Phi+e^{\Delta{Y}}\left({{H(\varphi)}\over{1+\beta}}-1\right)\Delta{N}\Phi\\
 &=(e^{\Delta{Y}}-1)(1+\Delta{U}_3)\Phi+(1+\Delta{U}_3)\Phi+e^{\Delta{Y}}\left({{H(\varphi)}\over{1+\beta}}-1\right)\Delta{N}\Phi\\
 &=(e^{\Delta{Y}}-1)\Delta{N}\Phi+(1+\Delta{U}_3)\Phi+e^{\Delta{Y}}\left({{H(\varphi)}\over{1+\beta}}-1\right)\Delta{N}\Phi\\
 &=(1+\Delta{U}_3-\Delta{N})\Phi+e^{\Delta{Y}}{{H(\varphi)}\over{1+\beta}}\Delta{N}\Phi.
   \end{split}  \end{equation*}
  In virtue of Proposition  \ref{Phi-N(G)}-(a), this proves that 
    \begin{equation}\label{special5}
       e^{\Delta{Y}}(1+\Delta{U}^{(\varphi)})\Phi\in{\mathcal{I}}^{o}_{loc}(N^{\mathbb{G}},\mathbb{G})\quad\mbox{if and only if}\quad e^{\Delta{Y}}{{H(\varphi)}\over{1+\beta}}\Delta{N}\Phi\in{\mathcal{I}}^{o}_{loc}(N^{\mathbb{G}},\mathbb{G}).
            \end{equation}
It is clear that the latter fact always hold  under the condition in the right-hand-side of (\ref{special4}), which is exactly  (\ref{IntegrabilityCondition4Esscher}), and this due to Proposition \ref{Phi-N(G)}-(d) (see (\ref{Integrability4BigPhi})). 
            Thus, by combining this resulting fact with (\ref{special5}), (\ref{special4}) and (\ref{special2}), the proof of the proposition follows immediately.
  \end{proof}

\end{appendices}


\end{document}